\documentclass[a4paper,11pt]{report}

\usepackage[left=2.5cm,right=2.5cm,top=2.5cm,bottom=2.5cm]{geometry}
\usepackage[pdftex]{graphicx}
\usepackage{amssymb, amstext, amsmath}
\usepackage{verbatim}
\newcommand{\Ha}{Hamiltonian}
\newcommand{\EoM}{equation of motion}

\newcommand{\HO}{harmonic oscillator}
\newcommand{\FP}{Fokker-Planck}
\newcommand{\CF}{continued fraction}
\newcommand{\Eq}{Equation}

\newcommand{\coupl}{c} 										
\newcommand{\CouHa}{V_{\rm{int}}}  						
\newcommand{\intCouHa}{\tilde{V}_{\rm{int}}} 				
\newcommand{\TranN}{W} 										
\newcommand{\TranUn}{W} 									
\newcommand{\Wbef}{w} 										
\newcommand{\iu}{\mathrm{i}} 									
\newcommand{\totHa}{H_{\rm{tot}} }  							
\newcommand{\sysHa}{H_{\rm{sys}}}  						
\newcommand{\bathHa}{H_{\rm{bath}}} 						
\newcommand{\bathCou}{B}

\begin{document}

\title{\textbf{STUDY OF CLASSICAL AND QUANTUM OPEN SYSTEMS}}

\author{Lee Chee Kong 
\\ \vspace{3cm} Supervisors: A/P Gong Jiangbin and Prof. Wang Jian-Sheng }
\date{\textit{Department of Physics\\ National University of Singapore}}

\pagenumbering{roman}
\maketitle
\tableofcontents

\begin{abstract}
This thesis covers various aspects of open systems in classical and
quantum mechanics. In the first part, we deal with classical
systems. The bath-of-oscillators formalism is used to describe an open
system, and the phenomenological Langevin equation is recovered. The
Fokker-Planck equation is derived from its corresponding Langevin
equation. The Fokker-Planck equation for a particle in a periodic
potential in the high-friction limit is solved using the
continued-fraction method. The equilibrium and time-dependent
solutions are obtained. Under strong periodic driving, we observe
significant non-linear effects in the dynamical hysteresis
loops. Shapiro steps appear in the time-average of the drift velocity
curves. Similar study is carried out for a dipole in an electric
field.

In the second part of the thesis, we begin the study of open quantum
systems by re-deriving the quantum master equation using perturbation
theory. The master equation is then applied to the bath-of-oscillators
model. The subtleties and approximations of the master equation are
discussed. We then use the master equation to solve the damped
harmonic oscillator. The equilibrium solution coincides with the
canonical distribution. The steady state response to DC and AC forces
is also studied. Driven systems are more challenging in quantum open
systems, and we manage to solve the quantum master equation with the
continued-fraction method. We obtain the frequency-dependent
susceptibility curves, which exhibit typical absorption and dispersion
profiles.

\end{abstract}

\pagenumbering{arabic}

\chapter*{Acknowledgments}
I thank Jos\'e Garc\'ia-Palacios for guiding me throughout this
project. I also thank  A/P Gong Jiangbin and Prof.~Wang Jian-Sheng for
being my supervisors. Above all, I thank my mum for her sacrifice to
make my higher education possible.

\chapter{Introduction}
Most coffee lovers have the frustrating experience of having their coffees cooled down before they could finish them. High school physics tells us that it is due to the heat transferred to the surroundings. In the language of statistical mechanics, it is because of the interaction with the environment. Every object, big or small, classical or quantum, is subject to this interaction. 
\\

The focus of this thesis is to study the effects of the environment on the statics and dynamics of our systems. The effects are two-fold: fluctuation and dissipation. Think of the pollen grains in water as observed by Robert Brown. Their trajectories exhibit random behavior due to the collisions with the water molecules. This randomness makes us unable to make exact predictions of the system evolution, we can only talk about its statistical properties. During the collisions, some of the pollen grains' momentum is transferred to the medium, causing them to lose energy.  Due of this dissipative process, the system can relax to a stationary state. These random and dissipative effects apply to any open systems, and we will address them in both classical and quantum regimes.
\\

This thesis is organized as follows. We deal with classical open systems in Chapter 2 to Chapter 4, while Chapter 5 and Chapter 6 are devoted to the study of open quantum systems. In Chapter 2, we start from a microscopic viewpoint  of classical open systems, and introduce two equivalent approaches to study them: the Langevin equation (trajectory) and the \FP\ equation (distribution). We then solve the \FP\ equation of a particle in periodic potential, and study its steady-state properties in time independent and time-dependent fields (Chapter 3). Similar study is done for a classical dipole in Chapter 4. 
\\

 In Chapter 5, we start the exploration of quantum open systems by discussing the reduced description of the open systems. Then using perturbation theory, we present a concise derivation of the so-called master equation: the \EoM\ of the reduced density matrix. In Chapter 6, we use the master equation to study a damped quantum \HO\ and make comparison with exact results when available.  
 \\
 
\begin{figure}[h!] 
  \begin{center}
    \includegraphics[width=4.5in]{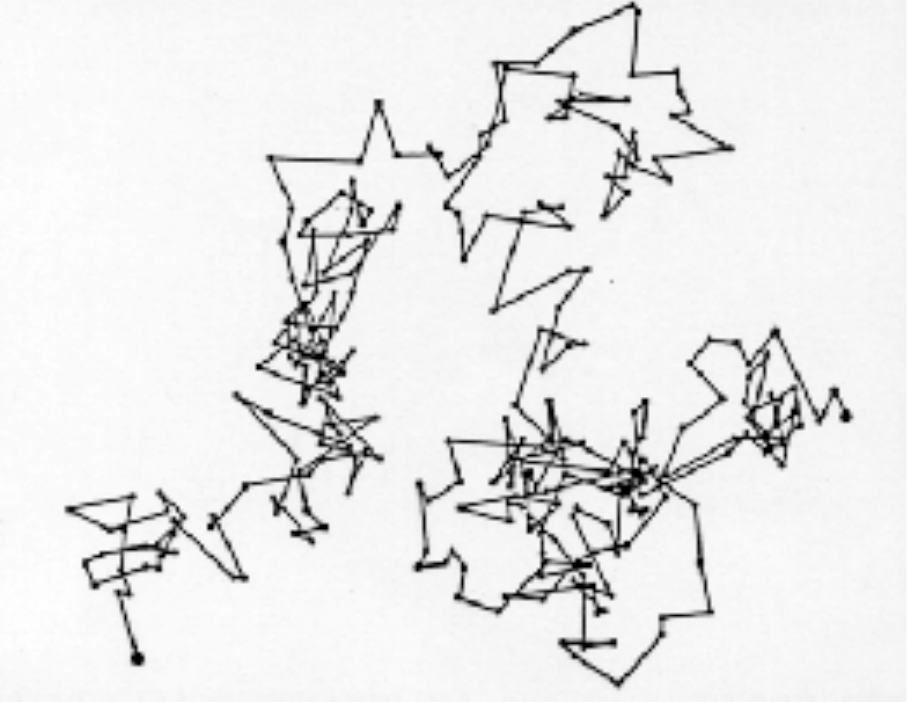}
    \caption{Trajectory of a Brownian particle (after J. P. Perrin, figure taken from http://cytoquant.com/41611/41667.html).}
  \end{center}
  \label{fig-label}
\end{figure} 

\chapter{Classical Open Systems: Langevin and Fokker-Planck Equations}

\section{Introduction}
In 1908, the French physicist Paul Langevin proposed a modified version of the Newton equation to describe the dissipative and random behavior of a Brownian particle \cite{coffey}. Consider a particle in one dimension for notational simplicity, the phenomenological \EoM\ reads
\begin{eqnarray}
	M\ddot{x}+\gamma M \dot{x}+V'(x)=\xi(t). \label{langevin}
\end{eqnarray}
The term $\gamma M \dot{x}$ causes dissipation, and $\gamma$ is the damping rate. The random force $\xi(t)$ originates from the impacts with the fluid molecules. It is assumed that the random force is Gaussian distributed,  and thus fully described by the first two moments:
\begin{eqnarray}
	\langle \xi(t)\rangle=0,\qquad       \langle \xi(t) \xi(t')\rangle=2 M \gamma k_{B} T \,\delta(t-t').
\end{eqnarray}
\Eq\ (\ref{langevin}) has a time-local frictional term, thus assuming the friction does not depend on the velocity of the past. In many cases of interest, the bath has a finite memory time, and the coresponding equation of motion is 
\begin{eqnarray}
	  M\ddot{x}+ M \int^t_{-\infty} \gamma(t-t')\,\dot{x}(t')\,\,dt' +V'(x)=\xi(t),
\end{eqnarray}
which is called the generalized Langevin equation. The damping rate is replaced by a memory function $\gamma(t)$, which comes from the finite noise correlation time $ \langle \xi(t) \xi(t')\rangle=2 M  k_{B} T \gamma(t-t')$.
\\

There are problems with these phenomenological equations: they are not time reversal-invariant and therefore contradict Newton's reversible laws. Furthermore, we cannot carry forward this formalism into the quantum regime since we do not know how to quantize a dissipative equation. In the next section, we will start from a ``microscopic'' point of view to describe fluctuations and dissipation using a time-honored framework in classical physics--- \Ha\ mechanics.  


\section{Bath-of-Oscillators Formalism}  

Here we will derive the Langevin equation by considering the \Ha\ equations of the  global system, system plus bath (from now on we will call the system of interest the ``system", the surroundings the ``bath").
\\

To obtain the equations of motion, we need the \Ha\
 \begin{eqnarray}
	 \totHa= \sysHa+ \bathHa + \CouHa.
\end{eqnarray}
This total \Ha\ $\totHa$ consists of  the system \Ha\ $\sysHa$, the bath \Ha\ $\bathHa$ and the interaction term $\CouHa\,$. We need to give content to the bath and coupling \Ha s.
\\

Following Zwanzig \cite{zwanzig} and others \cite{leggett-tunneling, leggett-QBM, weiss}, the bath is modeled as a set of \HO s.  The replacement of the true bath by a set of \HO  s is effectively equivalent to  the assumption that the coupling is weak such that the bath is only slightly perturbed away from its equilibrium configuration \cite{leggett-tunneling}. By the same argument, it is reasonable to assume that the system-bath coupling is linear with respect to the bath coordinates. We write the \Ha s as 
 \begin{eqnarray}
 	 \sysHa		&=& \frac{p^2}{2M}+V(x), \\
	 \bathHa		&=&	\sum_{\alpha=1}^{N}\Big( \frac{p_{\alpha}^2}{2m_{\alpha}}+
	 						\frac{1}{2}m_{\alpha}\omega_\alpha^2 x_{\alpha}^2\Big),  \\
	\CouHa		&=&	-F(x, p)\sum_{\alpha=1}^{N} \coupl_\alpha x_\alpha  + \Delta V(x,p),	
\end{eqnarray}
\noindent where $\alpha$ denotes the bath modes (which can be continuous), $\coupl_\alpha$ the coupling constants.  $F(x,p)$ can be any function of the system's coordinate and momentum $(x, p)$. It is worth noting that though the influence of the system on the bath is small, the opposite might not be true. 

\subsection{Counter-Term}
An additional term $\Delta V $ is added to the coupling \Ha\ to compensate the re-normalization caused by the term $Fx_\alpha$ \cite{ingold0}. In other words, we want to ensure that the global minimum of the \Ha\ is determined by the  bare potential $V(x)$ alone. To look for the minimum with respect to the bath, we need 
 \begin{eqnarray}
 	 \frac{\partial \totHa}{\partial x_\alpha}  = m_\alpha \omega_\alpha^2 x_\alpha -\coupl_\alpha F=0,
\end{eqnarray}	 
and obtain $x_\alpha=\frac{\coupl_\alpha}{m_\alpha \omega_\alpha^2 } F$. 
Using this result, we find the minimum with respect to the system coordinate 
 \begin{eqnarray}
 	 \frac{\partial \totHa}{\partial x}  = \frac{\partial V}{\partial x} -																						 														\sum_{\alpha=1}^{N}\frac{\coupl_\alpha^2}{m_\alpha \omega_\alpha^2 }  																					\frac{\partial F}{\partial x}F+\frac{\partial \Delta V}{\partial x}.
\end{eqnarray}
In order to satisfy $\frac{\partial \totHa}{\partial x}  = \frac{\partial V}{\partial x}$, we need the counter-term
 \begin{eqnarray}
 	\Delta V(x)= 	\sum_{\alpha=1}^{N}\frac{\coupl_\alpha^2}{2m_\alpha \omega_\alpha^2 } F^2.
\end{eqnarray}
\subsection{Bath-of-Oscillators Hamiltonian}
Finally gathering all the above results, we can write the total \Ha\ in the form
 \begin{eqnarray}
 	\totHa		&=&  \sysHa + \bathHa + \CouHa\,\,  \label{standard_form}
 \\ \nonumber
			     &=&   \sysHa +\sum_{\alpha=1}^{N} \Big[  \frac{p_{\alpha}^2}{2m_{\alpha}} + 	
				\frac{1}{2}m_\alpha \omega_\alpha^2\Big(x_\alpha -\frac{\coupl_\alpha}{m_\alpha \omega_\alpha^2}F \Big)^2   \Big].\end{eqnarray}
This model is frequently called the Caldeira-Leggett model in the context of quantum dissipative systems \cite{leggett-tunneling, leggett-QBM}.  This framework can also be used to study problems involving spin, in which the coupling is a function of spin variable, $F=F(\mathbf{s})$. In the next section, we will show that the \Ha\ (\ref{standard_form}) describes dissipation and fluctuations in the system.

\section{The Equation of Motion (Langevin) }
\subsection{Derivation}
Consider a dynamical variable $A(x,p)$ that only depends on the system's coordinate and momentum, the \Ha\ equations of $A(x,p)$, $x_\alpha$ and $p_\alpha$ are given by
\begin{eqnarray}
		&&\frac{dA}{dt}=\{A, \sysHa \}+ \sum_{\alpha=1}^{N} \frac{\coupl_\alpha^2}{2m_\alpha \omega_\alpha^2}\{A, F^2\}-	
						      	\sum_{\alpha=1}^{N}\coupl_\alpha x_\alpha\{A, F\}, \label{EoM-HO}\\
		&&\frac{dx_\alpha}{dt}= \frac{p_\alpha}{m_\alpha},      \quad \quad    \frac{dp_\alpha}{dt}=-m_\alpha \omega_\alpha^2x_\alpha +\coupl_\alpha F, 
\end{eqnarray}
where the Poisson bracket is
\begin{eqnarray}
	\{A,B\}=\frac{\partial A}{\partial x}\frac{\partial B}{\partial p} -\frac{\partial A}{\partial p}\frac{\partial B}{\partial x}. 
\end{eqnarray}
The solution to the bath coordinate is that of a forced \HO\ 
\begin{eqnarray}
		x_\alpha(t)=x_\alpha^h(t) +  \frac{\coupl_\alpha }{m_\alpha  \omega_\alpha}\int^t_{t_0}ds \sin[\omega_\alpha(t-s)]F(s),					\label{forced-HO}
\end{eqnarray}
where the homogeneous part is the free \HO\ evolution
\begin{eqnarray}
		x_\alpha^h(t)=x_\alpha(t_0)\cos [\omega_\alpha(t-t_0)]+
						\frac{p_\alpha(t_0)}{m_\alpha  \omega_\alpha}\sin[\omega_\alpha(t-t_0)].\label{free-HO}
\end{eqnarray}

Substituting Eq.~(\ref{forced-HO}) and Eq.~(\ref{free-HO}) into Eq.~(\ref{EoM-HO}) and performing integration by parts, we obtain
\begin{eqnarray}
		\frac{dA}{dt}=\{A, \sysHa \} - \{A, F \}\Big[\xi(t) -M\, \int^t_{t_0} ds \,\,\gamma(t-s)\frac{dF(s)}{ds} \Big], \label{EoM2}
\end{eqnarray}
where
\begin{eqnarray}
		\xi(t)             &=&  \sum_{\alpha=1}^{N} \coupl_\alpha \Big[x^h_\alpha(t)-
					             \frac{\coupl_\alpha}{m_\alpha  \omega_\alpha^2}F(t_0)\cos[\omega_\alpha(t-t_0)] \Big],\label{noise} \\
		\gamma(t) &=&   \frac{1}{M}\sum_{\alpha=1}^{N} \frac{\coupl_\alpha^2}{m_\alpha  \omega_\alpha^2}\cos(\omega_\alpha t).
\end{eqnarray}
We have obtained an equation similar to the Langevin equation for a general dynamic variable $A(x,p)$. The first term in Eq.~(\ref{EoM2}) is the free evolution. The second term gives rise to the fluctuations. The integral term keeps the memory of the previous states and causes dissipation. We will justify the interpretation of fluctuations and dissipation in the next subsection.

\subsection{Fluctuation and Dissipation}
\subsubsection*{Fluctuation}
\begin{figure}[b!] 
  \begin{center}
    \includegraphics[width=4.0in]{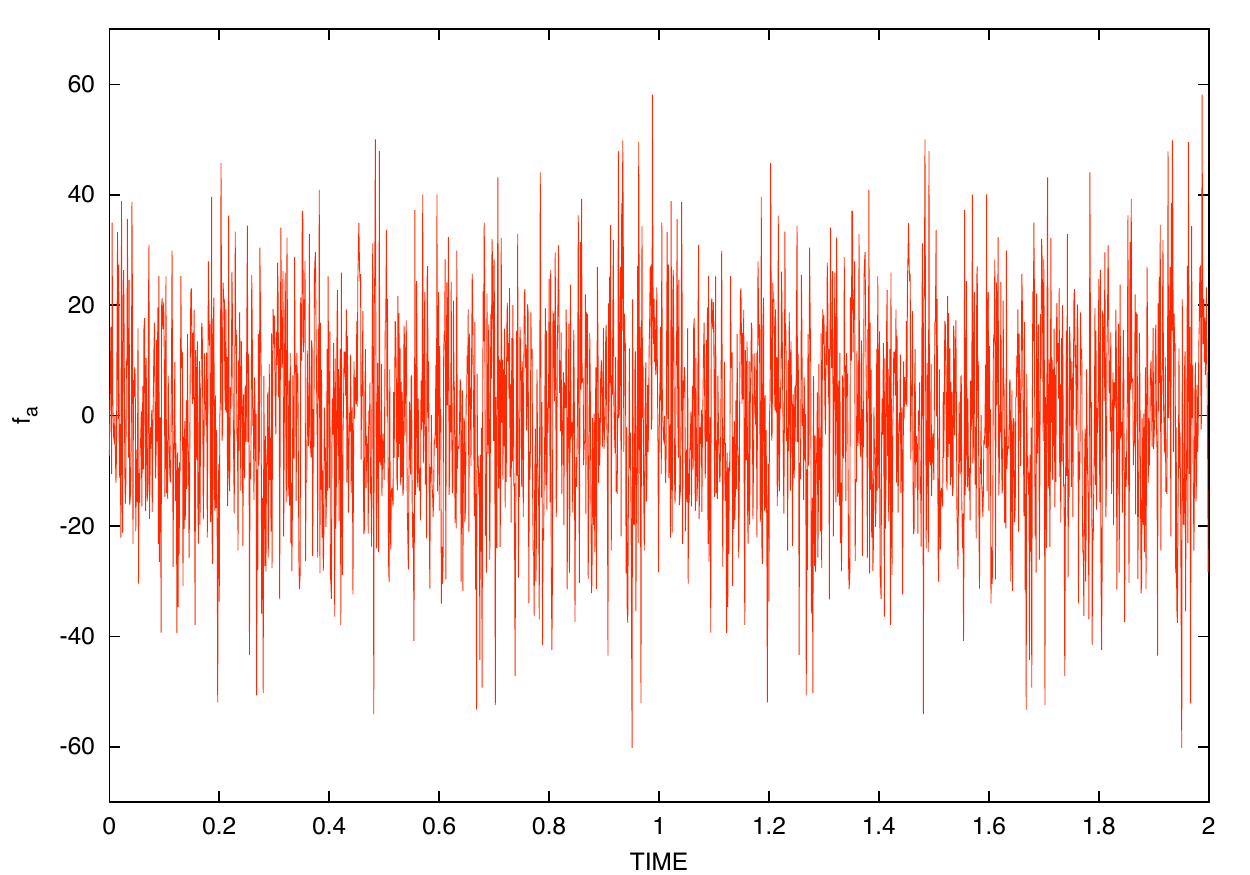}
    \caption{Computer simulation of the noise term, $\xi(t)$. It is obtained from summing (\ref{noise}) over 1000 oscillators having the canonical distribution as initial conditions. It does indeed look random.}
        \label{c2noise}
  \end{center}
\end{figure} 
The term $\xi(t)$ contains the free evolution of the bath oscillators and the initial state of the system. Being the sum of many terms with different frequencies and phases, it behaves like a random force (see Figure \ref{c2noise}). Let us assume the initial distribution of the bath follows the classical canonical distribution
\begin{eqnarray}
		\rho_{\rm{bath}}(t_0)= Z^{-1}\exp \Big\{-\beta\sum_{\alpha=1}^{N}
								\Big[ \frac{p_{\alpha}(t_0)^2}{2m_{\alpha}}+
	 						      \frac{1}{2}m_{\alpha}\omega_\alpha^2 \Big(x_{\alpha}(t_0)- \frac{\coupl_\alpha}{m_\alpha  \omega_\alpha^2}F(t_0)\Big)^2\Big]  \Big\}.
\end{eqnarray}
Drawing $x_{\alpha}(t_0)$ and $p_{\alpha}(t_0)$ from the canonical distribution, $\xi(t)$ becomes a random force with Gaussian distribution. Then it is fully characterized by the first two moments
\begin{eqnarray}
	\langle \xi(t)\rangle=0,\qquad       \langle \xi(t) \xi(t')\rangle=K(t-t'), \label{c2_noise}
\end{eqnarray}
where the correlator is related to the damping kernel as $K(t)=2 M\,k_B T \gamma(t)$.

\subsubsection*{Dissipation}
If we particularize the \EoM\ to the system Hamiltonian, the free evolution is zero and we are left with
\begin{eqnarray}
		\frac{d\sysHa }{dt} = - \{\sysHa, F \} \,\xi(t) + \{\sysHa, F \} \int^t_{t_0} ds \,\,\gamma(t-s)\frac{dF(s)}{ds}.
\end{eqnarray}
The first term averages to zero, loosely speaking. Otherwise, we assume $T=0$, where there is no thermal fluctuation and the first term is automatically zero. We are then left with the integral. For simplicity, we consider a free particle $\sysHa=p^2/2M$ with coordinate coupling $F=x$, and delta damping kernel $\gamma(t) = \gamma \,\delta(t)$, one finds
\begin{eqnarray}
		\frac{d\sysHa}{dt} \propto - \gamma \, p^2.
\end{eqnarray}
The negative rate of change indicates that dissipation indeed takes place. Eq. (\ref{c2_noise}) relates the noise or fluctuating force to the dissipation, and is known as the fluctuation-dissipation theorem.

\subsubsection*{Spectral Density}
It is convenient to introduce the spectral density
\begin{eqnarray}
	J(\omega)=\frac{\pi}{2} \sum_{\alpha=1}^{N} \frac{\coupl_\alpha^2}{m_\alpha  \omega_\alpha}\delta(\omega -\omega_\alpha).\label{spectral}
\end{eqnarray}
The correlator (\ref{c2_noise}) and the damping kernel can then be written as 
\begin{eqnarray}
	K(t)				&=&2k_BT\int^\infty_0\frac{d\omega}{\pi}\frac{J(\omega)}{\omega}\cos(\omega t); \label{damping_kernel} \\
	\gamma(t)		&=& \frac{1}{M} \int^\infty_0\frac{d\omega}{\pi} \frac{J(\omega)}{\omega} \cos(\omega t). \label{damping_kernel2}
\end{eqnarray}
For a bath with discrete modes, the spectral density is a set of delta peaks. However in a dissipative bath, the eigenfrequencies $\omega_\alpha$ form a continuum and $J(\omega)$ becomes a smooth function of $\omega$. 
\\

All the effects of the bath are incorporated into $J(\omega)$, which involves the frequencies and couplings. It these are not fully known, one proceeds to ``model" the bath, assuming different functional dependences. In the next section, we will discuss a system (Rubin model) where $J(\omega)$ can be explicitly computed, and this will provide us insights how to properly do such modeling.
\subsection{Examples: Brownian Particle and Spin}
In this section, we study the Langevin equations (\ref{EoM2}) for a Brownian particle (translational motion) and a Brownian spin (rotational motion).

\subsubsection*{Brownian Particle}
Consider a particle with Hamiltonian $\sysHa=p^2/2M+V(x)$, with its coordinate ($F=x$) coupled to the bath. Equation (\ref{EoM2}) for $A=x$ and $A=p\,$ then gives
\begin{eqnarray} 
		\frac{dx}{dt} =  p/M; \qquad   
		\frac{dp}{dt} = -V'(x)+\xi(t)-\int^t_{t_0}ds\,\gamma(t-s)\, p(s),
\end{eqnarray}
which is exactly the generalized Langevin equation introduced phenomenologically at the beginning of this chapter. \\

If the bath is Ohmic (Markovian limit), namely $J(\omega)=M\gamma \omega$, the damping kernel becomes a delta function and we recover the usual Langevin equation
\begin{eqnarray} 
		\frac{dp}{dt} = -V'(x)+\xi(t)-\gamma\, p,\label{c2_langevin}
\end{eqnarray}
with the bath correlator 
\begin{eqnarray} 
		\langle \xi(t) \xi(t')\rangle=2M\gamma k_B T \, \delta(t-t'). 
\end{eqnarray}

\subsubsection*{Brownian Spin}
Equation (\ref{EoM2}) is also valid for a spin with \Ha\ $\sysHa=\sysHa(s_x, s_y, s_z)$\footnote{ 
The underlaying canonical variables are $x\equiv \varphi $ and $p \equiv s_z$ \cite{gar2000}. But our formalism does not depend on the canonical variables, as we derive the \EoM\ for any $A(x,p)$ and we can set $A=s_i$ for $i=x,\,y,\,z$.  
}. Using the spectral density $J(\omega)=\lambda \omega$, we write directly the memoryless \EoM\ for $\mathbf{s}$
\begin{eqnarray} 
		\frac{d\mathbf{s}}{dt} &=& \{\mathbf{s}, \sysHa \} +  \{\mathbf{s}, F \}\Big[\xi(t)+ \lambda \frac{dF}{dt} \Big] \\ \nonumber
									&=&  \mathbf{s}\times \mathbf{B}_{\rm{eff}} + \mathbf{s}\times \boldsymbol{\zeta}(t)
											- \lambda \mathbf{s} \times \hat{\mathbf{A}} \frac{d \mathbf{s} }{dt},
\end{eqnarray}
where
\begin{eqnarray} 
		\mathbf{B}_{\rm{eff}}		&=& 	-\frac{\partial  \sysHa}{\partial  \mathbf{s}};\\
		\boldsymbol{\zeta}(t)		&=& 	\frac{\partial F}{\partial \mathbf{s}}\xi(t); \\
		 \hat{\mathbf{A}}			&=&	 \Big(\frac{\partial F}{\partial  \mathbf{s}} \Big)^{\dagger} \Big(\frac{\partial F} {\partial  \mathbf{s}} \Big). 
\end{eqnarray}
This is the Lagenvin equation for a spin. The first term arises from the free Hamiltonian and causes the spin to precess around the field direction. For example, the \Ha\ of a spin $\bf{s}$ in a magnetic field $\mathbf{B}$ is $\sysHa=-\mathbf{s}\cdot\mathbf{B}$ and $\mathbf{B}_{\rm{eff}}=\mathbf{B}$. The second and third terms describe the fluctuation and dissipation as discussed before. The damping term is a generalization of the phenomenological equations proposed by Gilbert and Landau-Lifshitz \cite{gar2000}. 
%
\subsection{Rubin Model \cite{weiss,ingold}}
Here we look at an instructive example where the oscillator-bath model represents the actual Hamiltonian. In the Rubin model (see Figure 2.2), a heavy particle of mass $M$ and coordinate $x$ is bilinearly coupled to a half infinite chain of \HO s with mass $m$ and spring constant $f=\frac{m \omega_R^2}{4}$. The total \Ha\ is 
\begin{eqnarray}
		\totHa=\frac{p^2}{2M}+V(x)+\sum_{n=1}^{\infty}\Big[ \frac{p_n^2}{2m} +\frac{f}{2}(x_{n+1}-x_n)^2 \Big]
				  +\frac{f}{2}(x-x_1)^2.
\end{eqnarray}
\begin{figure}[h!] 
  \begin{center}
    \includegraphics[width=5.5in]{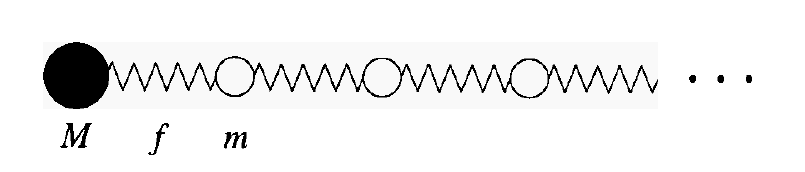}
    \caption{Pictorial Sketch of the Rubin model, figure taken from \cite{weiss}.}
  \end{center}
      \label{c2-Rubin}
\end{figure} 
To cast the above \Ha\ into the standard form (\ref{standard_form}), we make the following transformation to normal modes $X(k)$
\begin{eqnarray}
		x_n =\sqrt{2/\pi}\int^\pi_{0}dk \sin(n k) X(k).	
\end{eqnarray}
In normal mode representation, the \Ha\ reads
\begin{eqnarray}
		\totHa=\frac{p^2}{2M}+V(x)+\frac{f}{2}x^2 +
				   \int^\pi_0 dk\,\Big(  \frac{P^2(k)}{2m}+\frac{m}{2}\omega^2(k) X^2(k) - x\,\, c(k)X(k)     \Big).
\end{eqnarray}
The eigenfrequency $\omega(k)$ and the coupling function $c(k)$ are 
\begin{eqnarray}
		\omega(k)=\omega_R |\sin(k/2)|; \quad \quad c(k)=\sqrt{\frac{2}{\pi}}\,\frac{m\omega_R^2}{4}\,\sin(k).
\end{eqnarray}
Comparing with Eq.~(\ref{spectral}), the spectral density is found to be 
\begin{eqnarray}
		J(\omega)= \frac{\pi}{2}\int^\pi_0 dk\,\frac{c^2(k)}{m\omega(k)}\delta[\omega-\omega(k)]
					 =\frac{m\omega_R }{2} \omega\Big( 1-\frac{\omega^2}{\omega_R^2}\Big)^{1/2}\Theta(\omega_R-\omega),
\end{eqnarray}
where $\Theta(\omega)$ is the Heaviside step function. The frequency $\omega_R$ is the highest frequency in the bath and cuts off the spectral density $J(\omega)$ (see Figure 2.3). In fact, in any physical system, there always exists a cut-off frequency such that the contribution at high frequencies is suppressed. With the above spectral density, the damping kernel Eq.~(\ref{damping_kernel}) becomes
\begin{eqnarray}
		\gamma(t)=\frac{m}{2M}\omega_R\frac{J_1(\omega_R t )}{t},
\end{eqnarray}
where $J_1(\omega_R t )$ is the first order Bessel function. The memory time in the kernel is of the order of $1/\omega_R$. For large $\omega_R$ (corresponds to a stiff spring), the kernel becomes a sharply peaked function around zero (see Figure 2.4).
\begin{figure}[h!]
  \begin{center}
    \includegraphics[width=7.0in]{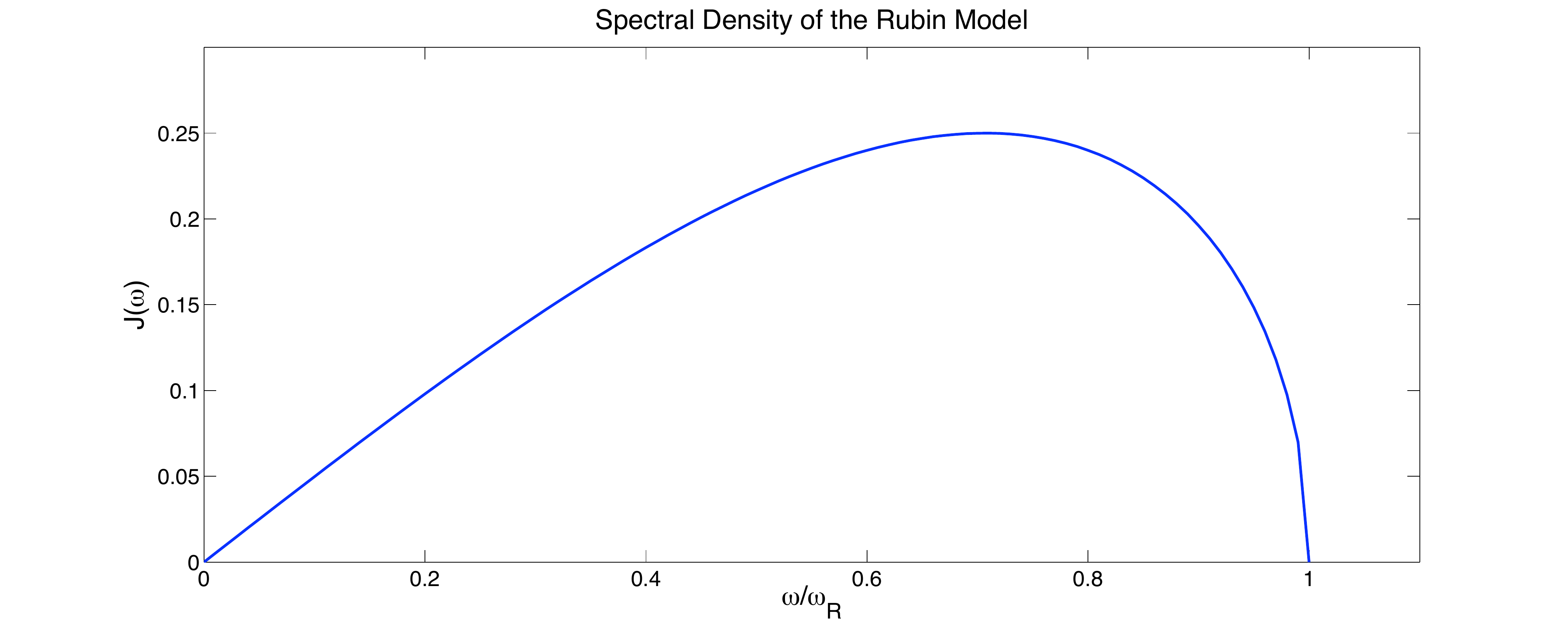}                   
      \caption{The spectral density of the Rubin model, $m\omega_R^2=1$.}
 \end{center}
 \label{fig-c2-rubin-spectral}
\end{figure} 
\begin{figure}[h!] 
  \begin{center}
    \includegraphics[width=5.0in]{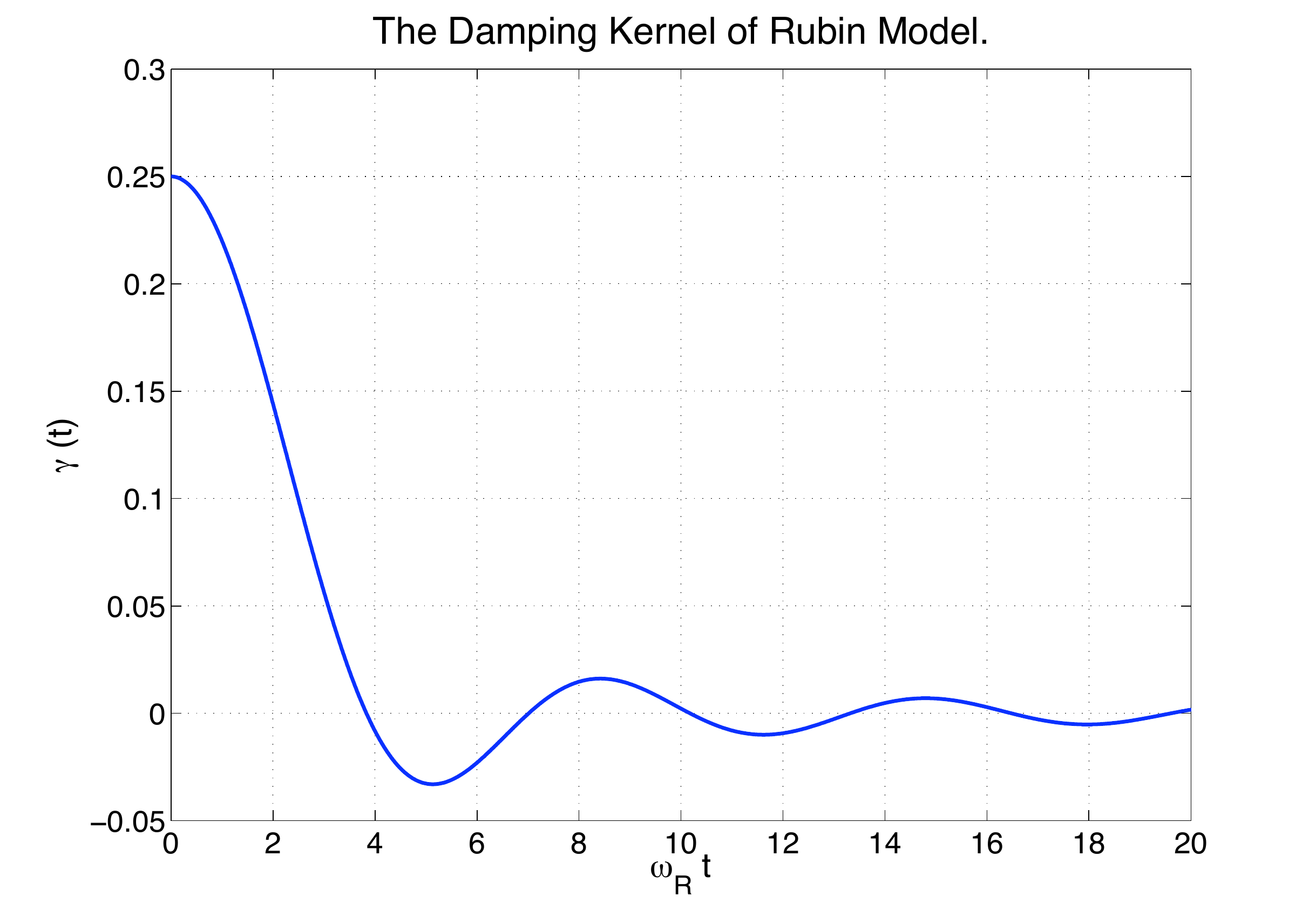}  
    \caption{The damping kernel of the Rubin model, $m=M=1$. If $\omega_R$ is large, the curve is sharply peaked.}
            \label{fig-c2-rubin-kernel}
   \end{center}
\end{figure}

The Rubin model not only gives us a concrete example where the oscillators represent a true bath, it provides us useful insights about the properties that are common to any physical bath. The continuum limit of the spectral density comes from the infinite number of degrees of freedom of the chain of oscillators. There is a natural cut-off to the spectral density at high frequency. We also see that, at large $\omega_R$, we approach the Ohmic limit [Markovian, $J(\omega)\propto \omega $], the damping kernel becomes short-lived and the harmonic chain exhibits little retardation.  

\section{\FP\ Equations}
We have studied the trajectory approach to the open systems based on the Langevin equation. Here we will look at the phase space distribution function and its evolution equation, the \FP\ equation. Both approaches are equivalent, provided the noise is delta correlated and has a Gaussian distribution; we will see why soon.

\subsection{Derivation \`a la Zwanzig \cite{zwanzig}}
To derive the \FP\ equation, we start with a general Langevin equation of a set of variables, $\mathbf{a}=\{a_j\}$, the \EoM\   in vector form is
\begin{eqnarray}
		\frac{d\mathbf{a}}{dt}=\mathbf{v}(\mathbf{a}) +\boldsymbol{\xi}(t), \label{Langevin-vector}
\end{eqnarray}
where the noise term $\boldsymbol{\xi}(t)$ is Gaussian distributed and has the following properties
\begin{eqnarray}
		\langle \boldsymbol{\xi}(t)\rangle=0, \qquad      
		 \langle\boldsymbol{\xi}(t) \boldsymbol{\xi}(t')^{\rm{T}}\rangle=2\,\mathbf{\hat{B}}\,\delta(t-t').
\end{eqnarray}

We are interested in the probability distribution of the dynamical variables, $\Wbef(\mathbf{a}, t)$. The normalization condition requires 
\begin{eqnarray}
		\int d\mathbf{a} \,\Wbef(\mathbf{a}, t) =1, \qquad \mbox{for all $t$.}
\end{eqnarray}
Similar to the conservation law in electromagnetism\footnote{
In electromagnetism, we have the continuity equation: $\nabla  \cdot \mathbf{J} + \frac{\partial \rho}{\partial t}=0$, where $\mathbf{J}$ and $\rho$ are  the current density and charge density respectively.}, we have the continuity equation
\begin{eqnarray}
		\frac{\partial \Wbef}{\partial t}+\frac{\partial }{\partial \mathbf{a}}\cdot \Big( \frac{d\mathbf{a}}{d t}\Wbef \Big)=0.
\end{eqnarray}
Replacing $ \frac{d \mathbf{a}}{d t}$ by Eq.~(\ref{Langevin-vector}), one has
\begin{eqnarray}
		\frac{\partial\Wbef}{\partial t}  &=& \mathcal {L}\Wbef \\
								  &=&- \frac{\partial }{\partial \mathbf{a}}\cdot \Big( \mathbf{v}(\mathbf{a})\Wbef+ \boldsymbol{\xi}(t)\Wbef \Big).
\end{eqnarray}
It is still a stochastic differential equation since it contains the noise. We are interested in the noise average of $\Wbef(\mathbf{a}, t)$. We denote the noiseless part of the operator
\begin{eqnarray}
		\mathcal {L}_0\Phi =-  \frac{\partial }{\partial \mathbf{a}}\cdot \Big( \mathbf{v}(\mathbf{a})\Phi \Big),
\end{eqnarray}
so that 
\begin{eqnarray}
		\frac{\partial \Wbef}{\partial t}  &=& \mathcal {L}_0 \Wbef-  \frac{\partial }{\partial \mathbf{a}}\cdot  \boldsymbol{\xi}(t)\Wbef. \label{EoM2-W}
\end{eqnarray}

The formal solution to the differential equation is (after setting initial time $t_0=0$) 
\begin{eqnarray}
		\Wbef(\mathbf{a},t)= \mbox{e}^{t\mathcal{L}_0}\Wbef(\mathbf{a},0)- 
					\int^t_0  ds \, \mbox{e}^{(t-s)\mathcal{L}_0} \frac{\partial }{\partial \mathbf{a}}\cdot \boldsymbol{\xi}(s)\Wbef(\mathbf{a},s).
\end{eqnarray}
Note that $\Wbef(\mathbf{a},t)$ only depends on the noise $\xi(s)$ at earlier time $s<t$. Substituting the above equation into  Eq. (\ref{EoM2-W}), we obtain
\begin{eqnarray}
		\frac{\partial \Wbef}{\partial t}  &=& \mathcal{L}_0 \Wbef(\mathbf{a},0)- 
							\frac{\partial }{\partial \mathbf{a}}\cdot \boldsymbol{\xi}(t)\mbox{e}^{t\mathcal{L}_0}\Wbef(\mathbf{a},0)\\ \nonumber
				&&+ \, \frac{\partial }{\partial \mathbf{a}}\cdot \boldsymbol{\xi}(t) 
				\int^t_0  ds \, \mbox{e}^{(t-s)\mathcal{L}_0} \frac{\partial }{\partial \mathbf{a}}\cdot \boldsymbol{\xi}(s)\Wbef(\mathbf{a},s).
\end{eqnarray}

Now we take the average over noise. The second term averages to zero. Since the noise is Gaussian, we can express any moments in terms of the first two moments. The last term contains two explicit noise factors, $\boldsymbol{\xi}(s)$ and $\boldsymbol{\xi}(t)$, and also the implicit noise factors in $\Wbef(\mathbf{a},s)$. Since $\Wbef(\mathbf{a},s)$ only depends on the noise at earlier time $s'<s$ ; the pairing with either  $\boldsymbol{\xi}(s)$ or $\boldsymbol{\xi}(t)$  gives zero contribution as the noise is delta correlated.  We only need to consider the pairing  $\langle\boldsymbol{\xi}(t) \boldsymbol{\xi}(s)\rangle=2 \,\mathbf{\hat{B}} \, \delta(t-s)$. Denoting the noise average of the distribution as $ \langle \Wbef(\mathbf{a},t)\rangle =W(\mathbf{a},t)$, the result is the \FP\ equation
\begin{eqnarray}
		\frac{\partial   W(\mathbf{a},t)}{\partial t}  = 
		-\frac{\partial }{\partial \mathbf{a}}\cdot 
		\Big[ \mathbf{v}(\mathbf{a})  W(\mathbf{a},t)\Big]
		+\frac{\partial }{\partial \mathbf{a}}\cdot \mathbf {\hat{B}}\cdot \frac{\partial }{\partial \mathbf{a}} W(\mathbf{a},t). \label{FP_eqn}
\end{eqnarray}
The first term on the right hand side is what one has in the absence of noise. The effect of the noise is introduced by the second term. This term has a diffusion structure, with a second order derivative, as in the standard diffusion equations. 
\subsection{Applications to Brownian Particle and Spin}
 As in the previous section, we will see the examples of a Brownian particle and a Brownian spin. 
 \subsubsection*{Brownian Particle}
 According to the recipe above, we can write down the \FP\ equation for a Brownian particle from its Langevin equation (\ref{c2_langevin}). The quantities that enter into the general \FP\ equation (\ref{FP_eqn}) are
\[\textbf{a}= \left( \begin{array}{ccccccc}
         x    \\
         p    
\end{array} \right);
\quad
\boldsymbol{\xi}(t)= \left( \begin{array}{ccccccc}
         0    \\
         \xi(t)    \\
\end{array} \right);
\quad
\textbf{v}(\textbf{a})= \left( \begin{array}{ccccccc}  
          p/M    \\
          -V'(x)- \gamma p    \\
\end{array} \right);
\quad
\mathbf{\hat{B}}= \left( \begin{array}{ccccccc}  
          0  &   0   \\
         0   & M \gamma k_B T    \\
 \end{array} \right).\] 
 
\noindent The resulting \FP\ equation is called the Klein-Kramers equation \cite{risken}
\begin{eqnarray} 
		\frac{\partial W}{\partial t}=\Big[-\frac{p}{M}\frac{\partial}{\partial x}+V'(x)\frac{\partial}{\partial p}+
											 \gamma \frac{\partial}{\partial p} \Big( p +M k_B T\frac{\partial}{\partial p} \Big) \Big]W. \label{c2-klein-kramers}
\end{eqnarray}

The first two terms arise from the Liouville operator $\{ \sysHa \,, W\}$, while the last two terms capture the effects of the interaction with the bath: damping and diffusion.
\\

 \subsubsection*{Brownian Spin}
The corresponding \FP\ equation for a Brownian spin is \cite{gar2000}
\begin{eqnarray} 
			\frac{\partial W( \mathbf{s},t)}{\partial t} = -\frac{\partial}{\partial \mathbf{s}} \cdot \Big\{ \mathbf{s} \times  \mathbf{B}_{\rm{eff}} 
			-\lambda  \mathbf{s}\times \hat{\mathbf{A}} \Big[  \mathbf{s} \times (\mathbf{B}_{\rm{eff}} -
			 k_BT \frac{\partial}{\partial \mathbf{s}})  \Big]
			\Big\}W( \mathbf{s},t).
\end{eqnarray}
We will return to this type of orientational diffusion equation in Chapter 4 when we study the Debye dipole.
%
\subsection{Solving the \FP\ Equations \cite{risken}}
In most cases, the \FP\ equation is not solvable analytically, and we have to resort to numerical methods. Here we will discuss the numerical method employed in this thesis. Let us consider a one-variable case; we can express the distribution function $W(x,t)$ in terms of an appropriate set of basis function $\{p_n(x) \}$ 
\begin{eqnarray}
		W(x,t)=\sum_{n}W_n(t) p_n(x). \label{W_basis}
\end{eqnarray}
The sum depends on the choice of basis function. Once we solve for the coefficients $W_n(t)$, we have full knowledge of the non-equilibrium distribution function. 
\\

The use of Eq. (\ref{W_basis}) casts the \FP\ equation into a set of recurrence relations 
\begin{eqnarray}
		\dot{W_n}= \cdots \,\,Q^{--}_n W_{n-2}+Q^{-}_n W_{n-1}+Q_n W_{n}+Q^{+}_n W_{n+1}+Q^{++}_n W_{n+2}
						\,\, \cdots\,\, ,
\end{eqnarray}
where Q's are some known constants, and we seek for short-ranged coupling. We will only encounter 3-term recurrence relations in this thesis, namely
\begin{eqnarray}
		\dot{W_n}=Q^{-}_n W_{n-1}+Q_n W_{n}+Q^{+}_n W_{n+1}.
\end{eqnarray}
This type of recurrence relations can be solved efficiently using the \CF\ method. The details of the \CF\ method are discussed in Appendix A.1.

\subsubsection*{\FP\ versus Langevin}
Though it is possible to run Langevin simulations to obtain the average of a dynamical variable, solving \FP\ equations with the \CF\ method requires much shorter computational time (few minutes on a laptop). The drawback is that we do not have any information about the trajectories. Though this drawback is offset by the fact that the distribution can also provide us valuable physical insights. 
\\

\section{Summary}
It is a long chapter, let us summarize what has been presented. We first described fluctuations and dissipation in an open system by modeling the bath as a set of harmonic oscillators. Using Hamiltonian mechanics, a Langevin-like equation of motion was obtained. Particularizing to the problems of particle and spin, we recovered the phenomenological Langevin equations. The example of Rubin Model provided us useful insights of the bath-of-oscillators model. 
\\

We then derived the \FP\ equation by making use of the continuity equation for the probability distribution. The \FP\ equations for particle and spin were obtained from their corresponding Langevin equations. Eventually, the use of the \CF\ method in solving \FP\ equations was discussed. In the next two chapters, we will demonstrate the use of this method in solving the \FP\ equations for a particle in a periodic potential and a dipole (spin), both in the large damping limit.

\chapter{ Translational Brownian Motion: Particle in a Periodic Potential}

\section{Introduction}

In this chapter, we apply the \CF\ method to solve the \FP\ equation for a Brownian particle in a periodic potential. This problem finds applications in the non-linear pendulums, superionic conductors, phased-locked loops in radio, Josephson tunneling junctions, etc. \cite{risken}. Similar works can be found in \cite{coffey2} and \cite{coffey3}, where the Langevin equation is used instead. 
\begin{figure}[h!]
		\centering
		\includegraphics[scale=.4]{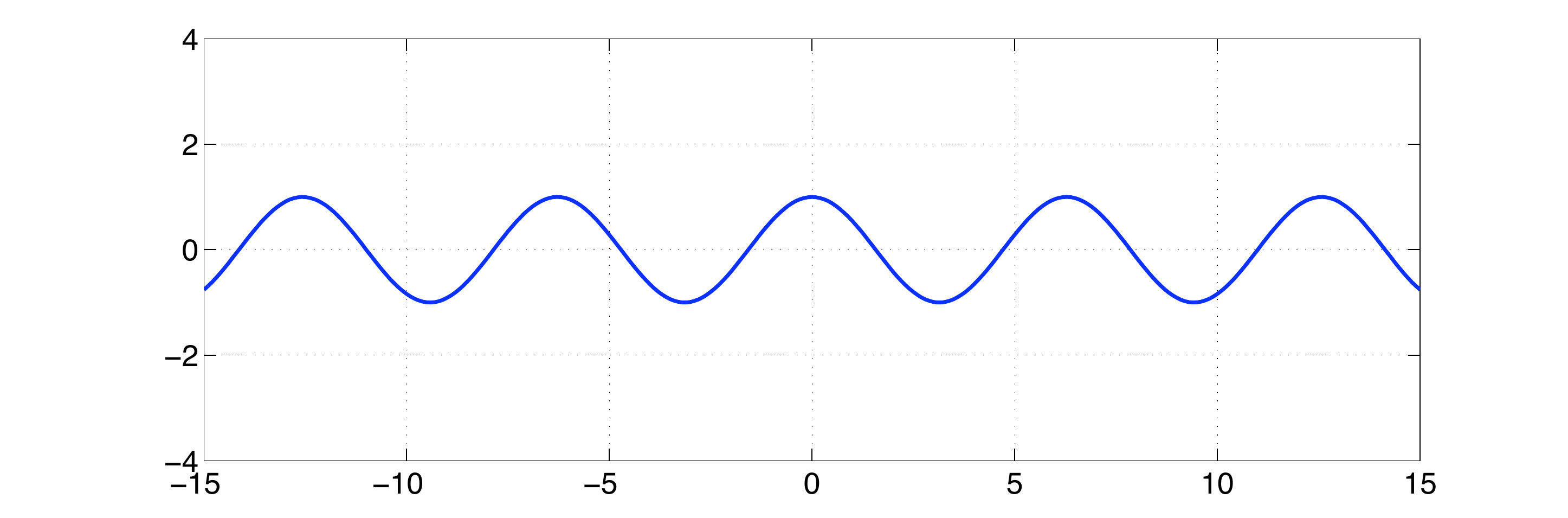}
		\includegraphics[scale=.4]{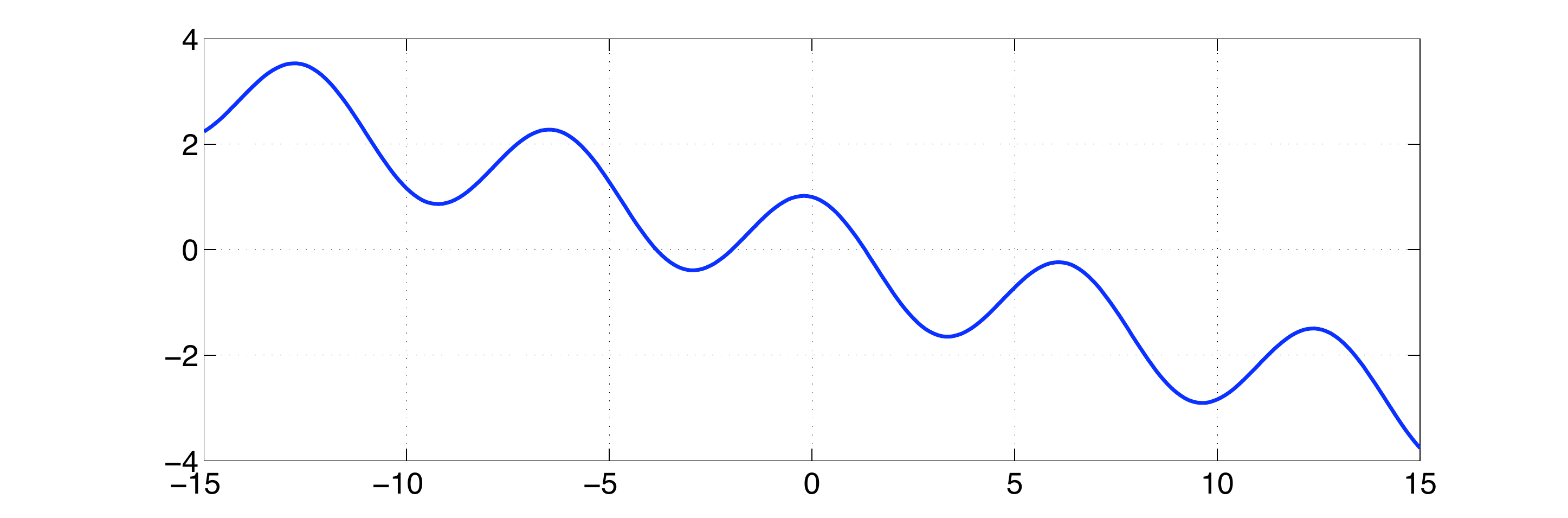}
	\caption{Periodic potential without (top) and with (bottom) biased force.}
	\label{fig:periodic_potential}
\end{figure}
\clearpage
Consider a one-dimensional case, the particle is kicked around by the Langevin force. When the Langevin force is large enough, the particle will travel from one potential well to the next, causing it to diffuse in both directions. If we apply an external force, the particle will diffuse in one direction preferably (see Figure \ref{fig:c3_biased-brownian-motion}), and we are interested in the drift velocity $\langle \dot{x} \rangle$.
\begin{figure}[h!]
		\centering
		\includegraphics[scale=1.0]{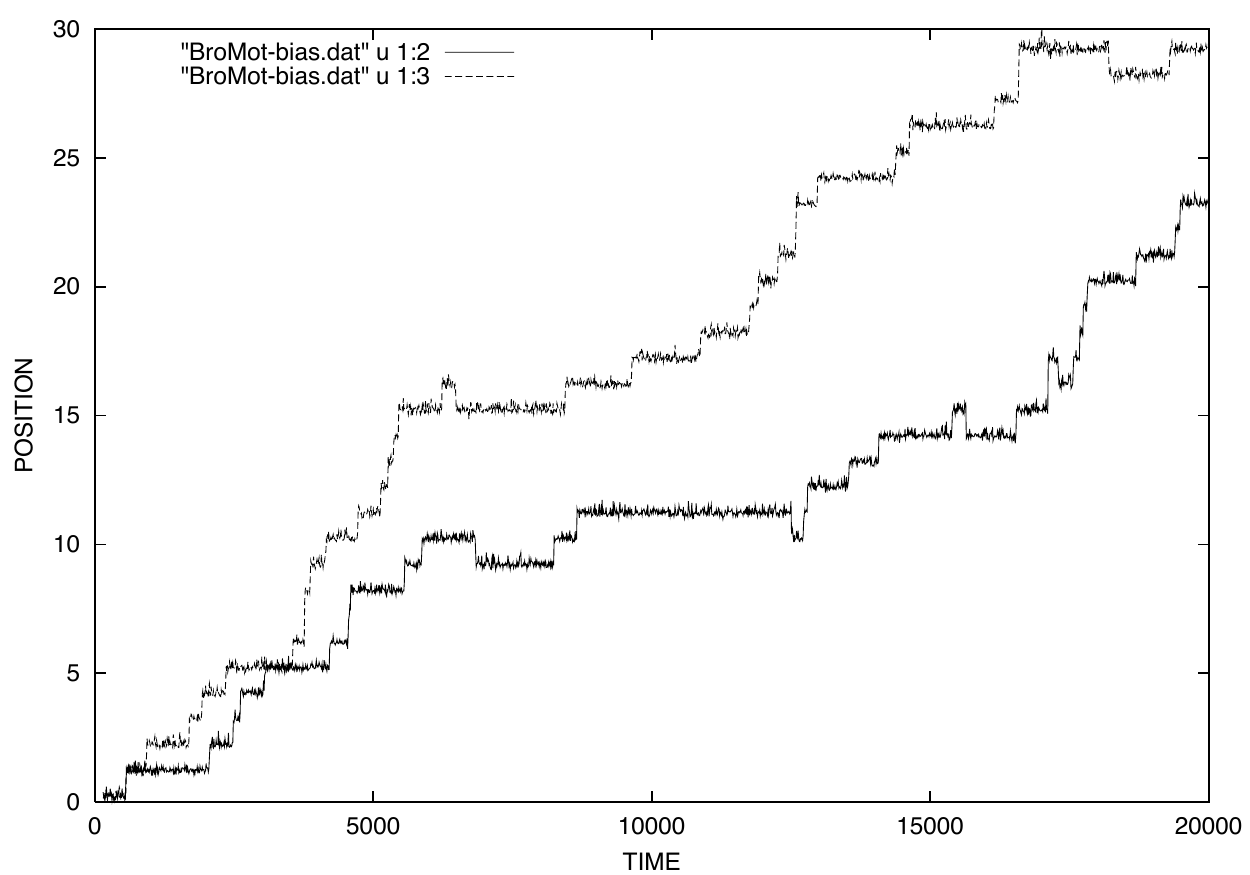}
	\caption{The trajectories of two independent Brownian particles in a periodic potential. The particles are subject to a constant force so that the random walk is biased to the force direction \cite{jose}.}
	\label{fig:c3_biased-brownian-motion}
\end{figure}

The Langevin equation can be written as
\begin{eqnarray}
		M\ddot{x}+\gamma M \dot{x}+V'(x)=F_{\rm{ext}}(t)+ \xi(t), \label{c3_Langevin}
\end{eqnarray}
where $F_{\rm{ext}}(t)$ is the external applied force\footnote{
Risken \cite{risken} discusses the application to super-ionic conductors. A super-ionic conductor consists of a a nearly fixed ion lattice in which some other ions are highly mobile. If an external field is applied to a one-dimensional model, neglecting the ion-ion interaction, the equation of motion is the same as Eq.~(\ref{c3_Langevin}).
}. We consider a periodic potential of the form
\begin{eqnarray}
		V(x)= -V_0 \cos\Big(\frac{2\pi}{L} x\Big),
\end{eqnarray}
where $2V_0$ is the height of the well. The minus sign is inserted for convenience.

\section{Strong Damping: Smoluchowski  Equation}
In the regime of high friction, the velocity of the particle reaches steady state rapidly, thus the inertial term, $M\ddot{x}$, can be omitted. The resulting Langevin equation reads
\begin{eqnarray}
		\gamma M \dot{x}(t)+  \frac{2\pi}{L} V_0\sin( \frac{2\pi}{L} x)=F_{\rm{ext}}(t)+\xi(t).
\end{eqnarray}
Following Risken \cite{risken}, we introduce the following dimensionless variables, 
\begin{eqnarray}
		\tilde{x} 		 = \frac{2\pi}{L}x    						;\quad
		\tilde{t} 		 = \frac{2\pi}{L}\sqrt{\frac{V_0}{M}} t    				;\quad
		\tilde{\gamma} = \frac{L}{2\pi}\sqrt{\frac{M}{V_0}}\gamma		; \quad
		\tilde{F}_{\rm{ext}}  = \frac{L}{2\pi} \frac{F_{\rm{ext}}}{V_0}   			;\quad
		\tilde{\xi}          = \frac{L}{2\pi} \frac{\xi}{V_0} 					;\quad
		\tilde{T}		 = \frac{k_B T}{V_0}.
\end{eqnarray}
The Langevin equation is transformed to
\begin{eqnarray}
		\tilde{\gamma}  \dot{\tilde{x}}+  \sin(\tilde{x})=\tilde{F}_{\rm{ext}}(\tilde{t})+\tilde{\xi}(\tilde{t}),
		 \label{c3_overdamped_langevin}
\end{eqnarray}
and the noise correlation function becomes
\begin{eqnarray}
		\langle \tilde{\xi}(\tilde{t}) \tilde{\xi}(\tilde{t}')\rangle=2 \, \tilde{\gamma} \,\tilde{T} \,\delta(\tilde{t}-\tilde{t}').
\end{eqnarray}
We will drop the tildes and it is understood that we are using the normalized units. The corresponding \FP\ equation, according to (\ref{FP_eqn}),  for the distribution $W(x,t)$ is given by
\begin{eqnarray}
	\gamma \frac{\partial W}{\partial t}=\frac{\partial }{\partial x}\Big[ \sin(x)-F_{\rm{ext}}+
												T \frac{\partial}{\partial x} \Big]W. \label{c3_smoluchowski}
\end{eqnarray}
This is a special case of the Smoluchowski equation, the \FP\ equation for an over-damped Brownian particle.

\section{Converting the Smoluchowski Equation into Recurrence Form}
In the steady state (long time limit), we expect the distribution to be periodic in space. In fact, in the problems of pendulums or Josephson junctions, the systems are indeed periodic (from  $0$ to $2\pi$). Then, we can express the distribution function as a Fouries series in space
\begin{eqnarray}
	 W(x,t)= \sum_{n=-\infty}^{\infty}W_n(t) \mbox{e} ^{\iu n x}, \label{c3_space_expansion}
\end{eqnarray}
and we will need to solve for the coefficients $W_n(t)$. Substituting  the expansion (\ref{c3_space_expansion}) into the Smoluchowski equation (\ref{c3_smoluchowski}), we obtain the three-term recurrence relation
\begin{eqnarray}
	 \gamma \dot{W}_n= Q^{-}_n W_{n-1} +Q_n W_{n}+ Q^{+}_{n}W_{n+1}, \label{c3_3RR}
\end{eqnarray}
where
\begin{eqnarray}
	 Q_n   			&=&  - \iu n F_{\rm{ext}}(t) -n^2 T;	\\ \nonumber
	 Q^{-}_n			&=&  + \frac{1}{2}n;		\\ \nonumber
	 Q^{+}_n			&=&  -\frac{1}{2}n. 
\end{eqnarray}
The zeroth term $W_0$, which is used as the ``seed" in the \CF\ method (Appendix A.1), is fixed by the normalization condition. We normalize the distribution function over a period,
\begin{eqnarray}
		 \int^{\pi}_{-\pi}W(x,t) \,dx &=&1; \,\,\,\,\,\,\,\,\,  W_{0} = \frac{1}{2\pi}. 
\end{eqnarray}

The drift velocity can be obtained by taking ensemble average of the Langevin equation (\ref{c3_overdamped_langevin}),
\begin{eqnarray}
	 \gamma \langle \dot{x} \rangle = F_{\rm{ext}} (t)- \langle \sin x \rangle.
\end{eqnarray}
Solving the recurrence relations, we get the distribution $W(x,t)$ and hence the average of any function involving $x$. From the expression above, we then can obtain the drift velocity $\langle \dot{x} \rangle$.
\\

We will consider applied force of the form
\begin{eqnarray}
	 F_{\rm{ext}}(t)= F + b\cos(\Omega t). 
\end{eqnarray}
The first term is a constant force which tilts the potential profile while the second term drives the system and allows us to study the dynamical properties.
\clearpage
\section{Stationary Response (DC)}
\begin{figure}[h!]
		\centering
		\includegraphics[scale=0.90]{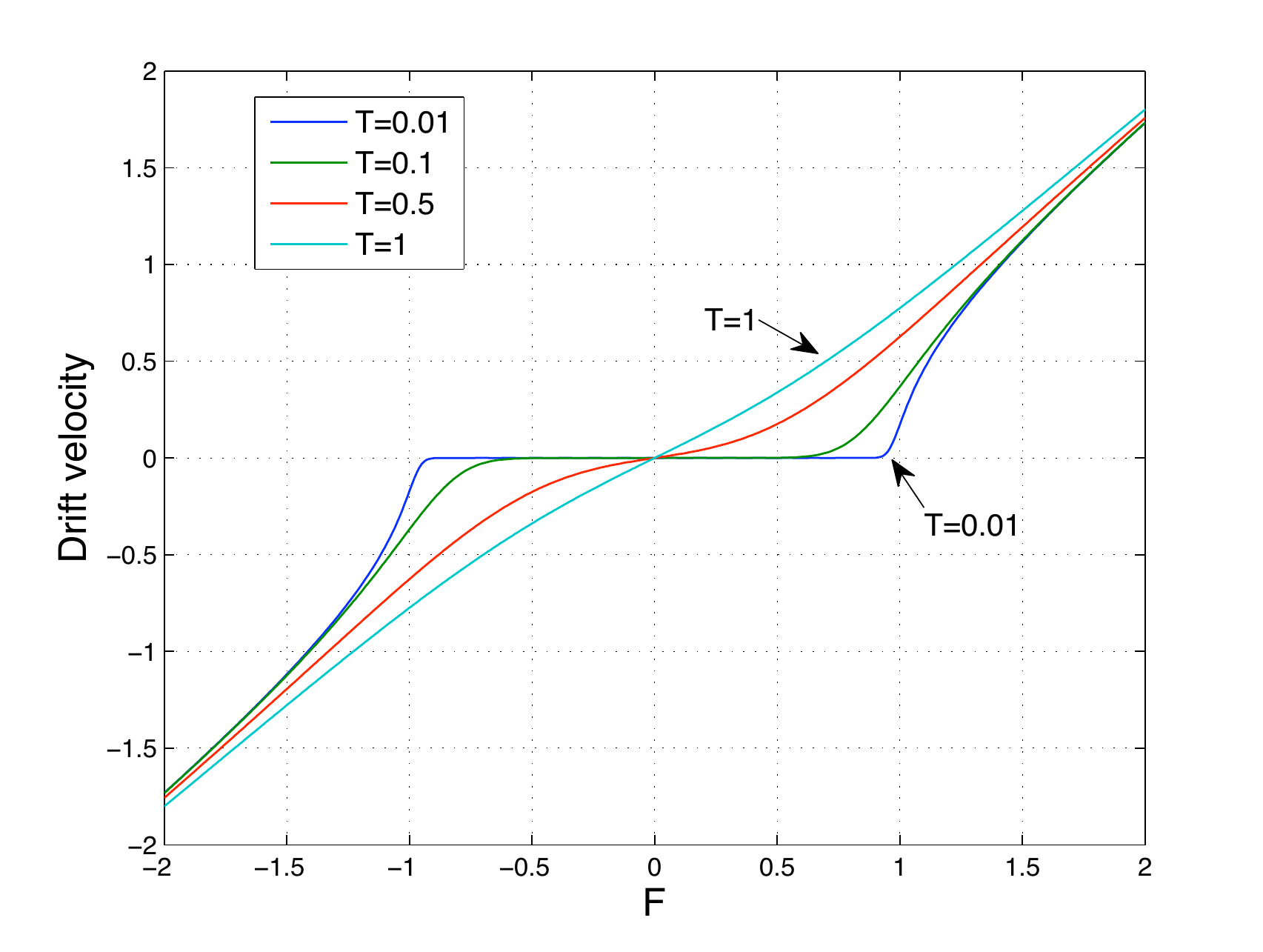}
	\caption{Drift velocity against constant force $F$ at various temperatures. The curves stay flat when the particle is trapped in the potential well. At large $F$ or $T$, the curves are almost linear to $F$. }
	\label{c3_drift}
\end{figure}

In the absence of AC driving $F_{\rm{ext}}(t)=F$, the stationary
solution $(\dot{W}_n=0)$ of the three-term recurrence relation
(\ref{c3_3RR}) can be obtained using the scalar \CF\ method (Appendix
A.1).  In Figure \ref{c3_drift}, the drift velocity is plotted against
the applied force, $F$, at different temperatures. There are regions
where the curves stay flat; the particle is ``locked" in the potential
well and there is not enough applied force or Langevin force (at low
temperature) to push it away from the well.  At large $F$ (or high
temperature), the effect of the potential well is less, and the
velocity grows linearly with the applied force. In between, we have
the depinning transition between the two regimes. At $T\approx0$, this
transition takes place when the force equals the cosine well depth
($F=1$), and breaks the minima structure.  

\section{System Under AC Driving}
In the presence of driving, the system will never reach a stationary state. Instead, in the long time dynamics, we expect it to be oscillating with the same period as the driving. This is similar to the case of a driven damped oscillator, where the driving frequency is the only time scale in the long time dynamics. Therefore, we can take care of the time dependence of the distribution function by expanding it into a Fourier series in time $\{\mbox{e}^{ik\Omega t}\}$. Together with the Fourier series in space $\{\mbox{e}^{inx}\}$, we have 
\begin{eqnarray}
	 W(x,t)= \sum_{n=-\infty}^{\infty} \sum_{k=-\infty}^{\infty}W_n^{(k)} \mbox{e} ^{\iu n x}\mbox{e} ^{\iu k \Omega t}, \label{c3_Fourier_expand}
\end{eqnarray}

In the driven system, we are interested in the susceptibility $\chi$, defined as
\begin{eqnarray}
	 \gamma \langle \dot{x} \rangle_{\Delta} &=& \gamma \langle \dot{x} \rangle(t) -\gamma  \langle \dot{x} \rangle_{0}\\
	                &=&\sum_{k=1}^{\infty} \Big(\frac{b}{2}\Big)^{k} \Big[ \chi ^{(k)}\mbox{e}^{+\iu k\Omega t}+ \chi ^{*(k)}\mbox{e}^{-\iu k\Omega t} \Big],
\end{eqnarray}
where $\langle \dot{x} \rangle_{0}$ denotes the time-independent part of the drift velocity. We will adopt the following convention,
\begin{eqnarray}
	 \chi^{(k)}= \chi'^{(k)} -\iu \chi ''^{(k)}.
\end{eqnarray}
In the regime of linear response $(b\rightarrow 0)$, we only have the first order term (we omit the superscript), and the linear susceptibility is
\begin{eqnarray}
	 \gamma  \langle \dot{x} \rangle _{\Delta}  &=&\frac{b}{2} \Big( \chi \, \mbox{e}^{+\iu \Omega t}+ \chi ^{*}\mbox{e}^{-\iu \Omega t} \Big).
\end{eqnarray}
The interpretation is easier in linear response: susceptibility is the coefficient of the time dependent part (of a dynamical quantity) that is oscillating at the driving frequency. The real part is in phase with the driving, while the imaginary part is the out of phase contribution.
%
\subsection{Perturbative Treatment}
In the case of weak driving, we can expand the distribution function in Taylor's series of the driving amplitude $b$,
\begin{eqnarray}
	 W(x,t)=\sum_{q=0}^{\infty}\,b\,^q \Big[ \sum_{n=-\infty}^{\infty}\, \sum_{k=-\infty}^{\infty}\, \,W_n^{(k)}[q] \,\mbox{e} ^{\iu n x}\,\mbox{e} ^{\iu k \Omega t}\Big] , \label{c3_perturb_expand}
\end{eqnarray}
and we need to solve for the coefficients $W_n^{(k)}[q] $. In this problem, the driving amplitude should be $b< 0.2$ for this expansion to hold.
\\

The expansion (\ref{c3_perturb_expand}) turns the Smoluchowski equation (\ref{c3_smoluchowski}) into
\begin{eqnarray}
	    \tilde{Q}^{-}_n \,W_{n-1}^{(k)}[q] \,+\, \tilde{Q}_n^{k} \,W_{n}^{(k)}[q] \,+\, \tilde{Q}^{+}_{n}\,W_{n+1}^{(k)}[q] =-f^{(k)}_n[q],
	    \label{c3_perturb_3RR}
\end{eqnarray}
where
\begin{eqnarray}
	 \tilde{Q}^{k\mbox{ }}_n  	&=&  - \iu n F -\iu k\gamma\Omega -n^2 T;		\\ \nonumber
	 \tilde{Q}^{-}_n			&=& +\frac{1}{2}n;				 \\ \nonumber
	 \tilde{Q}^{+}_n			&=&  -\frac{1}{2}n; 						\\ \nonumber
	 f^{(k)}_{n}[q]				&=& -\frac{\iu }{2}\, n\Big[ W^{(k+1)}_n[q-1]+W^{(k-1)}_n[q-1]\Big].
\end{eqnarray}
The perturbative structure is clear in Eq. (\ref{c3_perturb_3RR}). The
solution of the previous order equation $[q-1]$ enters into the
inhomogeneous part of the next order $[q]$. The zeroth order equation
is homogeneous, and its solution enters the right hand side of first
order equation, and so on. This set of iterative equations can be
computed up to any order, until the solution converges.
\begin{figure}[h!]

	\centering
	\includegraphics[scale=0.65]{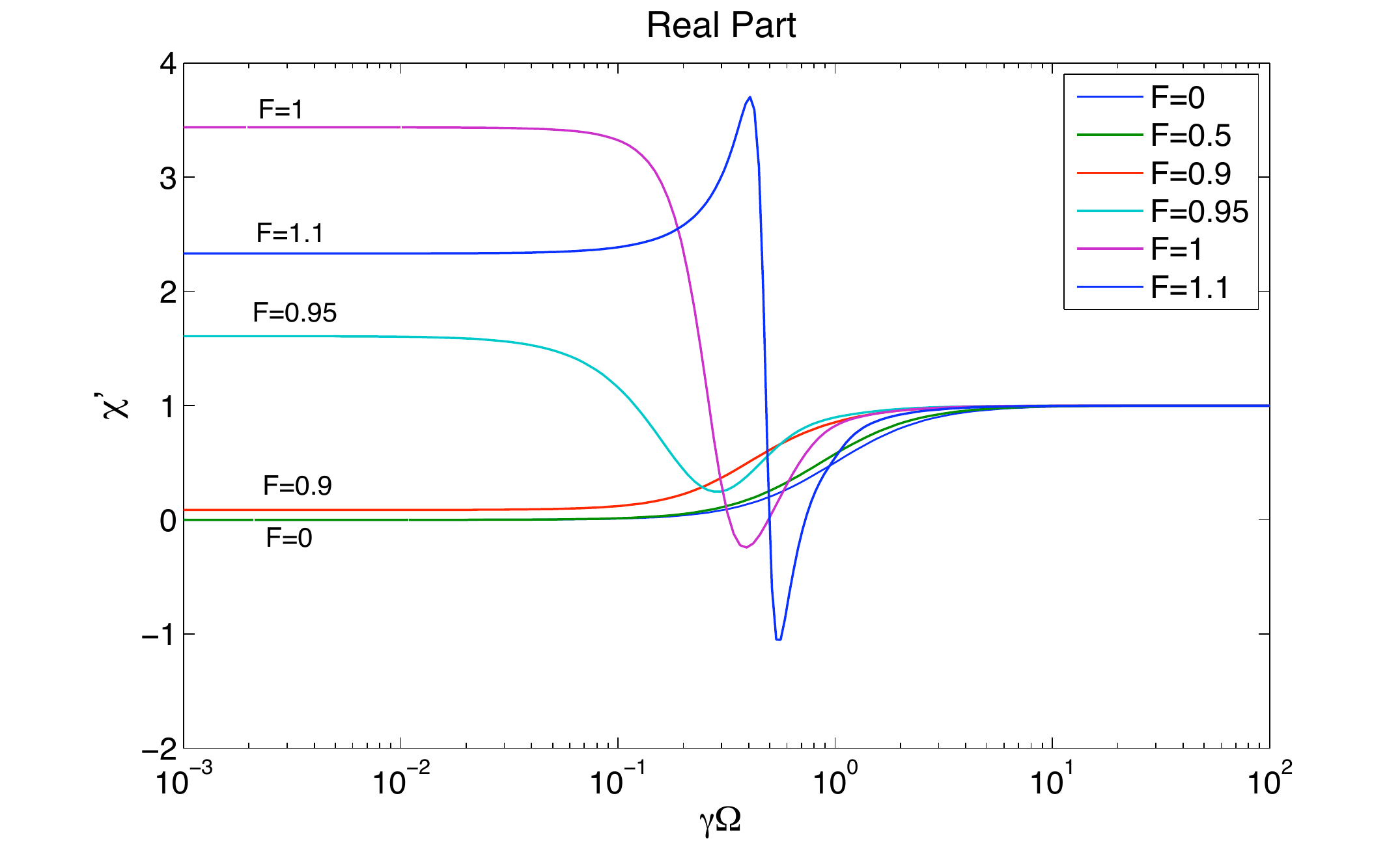}

\hspace{0.4cm}

	\centering
	\includegraphics[scale=0.65]{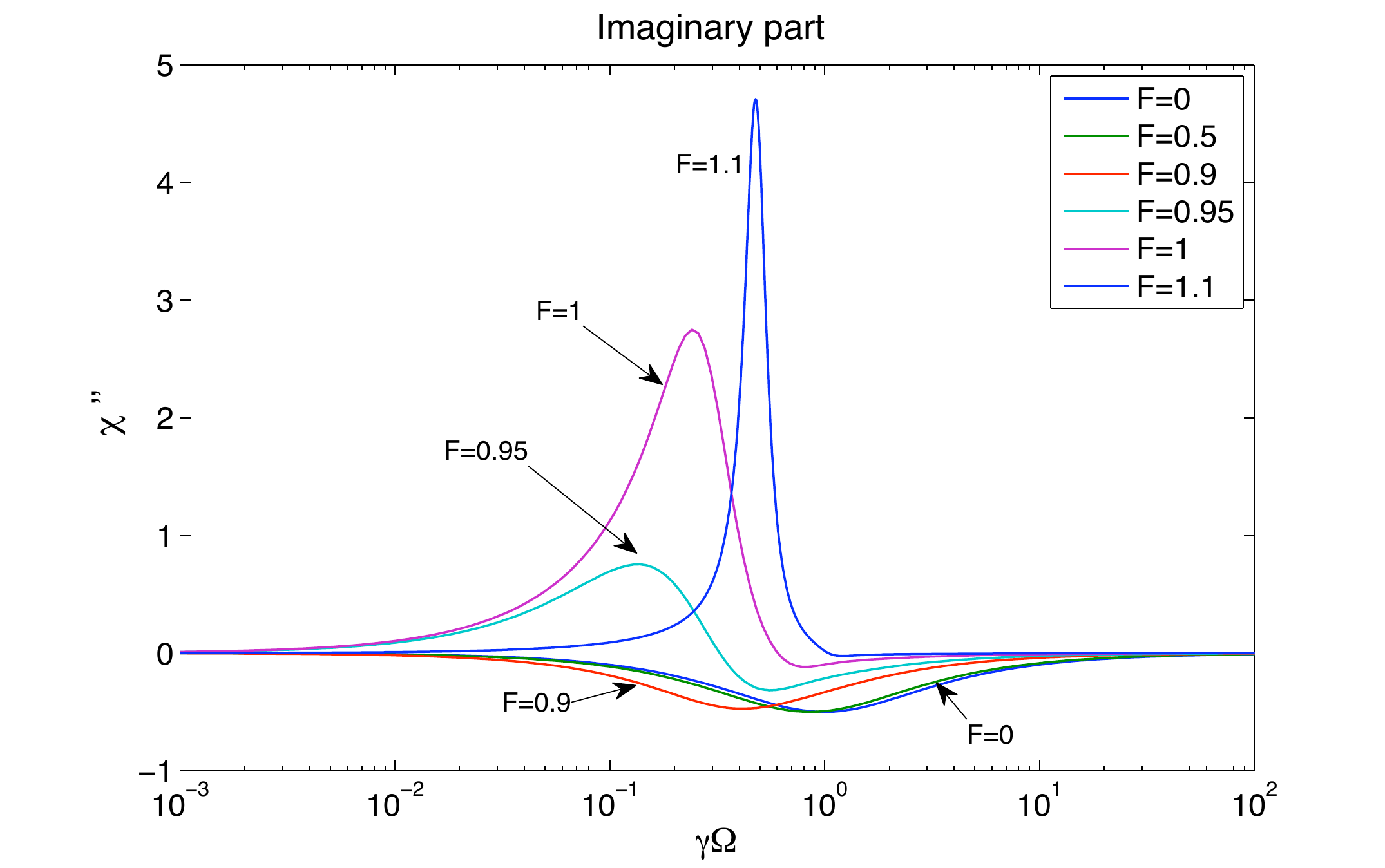}

\caption{The linear susceptibility at $T=0.01$. The curves of $F=0$ and $F=0.5$ are barely distinguishable. The profiles change significantly for $F> 0.9$ as the particle is no longer trapped in the potential well and this corresponds to the sloped region of the curves in Figure \ref{c3_drift}.}
\label{fig:c3_linear_sus}
\end{figure}

We will first start looking at the linear response, i.e. we only solve
up to the first order. The linear susceptibility is plotted in Figure
\ref{fig:c3_linear_sus}. Most of the interesting behaviors happen
around the resonant frequency (the frequency of oscillations near the
bottom of the wells, in our units, $\gamma \Omega=1$). It is the same
situation as the response of a driven damped oscillator. The increment
of the constant field, up to $F\approx0.9$, does not change the
susceptibility profile much; the particle is locked in the potential
well for small fields. Note that we have used a low temperature
$T=0.01$, such that the locking behavior is evident (recall the
deppining behaviors in Figure \ref{c3_drift}). When the field is
increased until it reaches the depinning regime, the profile changes
considerably. It is because the particle gains enough energy to hop
from one well to the next, it is no longer trapped.

Let us look at the low frequency region (adiabatic driving), where the system responds in phase with the field. At small force (up to $F\approx0.9$), the response is zero.  It can be explained by looking at the flat portion of the drift velocity curves in Figure \ref{c3_drift}. Turn on the driving, the system moves back and forth along the $x$-axis, but there is no vertical movement. At large force, the drift velocity curve is sloped, thus giving non-zero response.
\\

The real part of the susceptibility is proportional to the power dissipated into the medium due to the driving. The power is calculated by multiplication of the velocity and the force, $P=F(t)v(t)$. In our context, 
\begin{eqnarray}
		P &\propto& b \cos(\Omega t) \frac{b}{2} \Big( \chi \mbox{e}^{+\iu \Omega t}+ \chi ^{*}\mbox{e}^{-\iu \Omega t} \Big) \\
		&\propto& b^2 \Big[\chi' \cos^2(\Omega t)+ \chi'' \sin(\Omega t) \cos(\Omega t)\Big].		\nonumber
\end{eqnarray}
Taking the time average over a cycle, one finds $P\propto \chi'$.

\subsection{Exact Treatment}
Now we solve the \FP\ equation exactly, using the matrix  \CF\ method. This method is more computationally expensive, since it involves matrix multiplication and inversion.
\\

We define the vector
%
\[\mathbf{W}_n= \left( \begin{array}{ccccccc}
         \vdots       \\
         W^{(-2)}_n    \\
         W^{(-1)}_n    \\
         W^{(0)}_n    \\
         W^{(+1)}_n    \\
         W^{(+2)}_n    \\
         \vdots       \\
\end{array} \right).\] 
Substituting the expansion (\ref{c3_Fourier_expand}) into the Smoluchowski equation (\ref{c3_smoluchowski}), we obtain
\begin{eqnarray}
	    \mathbf{\hat{Q}}^{-}_n \, \mathbf{W}_{n-1} +  \mathbf{\hat{Q}}_n  \mathbf{W}_{n} + \mathbf{\hat{Q}}_{n}^{+}\,\mathbf{W}_{n+1} =0,
\end{eqnarray}
where
\begin{eqnarray}
	 \mathbf{\hat{Q}}^{\mbox{ }}_n  		&=&  (- \iu n F -n^2 T)\mathbf{I} -\iu \gamma\Omega \mathbf{\hat{A}} -\frac{\iu}{2}nb\,\mathbf{\hat{B}};		\\ \nonumber
	 \mathbf{\hat{Q}}^{-}_n					&=&  + \frac{1}{2}n\mathbf{I};				 \\ \nonumber
	 \mathbf{\hat{Q}}^{+}_n				&=&  -\frac{1}{2}n\mathbf{I},						\\ \nonumber
\end{eqnarray}
and
%
 \[ \mathbf{\mathbf{\hat{A}} }= \left( \begin{array}{cccccccc}  
         \ddots  &     &          &     	&  	 	 &   	   &       &   \\
         		     & -2 &          &     	&  	 	 &     	   &       &   \\
          		     &     &  -1    &    	&  	  	&     	   &       &   \\
		          &     &         &    0   &  	  	&     	   &       &   \\
			    &     &         &    	     &  	 +1 &     	   &       &   \\
			     &     &         &    	&  	  	&     +2  &       &   \\
                     &     &         &       &   	     &           &       &   \ddots 
 \end{array} \right);
 \quad
 \mathbf{\mathbf{\hat{B}} }= \left( \begin{array}{cccccccc}  
          0    &   1    &     	     &  	 	    &   	   		&       &   \\
           1   &    0    &    1 	     &  	 	    &     	      &       &   \\
                 &    1    &    0       &  	 1 	   &     	  		 &       &   \\
		    &         &    \ddots & \ddots & \ddots  &       &   \\
		   &         &    	    &  	  1	    &      0    &    1   &   \\
		    &         &    	    &  	  	    &      1    &    0   &  1 \\
               &         &             &   	          &      	   &      1   &  0 \\
 \end{array} \right).\] 
$\mathbf{I}$ is the identity matrix. The recurrence relations can be solved by the matrix \CF\ method (Appendix A.1).
 \\
 
\begin{figure}[h!]  
		\centering
		\includegraphics[scale=0.50]{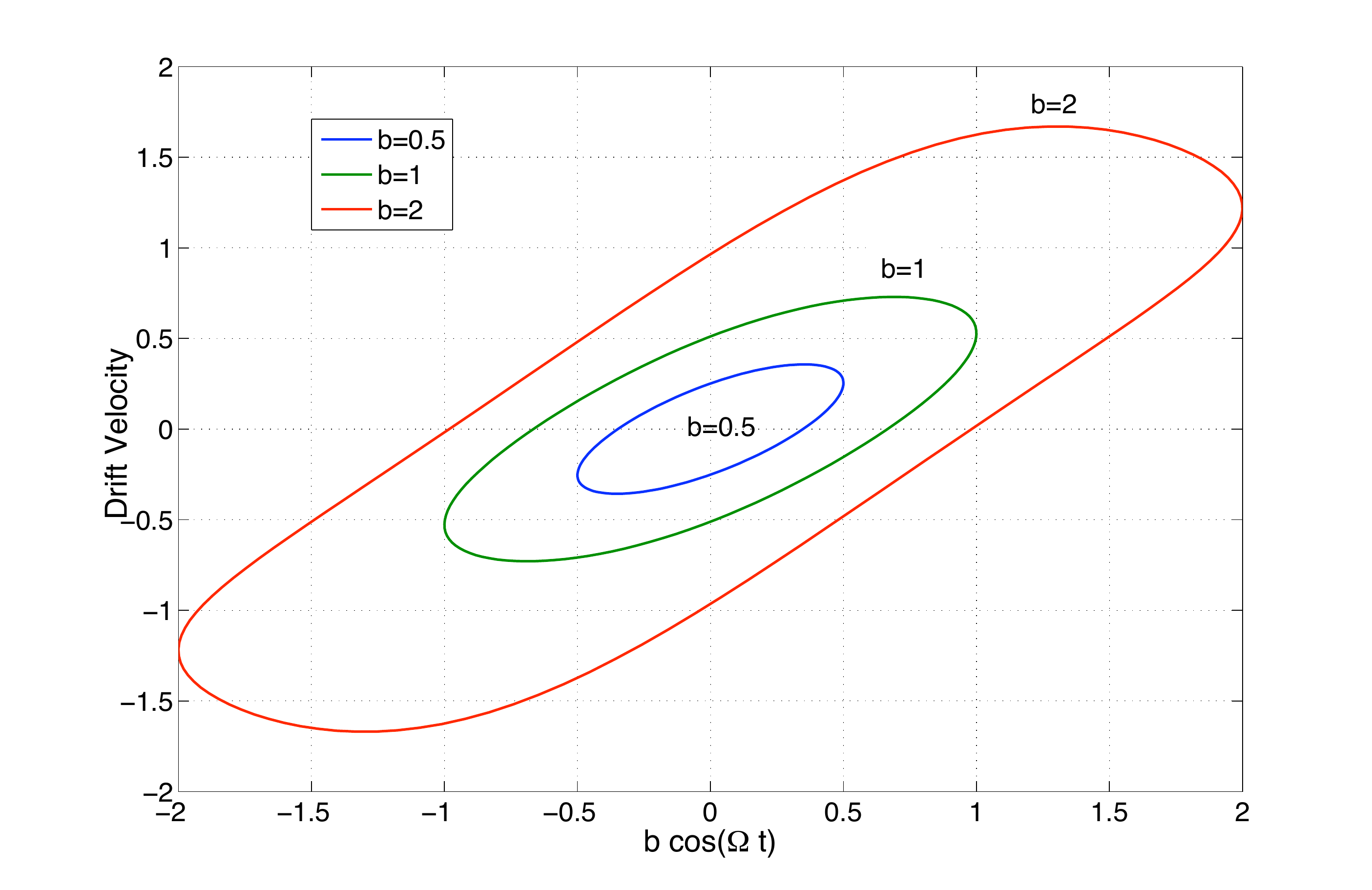}
		\includegraphics[scale=0.50]{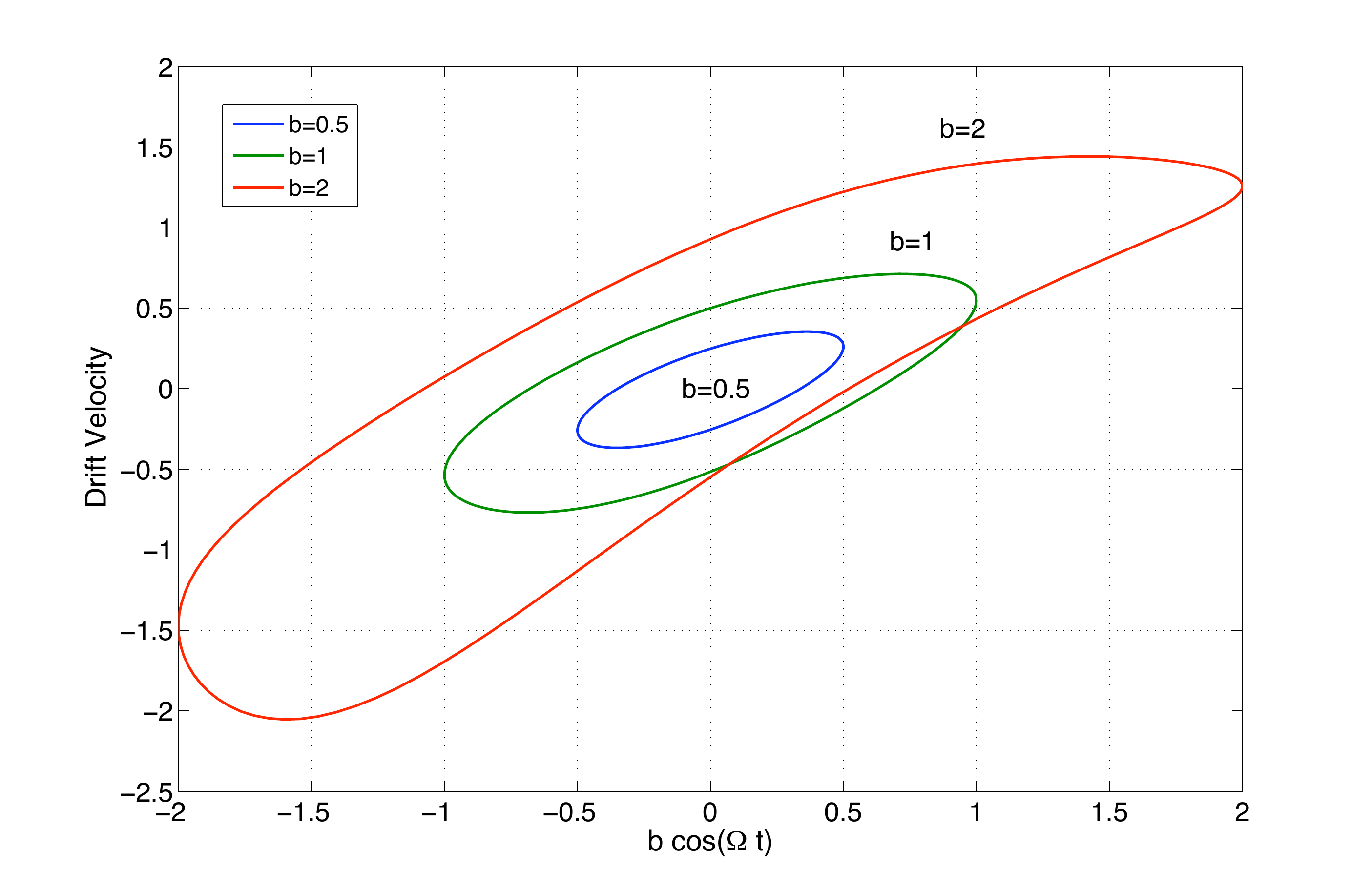}
	\caption{Top: Hysteresis loops at  $\gamma\Omega =1$ and $T=0.01$ without biased force. At large driving force $b$, the elliptical shape is distorted. 
	Bottom: The biased force $F_0=0.2$ breaks the left-right symmetry.}
	\label{fig:c3_hysteresis}
\end{figure}

We plot the real time dynamical loops (hysteresis loops): the curves of the drift velocity against the real time driving field, $b \cos(\Omega t)$, in Figure \ref{fig:c3_hysteresis}. At small driving, the loops are elliptical since the response is linear to the driving field. When the driving field increases, the contribution from the (non-linear) higher harmonic susceptibilities, $\chi^{(2)}$, $\chi^{(3)}$, $\chi^{(4)}$, etc., becomes significant and the loops are distorted. The area of the loop is proportional to the power dissipated. 
\\

\begin{figure}[h!]
	\begin{minipage}[b]{0.5\linewidth}
		\centering
		\includegraphics[scale=0.55]{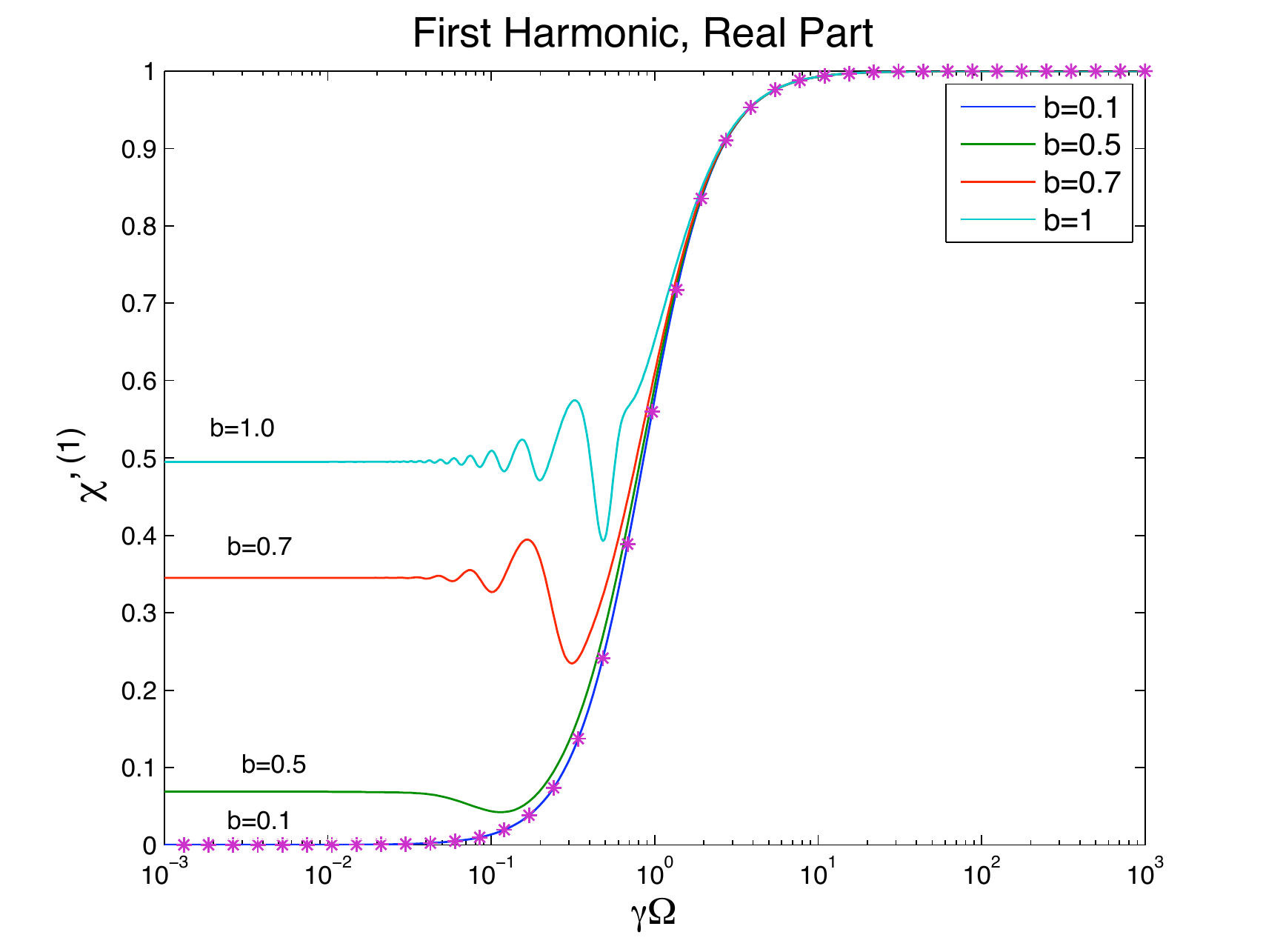}
	\end{minipage}
	\hspace{0.5cm}
	\begin{minipage}[b]{0.5\linewidth}
		\centering
		\includegraphics[scale=0.55]{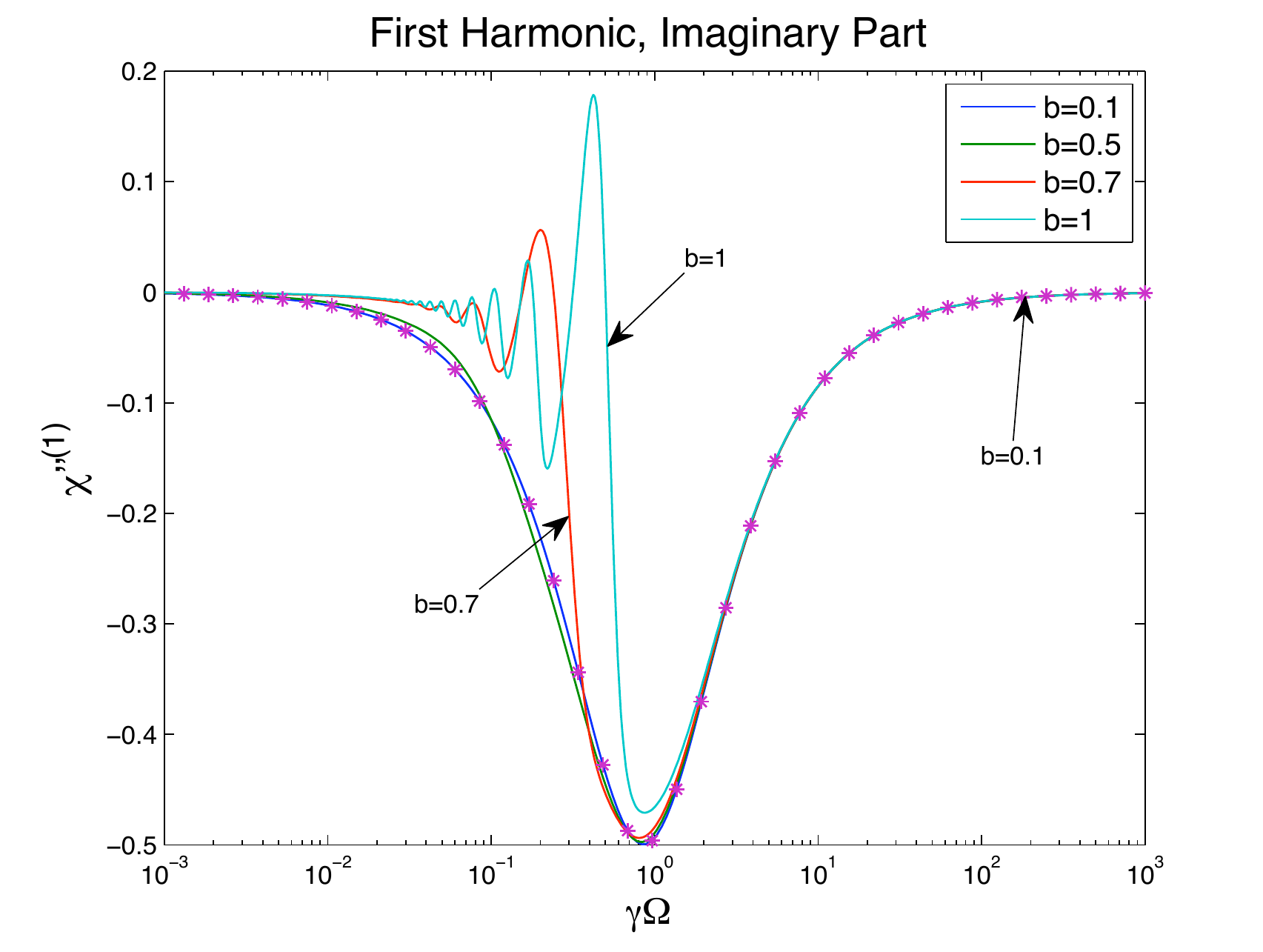}
	\end{minipage}
%
	\begin{minipage}[b]{0.5\linewidth}
		\centering
		\includegraphics[scale=0.55]{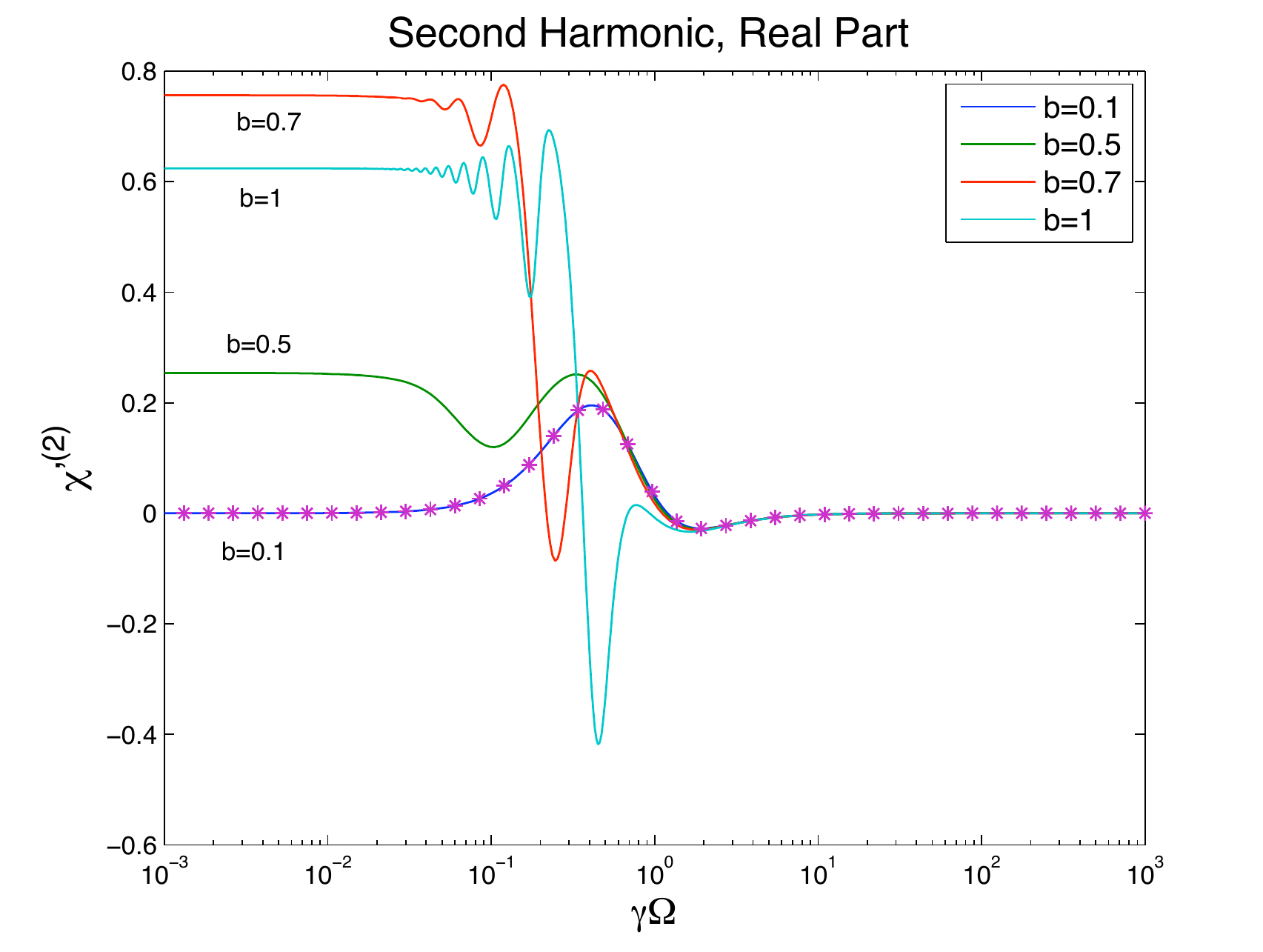}
	\end{minipage}
	\hspace{0.5cm}
	\begin{minipage}[b]{0.5\linewidth}
		\centering
		\includegraphics[scale=0.55]{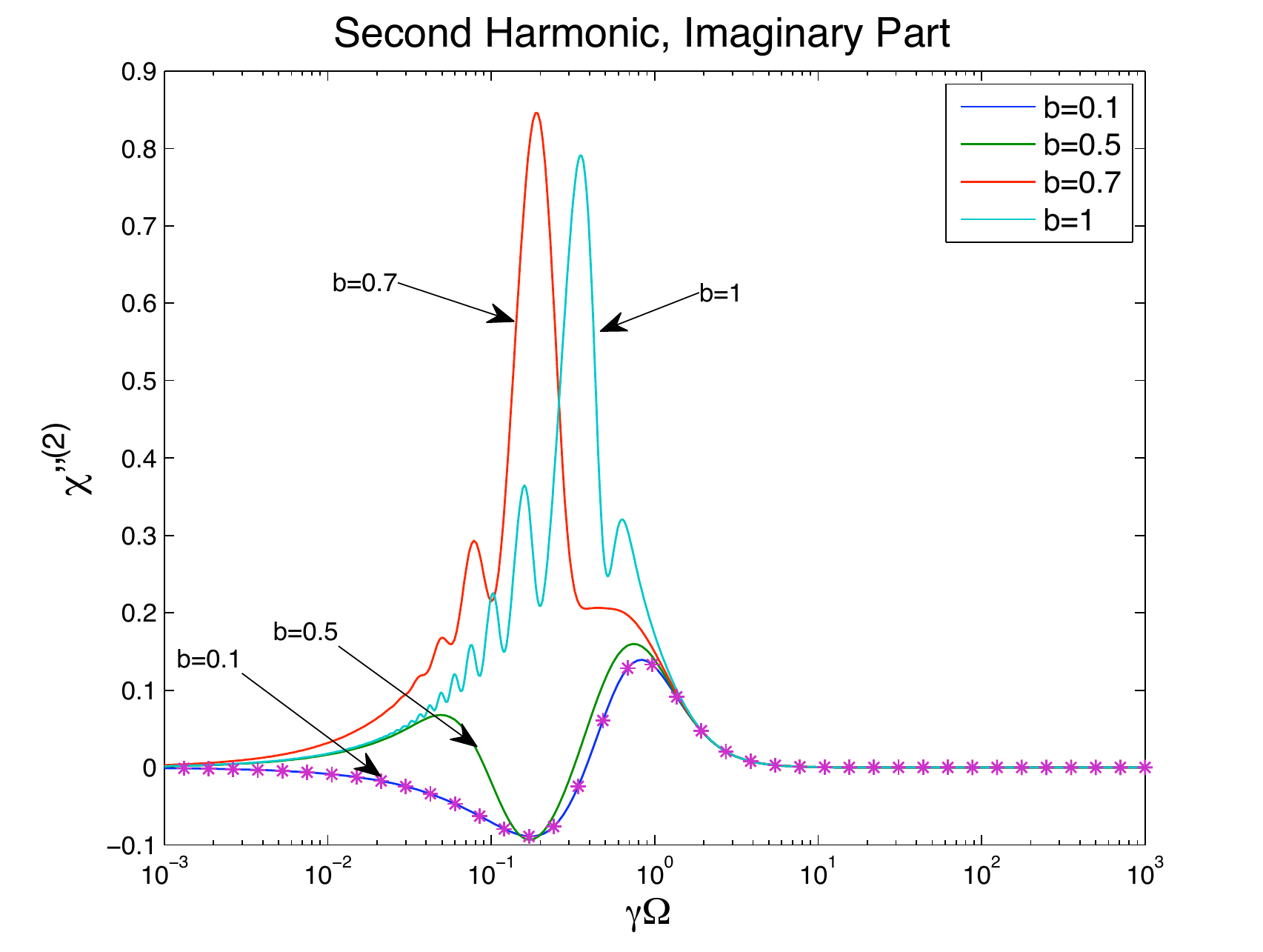}
	\end{minipage}
	\begin{minipage}[b]{0.5\linewidth}
		\centering
		\includegraphics[scale=0.55]{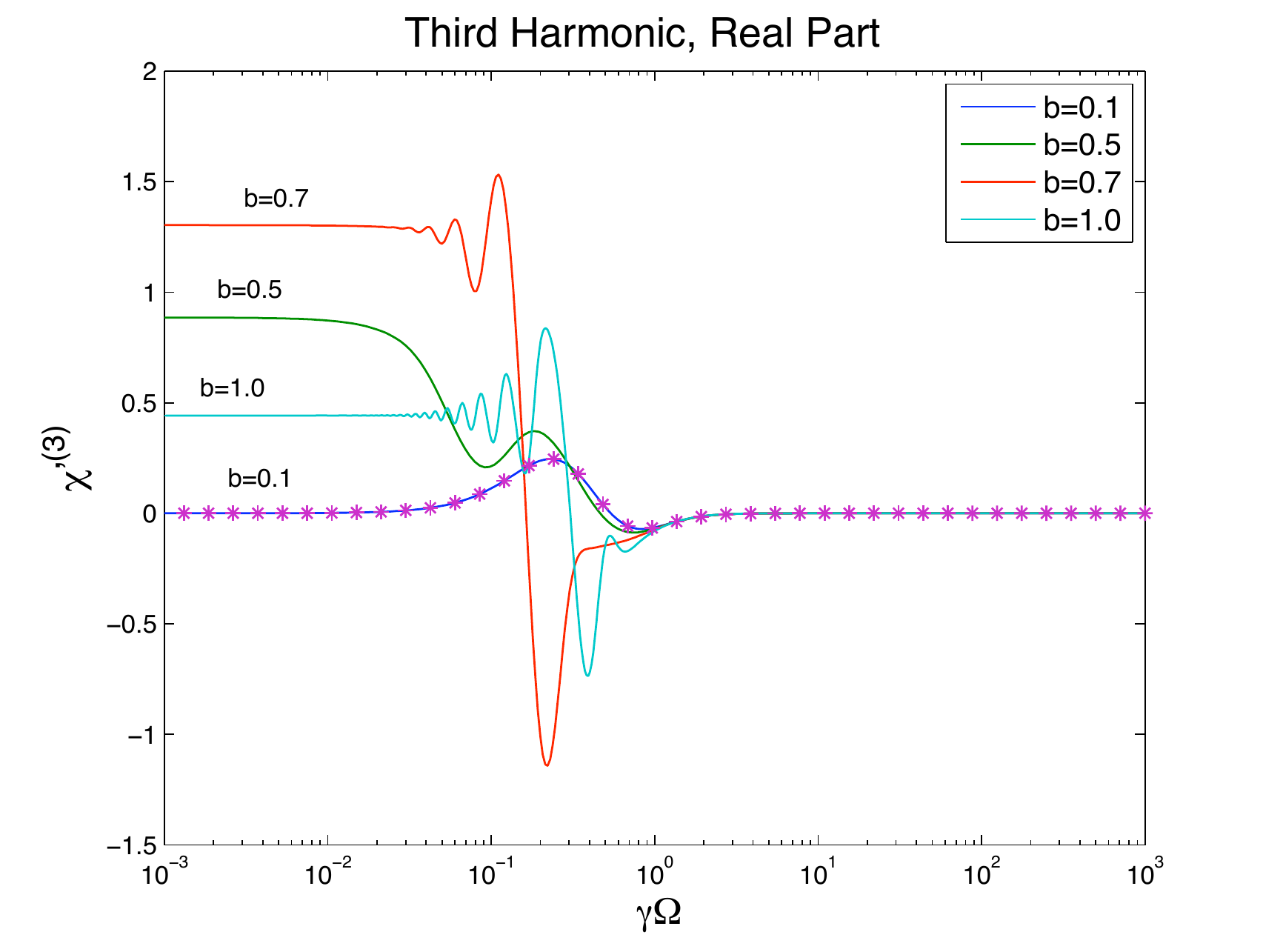}
	\end{minipage}
	\hspace{0.5cm}
	\begin{minipage}[b]{0.5\linewidth}
		\centering
		\includegraphics[scale=0.55]{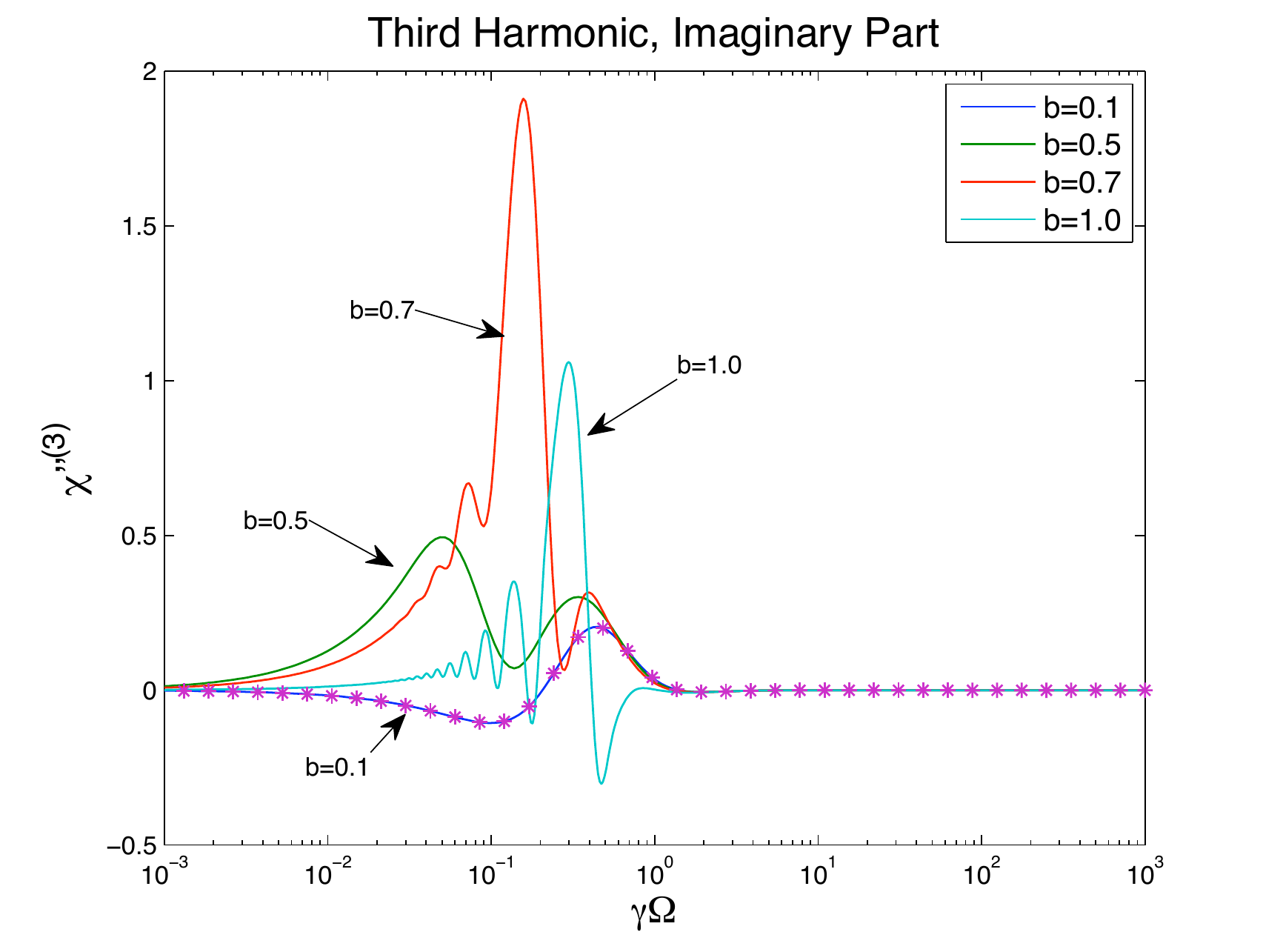}
	\end{minipage}
	\caption{The first three harmonics of the  susceptibility, $F=0.5$ and $T=0.01$. Symbols are the results from the perturbative treatment. Oscillatory behaviors can be observed around $\gamma \Omega \approx 0.1-1$. }
	\label{fig:c3_sus_harmonics}
\end{figure}

To see the contribution of the non-linear susceptibilities, we plot the first three harmonics in Figure \ref{fig:c3_sus_harmonics}. As opposed to the linear response, the susceptibilities at low frequency are lifted from zero at large driving field. It is because the constant and driving forces combined are large enough to kick the particle away from the potential well, the particle is no longer trapped.  We also observe oscillatory behaviors of the susceptibility curves in the frequency range $\gamma \Omega \approx 0.1- 1$. We will relate the oscillations with the celebrated Shapiro steps discussed below.
%

\clearpage
\begin{figure}[] 
	\centering
	\includegraphics[scale=0.60]{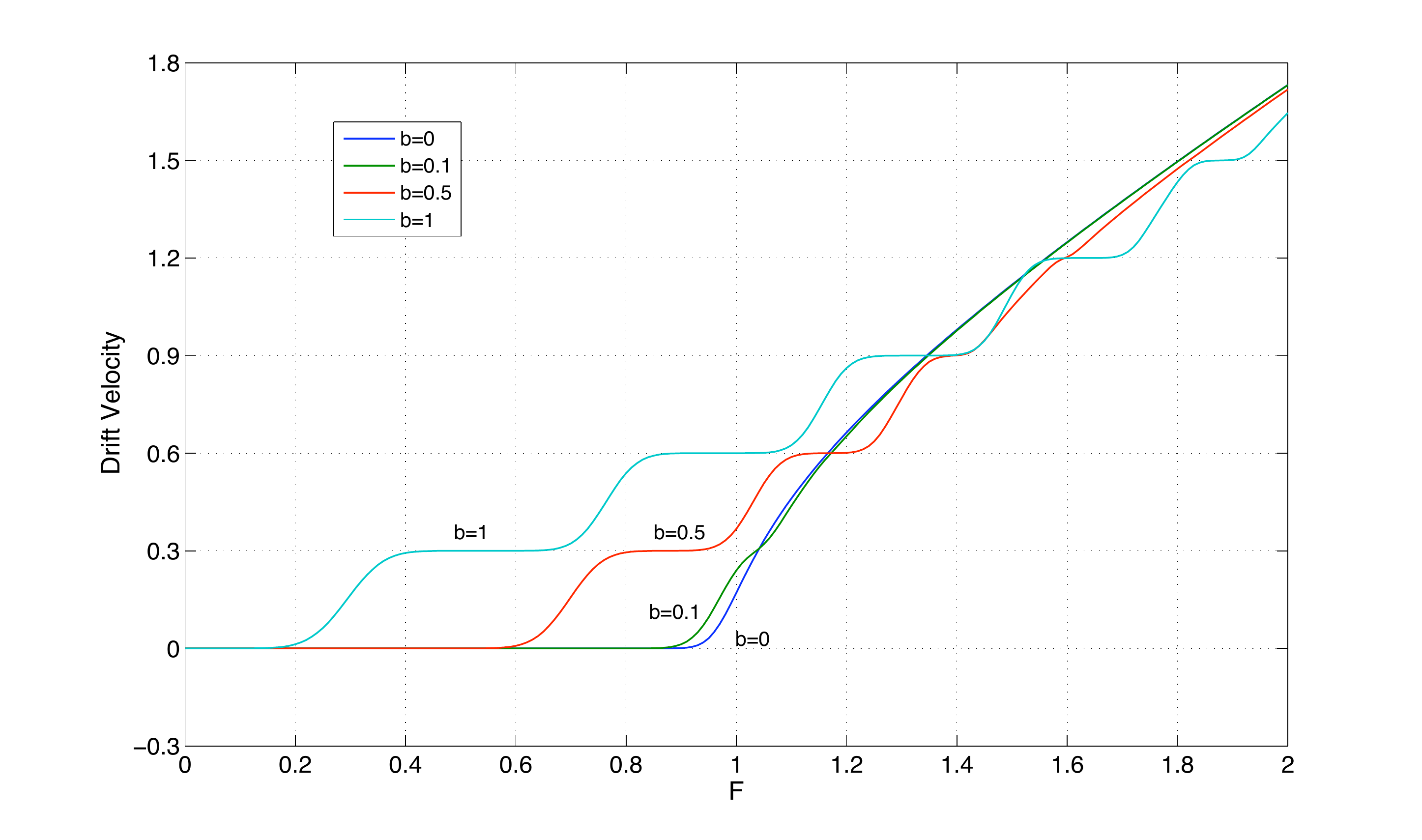}
	\caption{The time-average of the drift velocity at $\gamma \Omega =0.3$ and $T=0.01$. Shapiro steps occur at $ \langle \bar{\dot{x}} \rangle = n \gamma\Omega$, and their widths increase with driving force.}
	\label{fig:c3_time_average}
\end{figure}

We plot the time averaged drift velocity $ \langle \bar{\dot{x}} \rangle $ in Figure \ref{fig:c3_time_average}. The curves exhibit Shapiro steps at the multiples of the driving frequency, $ \langle \bar{\dot{x}} \rangle = n \gamma\Omega$, where $n=1, 2......$ (see Ref.~\cite{jung} for a discussion). The system is locked at the resonant frequency $\gamma \Omega= \langle \bar{\dot{x}} \rangle/n$.  The appearance of the Shapiro steps is due to the fact that the quantity $ \langle \bar{\dot{x}} \rangle $ becomes stable with respect to a small change in external parameter ($F$ here) at resonant frequency \cite{shapiro}. One needs a finite change to push the system away from this stability. This phenomenon is well known in the context of Josephson junctions. 
\\

\begin{figure}[] 
	\centering
	\includegraphics[scale=0.50]{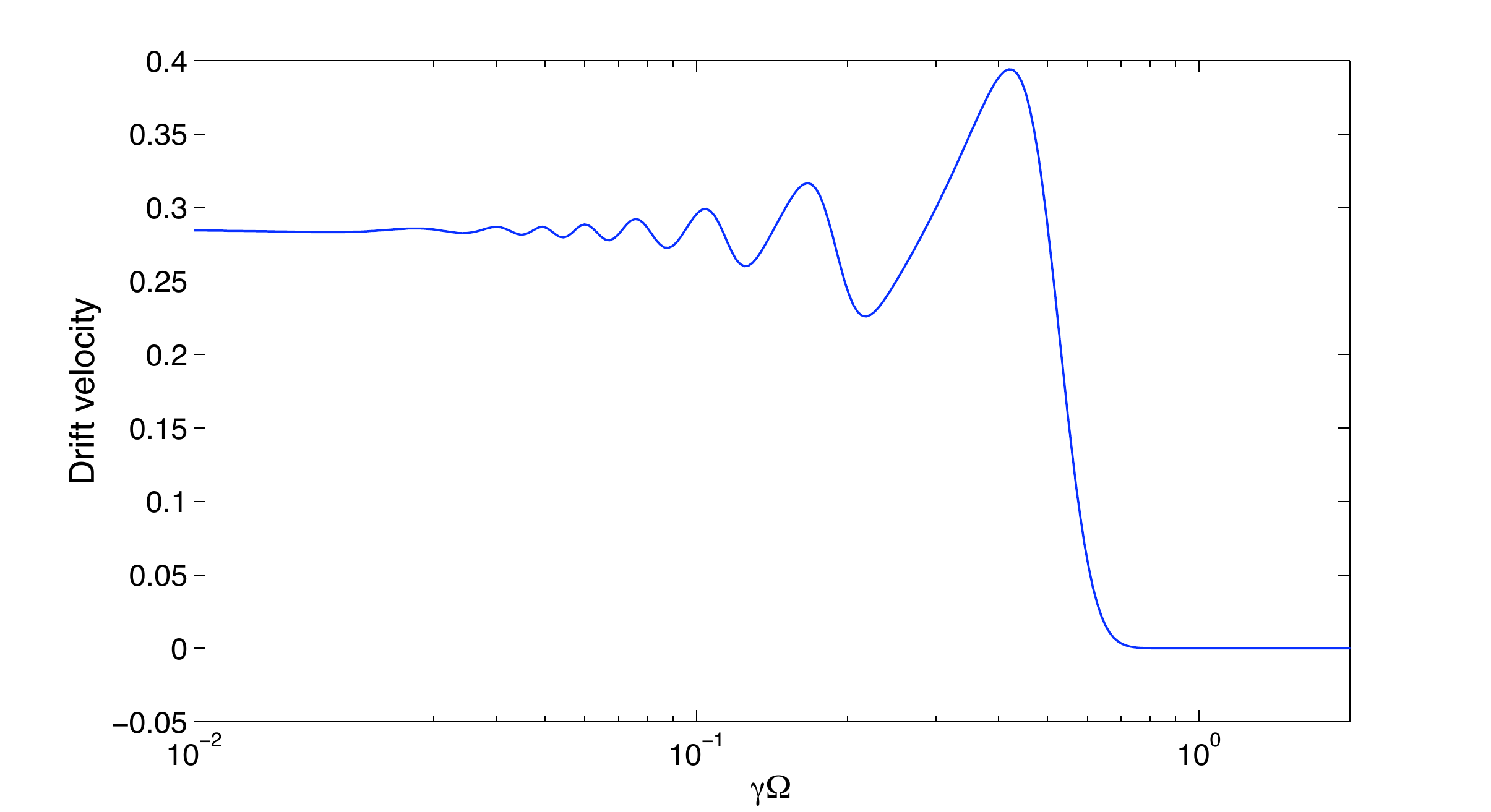}
	\caption{The time-average of the drift velocity versus frequency at $b=1$, $F=0.5$ and $T=0.01$. One can relate the oscillations with the oscillations observed in the susceptibility curves in Figure \ref{fig:c3_sus_harmonics}.}
	\label{fig:c3_time_average_freq}
\end{figure}

To relate it with the oscillatory behavior observed in the susceptibility curves, we look at a particular value of $F$ and vary the driving frequency, the curve in Figure \ref{fig:c3_time_average} will rotate back and forth. The curve is flat when the frequency is resonant and sloped when it is not  (see Figure \ref{fig:c3_time_average_freq}). This ``modulation" might explain the oscillatory features in the susceptibility curves. 
%
\section{Summary}
In this chapter, we showed explicitly how the \CF\ method is used to solve a \FP\  equation. We obtained both the time-independent and time-dependent solutions. We also presented two approaches of using the \CF\ methods: perturbative and exact. Strong non-linear effects are observed in the dynamical hysteresis loops under strong driving.  We could include the inertia term $M\ddot{x}$ and solve the Klein-Kramers equation (\ref{c2-klein-kramers}). This would require one more matrix index for the expansion in momentum, which entails larger computational efforts, see Refs. \cite{risken} and \cite{jung}.

\chapter{Rotational Brownian Motion: Debye Dipole}
\section{Introduction}

\begin{figure}[h!]  
  	\begin{center}
   		 \includegraphics[width=2.0in]{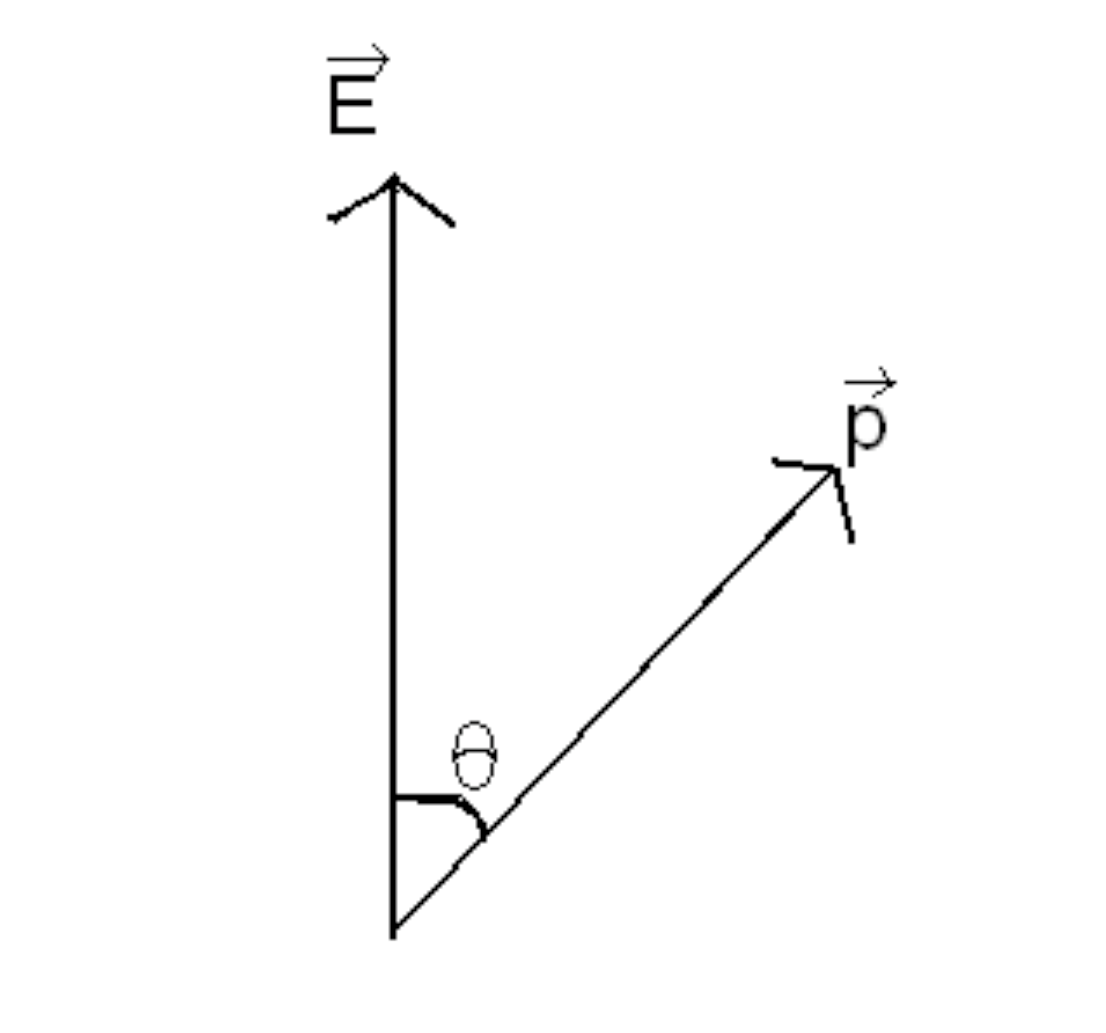}
   		 \caption{Pictorial skectch of a dipole, $\vec{p}$ in an electric field, $\vec{E}$  .}
	  \end{center}
	  \label{Fig:c4_dipole}
\end{figure}

Here we will study a \FP\ equation of different structure, which
involves a dipole. We investigate the problem of non-interacting
dipoles subject to DC and/or AC fields.  This problem was first
studied by Peter Debye in the 1920's, and constitutes the first
example of rotational Brownian motion. In the Debye model, he assumed
high friction and isotropy, i.e. the Brownian motion exhibits no
preferential direction. The orientation of the dipoles solely depends
on the angle between the electric field and the dipole vector,
$\theta$ in Figure \ref{Fig:c4_dipole}. This problem not only finds
applications in dielectric relaxation, but also rotational relaxation
of ferromagnetic nanoparticles \cite{coffey}. 

\section{Fokker-Planck Equation}

\subsection{Debye Orientational Diffusion Equation}
In the high friction limit, the \FP\ equation describing the dipoles is \cite{jose}
\begin{eqnarray} 
	\zeta \frac{\partial W(\theta, t)}{\partial t} &=& 
	\frac{1}{\sin \theta} \frac{\partial}{\partial \theta}\Big[ \sin \theta \Big(k_{B}T\frac{\partial}{\partial \theta}+pE(t)\sin \theta \Big)\Big]W(\theta, t),\label{c4_FP0}
\end{eqnarray}
where $\zeta$ is the viscosity coefficient. We define the Debye relaxation time and the dimensionless field parameter
\begin{eqnarray} 
	 \tau_D=\frac{\zeta}{2k_B T}\,; \,\,\,\,\,\,\,\,\, \alpha(t)=\frac{p E(t)}{k_B T}.
\end{eqnarray}
Making the transformation $x=\cos\theta$, the resulting \FP\ equation reads
\begin{eqnarray}
	2 \tau_D\frac{\partial W(x,t)}{\partial t}  &=& \frac{\partial}{\partial x}\Big[\Big(1-x^2\Big)\Big(\frac{\partial}{\partial x}- \alpha (t)\Big)\Big ]W(x,t). \label{c4_FP1}
\end{eqnarray}
%
\subsection{Method of Solution}
Because of the range of $x$ ($-1$ to $+1$), the natural choice of the basis function here is the Legendre polynomials $P_n(x)$\footnote{\noindent The Legendre polynomials can be expressed as the Rodrigues' formula
		\begin{center}
			$P_{n}(x)=\frac{1}{2^{n} n!}\frac{d^{n}}{dx^{n}}\Big[ (x^2-1)^n \Big].$\\
		\end{center}
\noindent They obey the orthogonality relation 
		\begin{center}
			$\int^{+1}_{-1}P_n(x)P_m(x)dx=\frac{2}{2n+1}\delta_{mn}$. 
		\end{center}
},
%
%
\begin{eqnarray}
	W(x,t)=\sum_{n=0}^{\infty} W_{n}(t)P_n(x). \label{c4_expand}
\end{eqnarray}
Substituting Eq. (\ref{c4_expand}) into Eq. (\ref{c4_FP1}), we obtain the three-term recurrence relation
\begin{equation} 
	2\tau_D \dot{W}_n=Q_{n}^{-}W_{n-1}+Q_{n}W_{n}+Q_{n+1}^{+}W_{n+1}, \label{c4_recurrence}
\end{equation}
where
\begin{eqnarray}  
	Q_{n}			&=&	-n(n+1)  \label{Q_values}; \\
	Q_{n}^{-}	&=& 	+\alpha(t) \frac{n(n+1)}{2n-1};\\ \nonumber
	Q_{n}^{+}	&=&	-\alpha(t) \frac{n(n+1)}{2n+3}.  \nonumber
\end{eqnarray}
Similar to the previous chapter, we consider the combination of DC and AC fields,
\begin{eqnarray}  
	E(t)&=&E_0+E_{\rm{d}}\cos(\Omega t); \\
	\alpha(t)&=& \alpha_0+b\cos(\Omega t); \,\,\,\,\,\,\,\, \alpha_0= \frac{pE_0}{k_B T}; \,\,\,\,\,\,\,\, b=\frac{p E_d}{k_BT}.
\end{eqnarray}
The object of interest here is the average orientation $\langle x \rangle= \langle \cos\theta \rangle$.
\\
\section{Equilibrium Properties}
Without driving, we solve for the equilibrium solution to the recurrence relation (\ref{c4_recurrence}) with the scalar \CF\ method (Appendix A.1), and study its average orientation. In fact, we can obtain the analytic result using elementary statistical mechanics. The canonical distribution of such a system is given by
\begin{equation}  
	\rho_c= \frac{\mbox{e}^{-\beta H}}{Z},\label{c4_canonical}
\end{equation}
where the Hamiltonian is $H=-\mathbf{p}\cdot \mathbf{E}$ and $Z$ the partition function\footnote
{
One can easily show that Eq.~(\ref{c4_canonical}) is the stationary solution by substituting it into the Fokker-Planck equation (\ref{c4_FP0}).
}. The average orientation is defined as
\begin{equation}  
	\langle \cos \theta \rangle = \int d\Gamma \, \rho_c \cos\theta ,
\end{equation}
where the integration is carried over the solid angle. The result is the Langevin function 
\begin{equation}  
	  \langle \cos \theta \rangle=L(\alpha_0)=\coth(\alpha_0)-\frac{1}{\alpha_0}.  \label{c4_langevin_fn}
\end{equation}

\begin{figure}[h!]  
  	\begin{center}
   		 \includegraphics[width=6.0in]{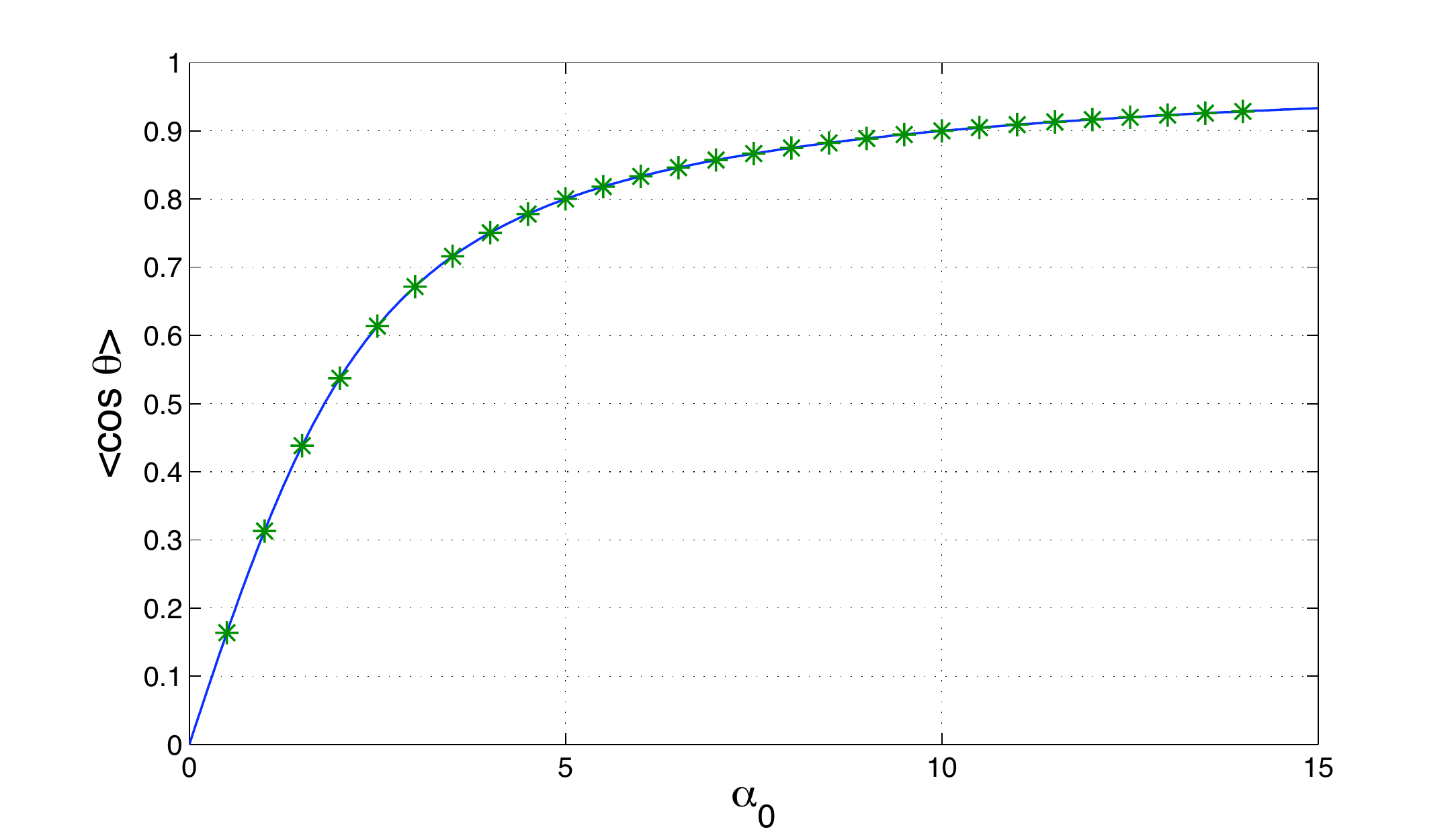}
	  \end{center}
	  \caption{The average orientation of a dipole, the solid line is the numerical results while the symbols are from the Langevin function Eq. (\ref{c4_langevin_fn}).}
 	 \label{fig:Langevin}
\end{figure}

The Langevin function is plotted together with the numerical results
in Figure \ref{fig:Langevin}. At small field, the average orientation
grows linearly with $\alpha_0$ since $L(\alpha_0)\approx \alpha_0/3 -
\alpha_0^3/45+......$ . At large field, we reach the saturated region
where $\langle \cos \theta \rangle \approx 1$, the dipoles are almost
fully aligned with the field. The comparison with the Langevin
function serves as a test for our numerical method.
%

\section{Driven Dipole}
As in the previous chapter, we focus on the time periodic solution and perform the Fourier time expansion of the distribution function of a driven system,
\begin{eqnarray}  
	W(x,t)=\sum_{n=0}^{\infty} \sum_{k=-\infty}^{\infty}W^{(k)}_n e^{\iu k \Omega t}P_{n}(x).
\end{eqnarray}
We will not discuss the perturbative treatment and the exact treatment again, they are almost identical as in Chapter 3. 
\\

Again, the susceptibilities $\chi$ are defined as 
\begin{eqnarray}  
	 \langle x \rangle _\Delta &=& \langle x(t) \rangle - \langle x \rangle_0 \nonumber \\
							&=&\sum_{k=1}^{\infty} \Big (\frac{b}{2}  \Big) ^{k}\Big[ \chi^{(k)}\mbox{e}^{+\iu k \Omega t} + \chi^{*(k)}\mbox{e}^{-\iu k 											\Omega t}\Big],
\end{eqnarray}
where $\langle x \rangle_0$ is the time-independent part. 
In the regime of linear response (weak driving), we only keep the linear susceptibility
\begin{eqnarray}  
	 \langle x \rangle _\Delta &=&  \frac{b}{2} \Big[ \chi \, \mbox{e}^{+ \iu  \Omega t} + \chi ^{*}\,\mbox{e}^{-\iu k \Omega t}\Big].
\end{eqnarray}
%
\subsection{Linear Response}

From linear response theory, the susceptibility  at zero DC field is given by the Debye relaxation formula \cite{jose}
\begin{eqnarray}  
	\chi  &=& \frac{1}{3}\, \,\frac{1}{1+ \iu \Omega \tau_D }.
\end{eqnarray}
%
%
\begin{figure}[b!]
\begin{center}
	\includegraphics[scale=0.55]{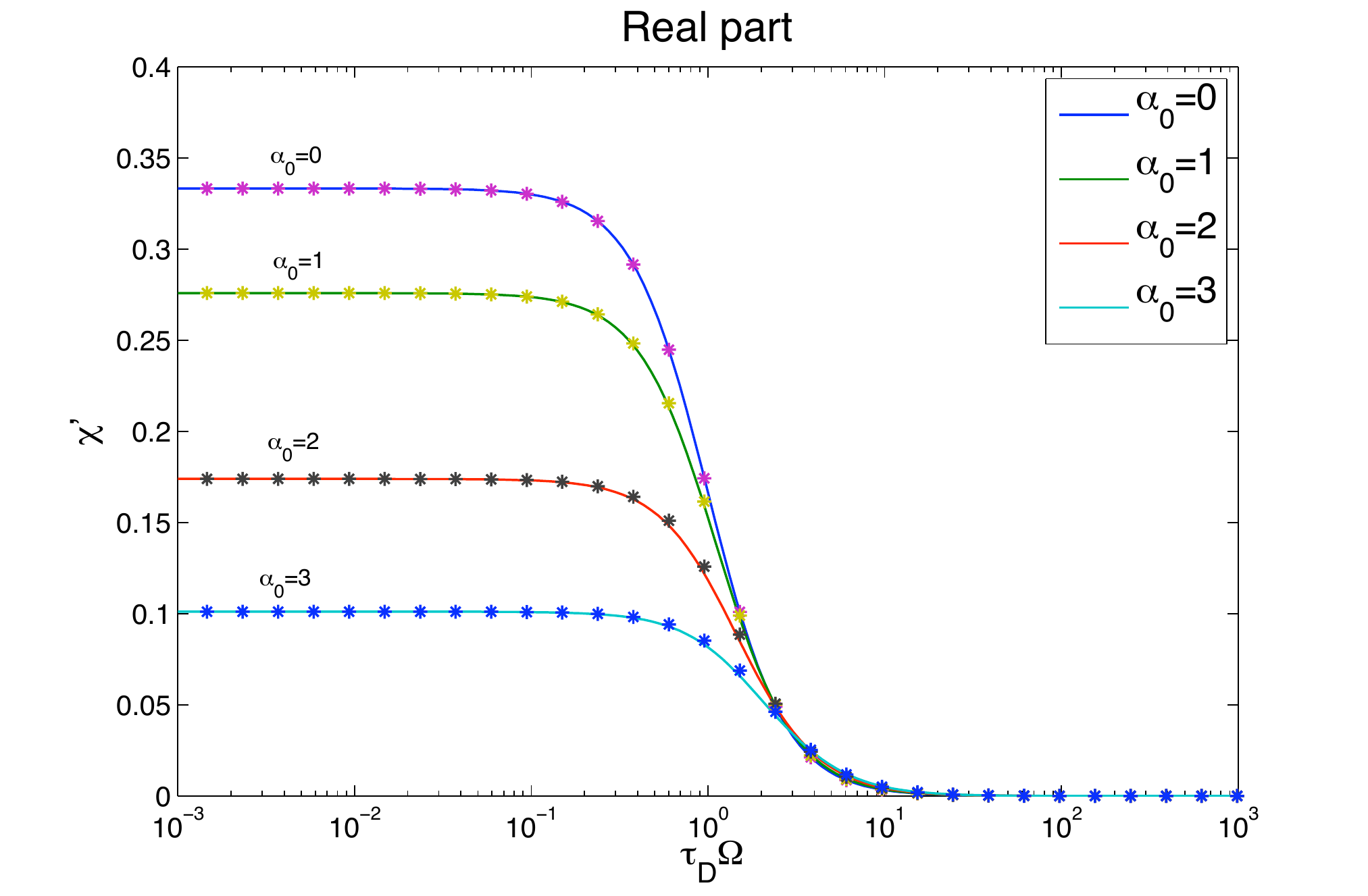}
	\includegraphics[scale=0.55]{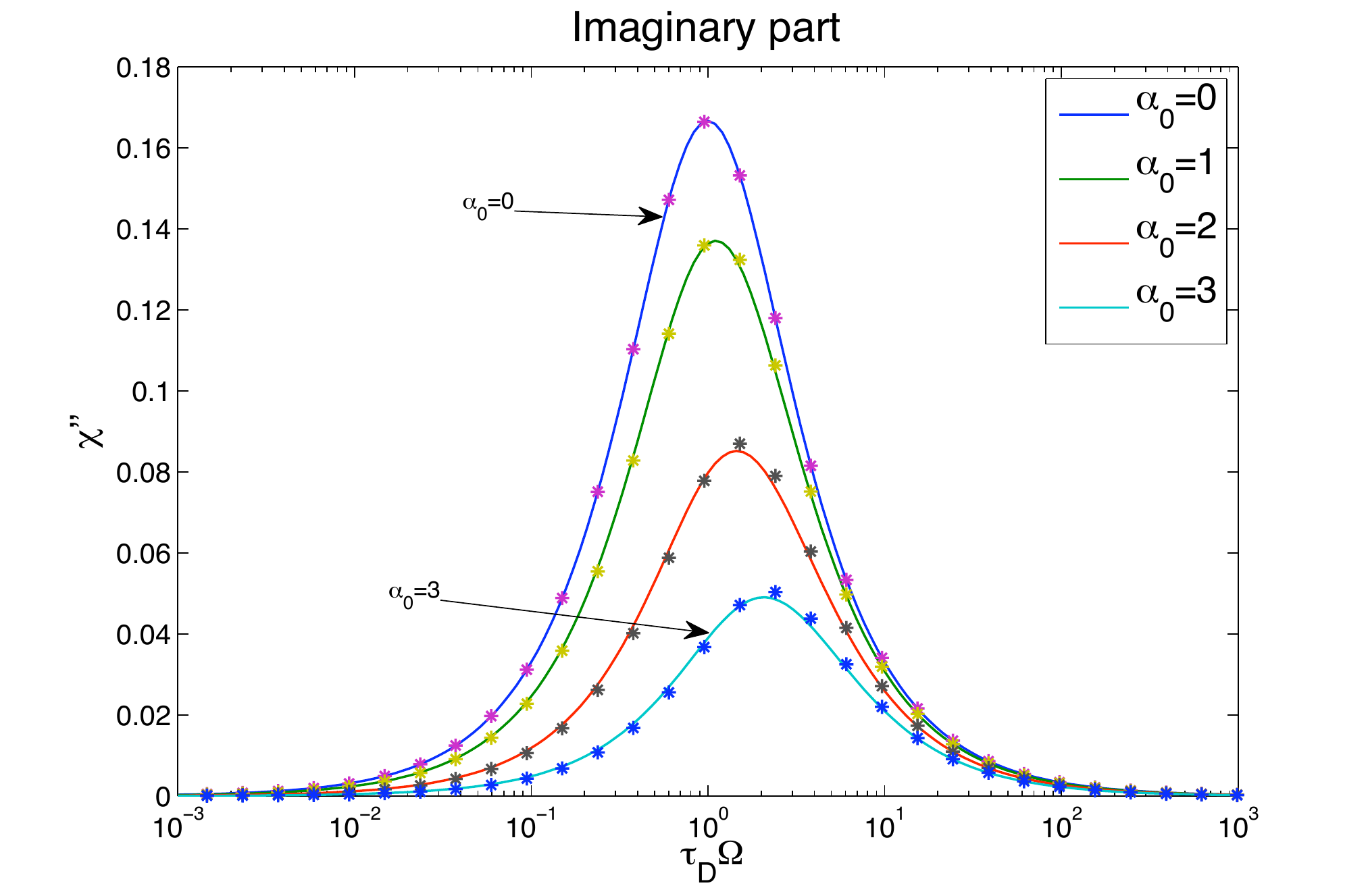}
	\caption{The linear susceptibility. The symbols are from the expression (\ref{c4_eff_debye}) while the solid lines are the 			             numerical results. Both real and imaginary parts decrease at large constant field, $\alpha_0$.}
	\label{fig:effective_sus}
\end{center}
\end{figure}
The expression is compared with the numerical results in Figure
\ref{fig:effective_sus} as a check of our numerical method, the top
curves ($\alpha_0=0$) of both panels. We also try a heuristic
expression for the linear susceptibility at non-zero DC field,
\begin{eqnarray}  
	\chi  &=&L'(\alpha_0)\frac{1}{1+ \iu\, \Omega \,\tau_{\rm{eff}} }, \label{c4_eff_debye}
\end{eqnarray}
where the effective relaxation time is 
\begin{eqnarray}  
	\tau_{\rm{eff}}=\tau_{D}\, \frac{\alpha_0}{L(\alpha_0)}\Big[1-\frac{2}{\alpha_0}L(\alpha_0) -L^2(\alpha_0)\Big]. \label{c4_effective_sus}
\end{eqnarray}
(The effective time is defined as the initial slope of the relaxation
curve when there is a small change of the applied field \cite{coffey4}.)

We compare the expression with the numerical results in Figure
\ref{fig:effective_sus} and good agreement is observed. The reason the
derivative of the Langevin function, $L'(\alpha_0)$, enters can be
justified. Without driving, the average orientation takes the value of
$L(\alpha_0)$. Turn on the driving, we move back and forth along the
Langevin curve and the vertical movement depends on the slope
$L'(\alpha_0)$ of the curve. $\chi'(\Omega=0)$ decreases with
increasing constant field because of the decreasing slope of the
Langevin function. In the limit when the dipoles are nearly fully
aligned by the large constant field, it is more difficult to rotate
them, and their response drops.
\begin{figure}[h!]  
	\centering
		\includegraphics[scale=0.46]{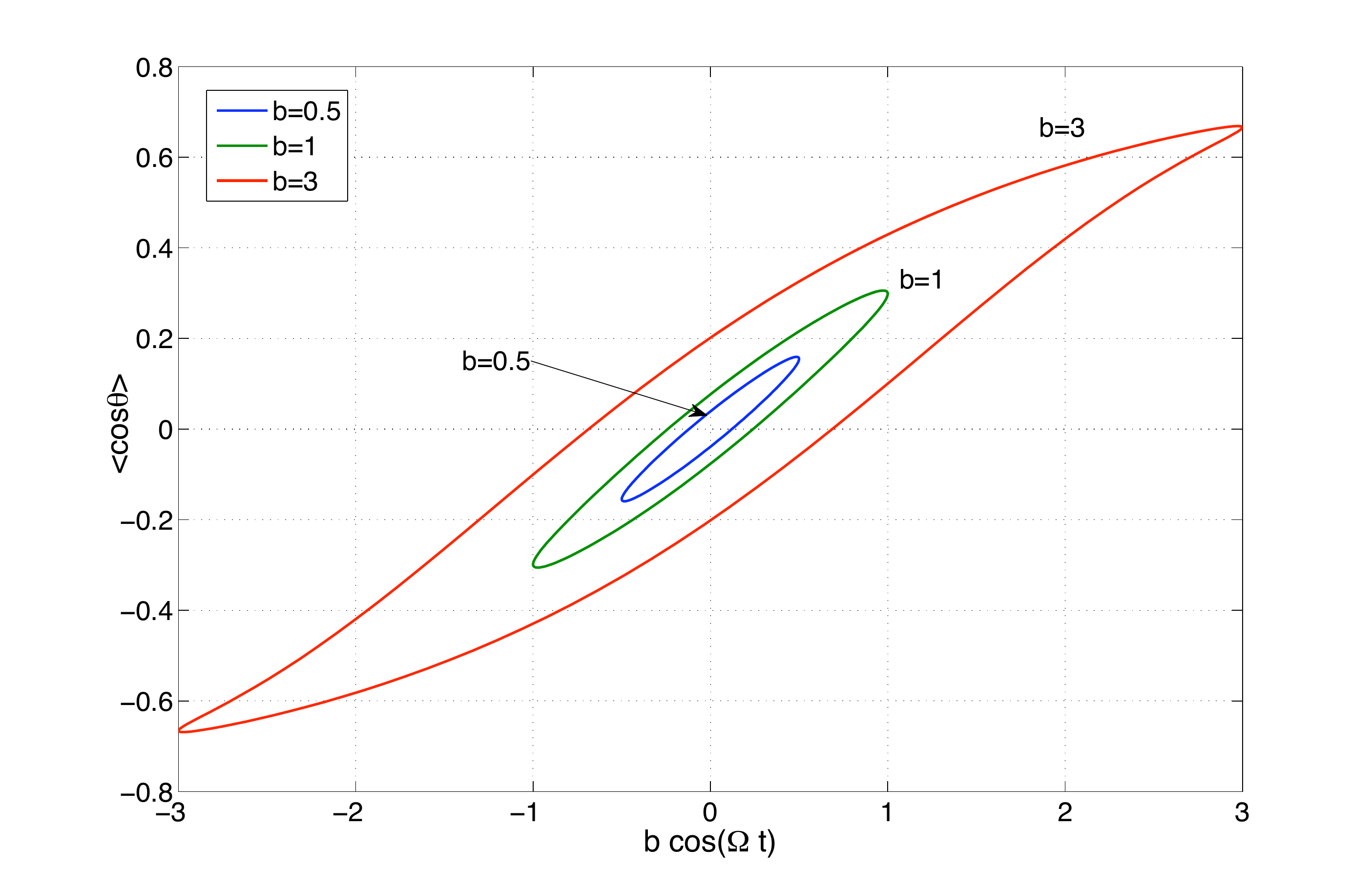}
		\includegraphics[scale=0.46]{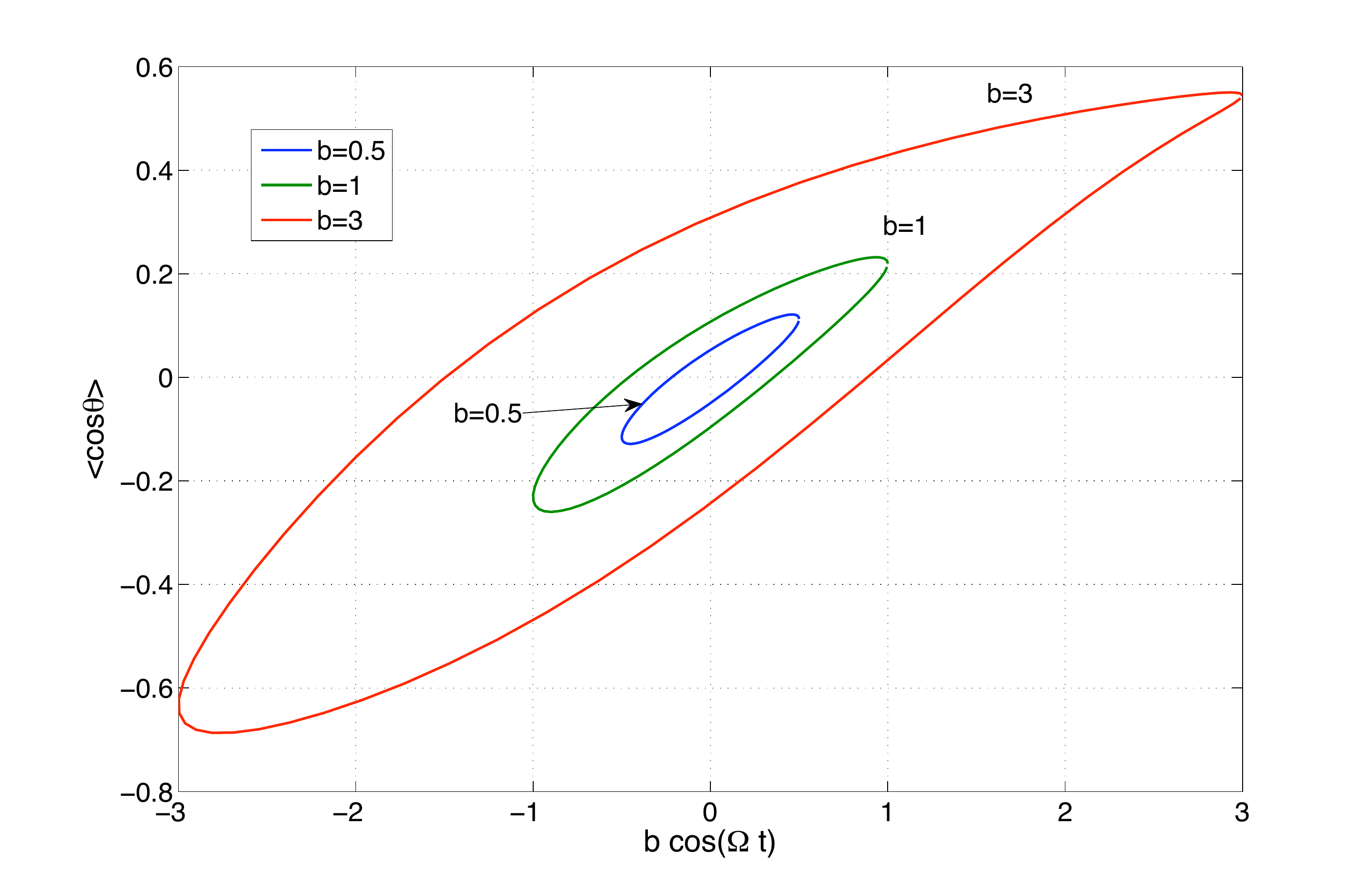}
	\caption{Hysteresis loops at  $\tau_D\Omega=0.5$ without (top) and with (bottom) constant field. The elliptical loops are deformed at large driving field, $b=\frac{pE_d}{k_BT}$. In the bottom figure, the constant field, $\alpha_0=0.5$, breaks the left-right symmetry.  }
	\label{fig:c4_hysteresis}
\end{figure}

Note that the power dissipated is proportional to the imaginary part
of the susceptibility, as opposed to the real part in the transport
problem.  It is due to the fact that we need to take the  time
derivative of $ \langle x \rangle_\Delta$ to get the ``velocity'',
giving $\langle\dot{x} \rangle_\Delta \propto \iu \Omega\,\, \langle x
\rangle _\Delta$. The imaginary unit exchanges the role of the real
and imaginary parts in the dissipation.

\newpage%
\subsection{Beyond Linear Response}
Beyond linear response, we solve for the polarization at arbitrary AC
field, and the hysteresis loops are plotted in Figure
\ref{fig:c4_hysteresis}. The deformation of the elliptical loops can
clearly be seen at large driving, due to the contribution of higher
harmonic susceptibilities. The loops develop a spike-like structure at
the tips, as in the custom hysteresis loops of magnetism.
%
\begin{figure}[tbh!]
	\begin{minipage}[b]{0.5\linewidth}
		\centering
		\includegraphics[scale=0.39]{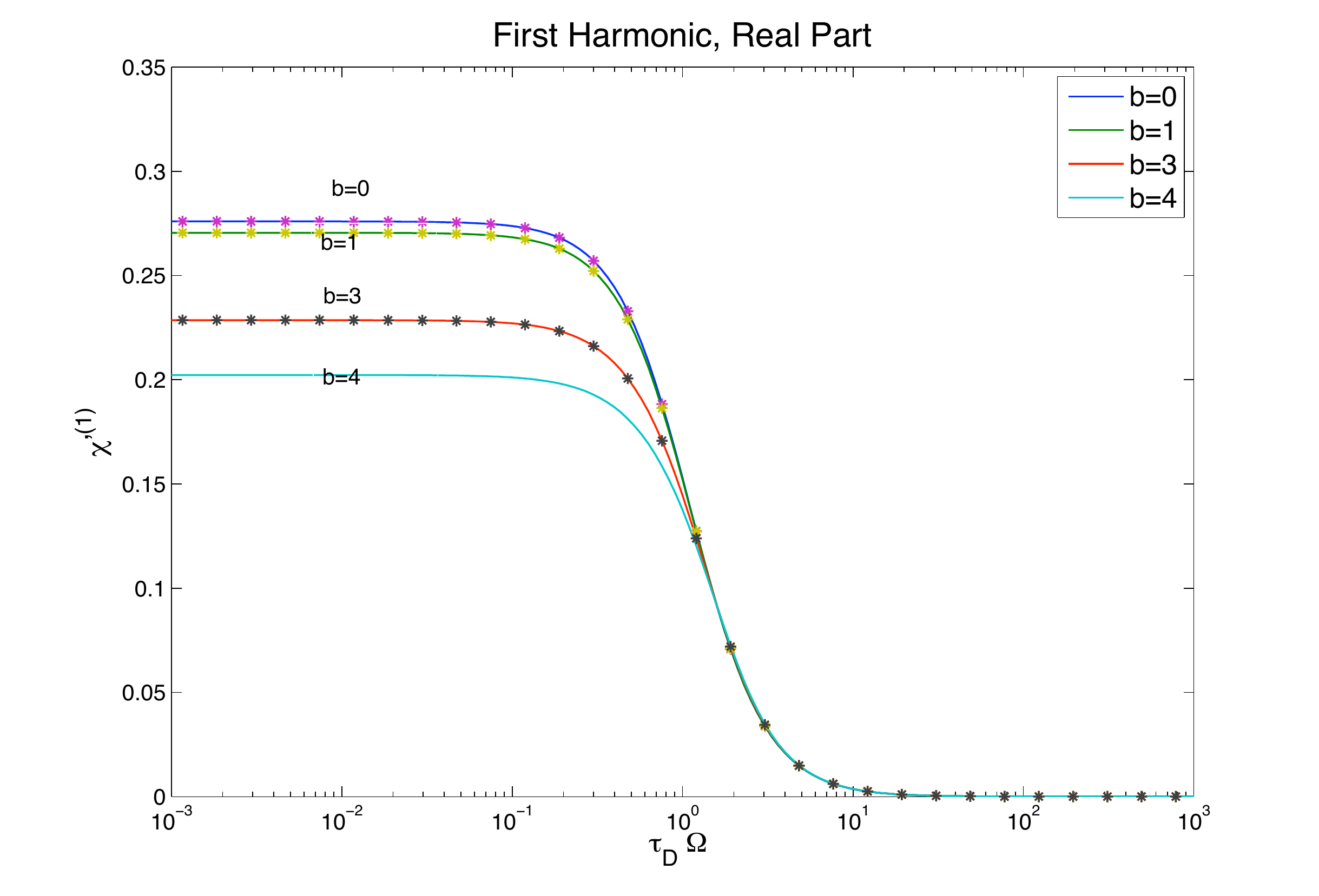}
	\end{minipage}
	\hspace{0.7cm}
	\begin{minipage}[b]{0.5\linewidth}
		\centering
		\includegraphics[scale=0.39]{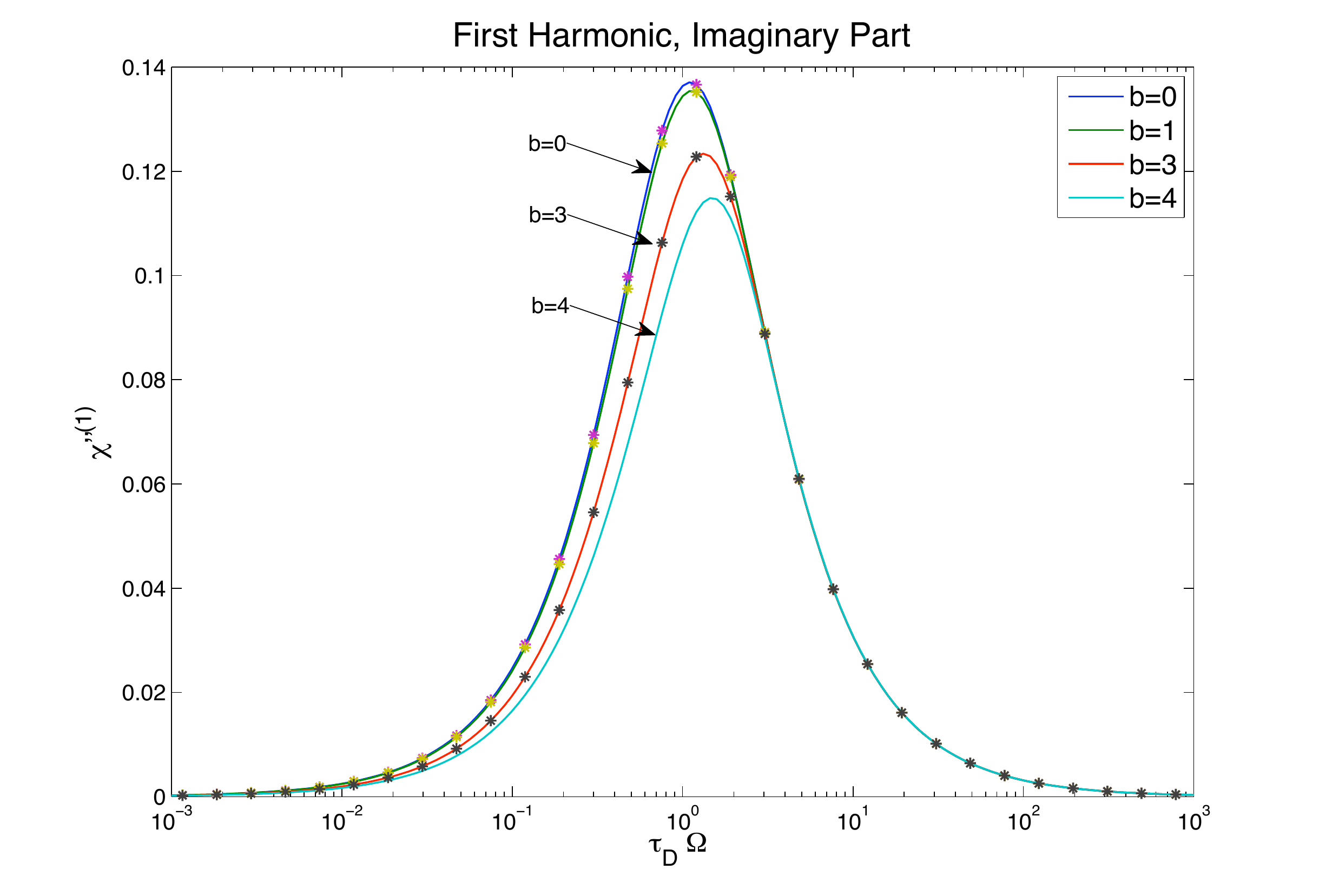}
	\end{minipage}

	\begin{minipage}[b]{0.5\linewidth}
		\centering
		\includegraphics[scale=0.39]{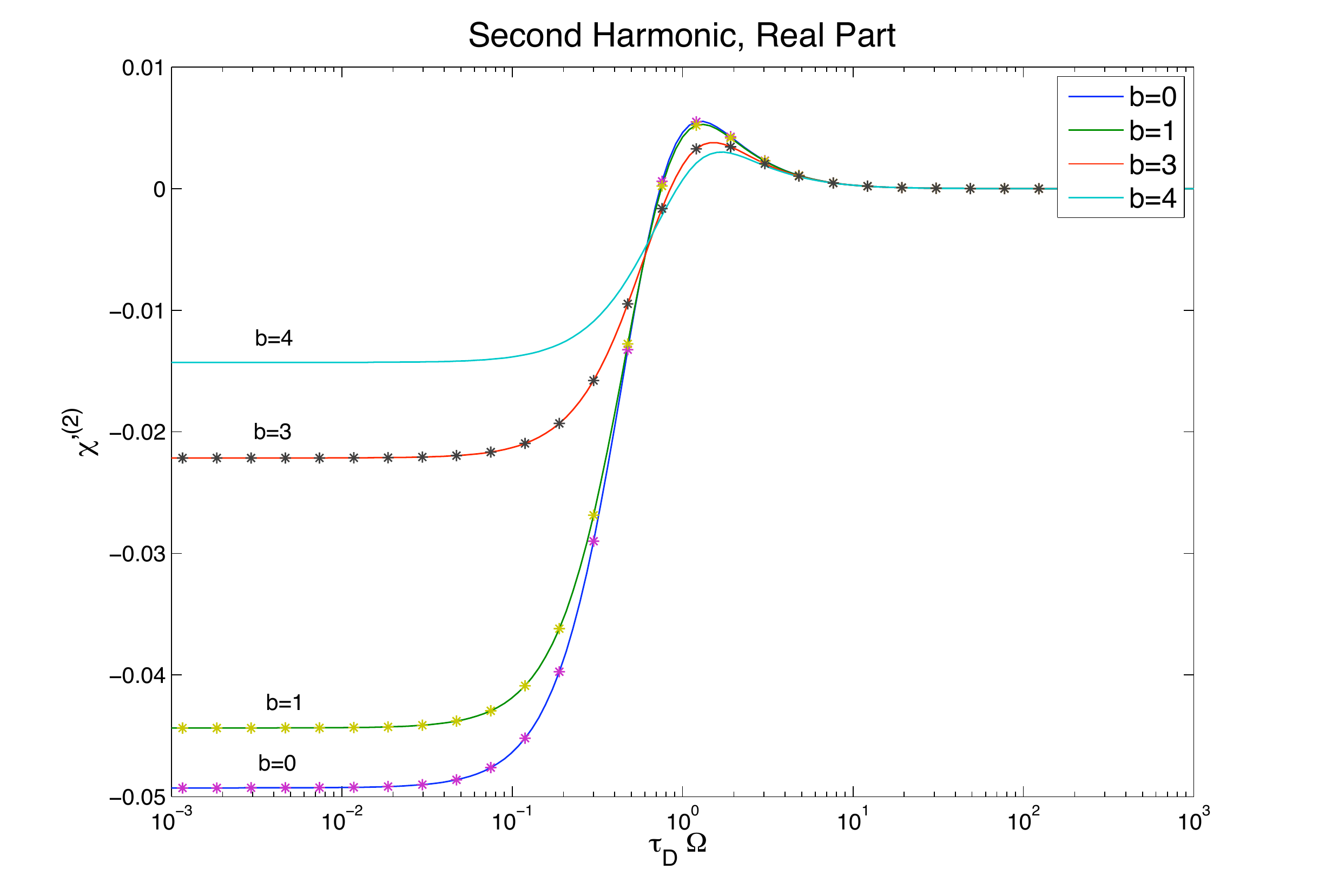}
	\end{minipage}
	\hspace{0.7cm}
	\begin{minipage}[b]{0.5\linewidth}
		\centering
		\includegraphics[scale=0.39]{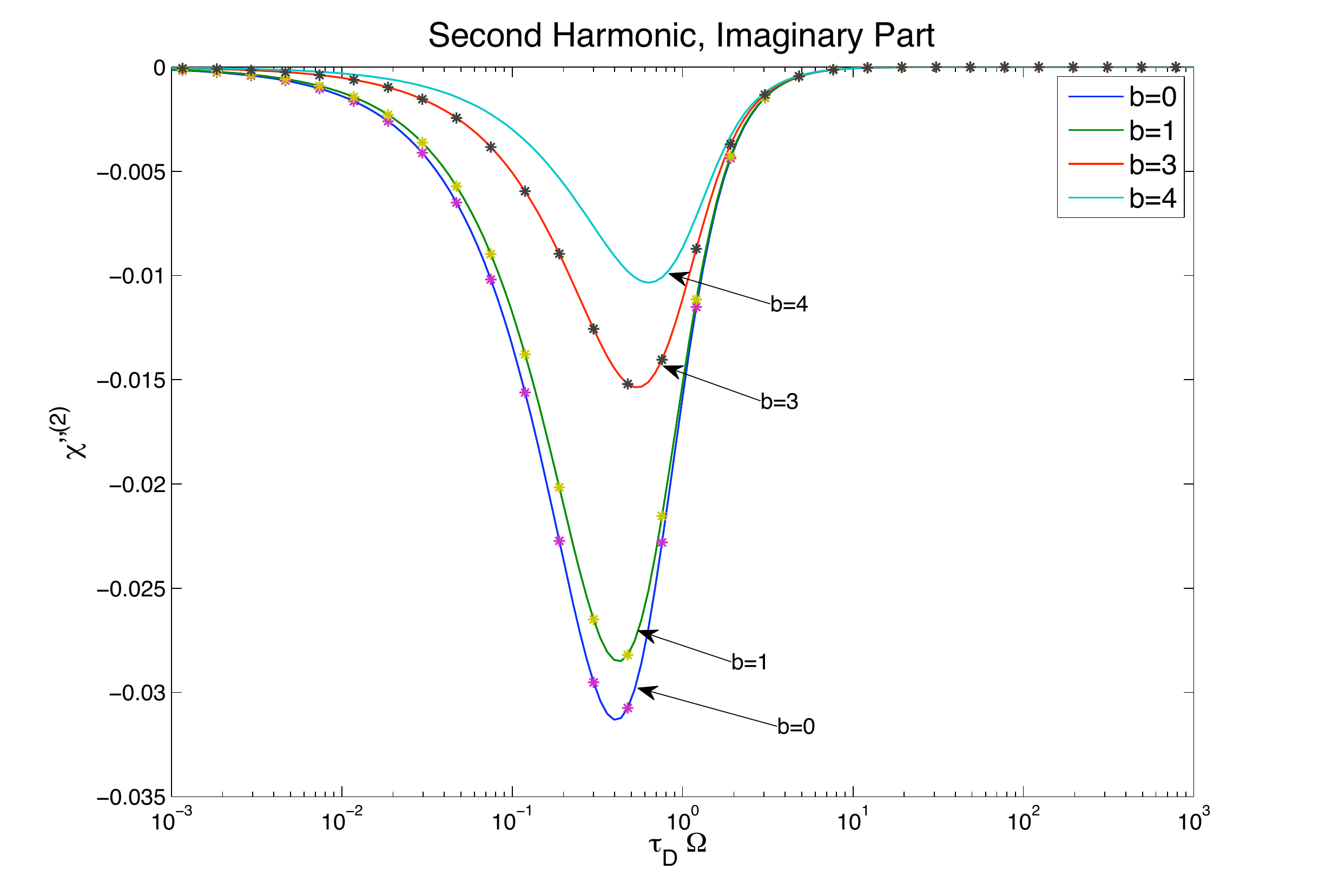}
	\end{minipage}

	\begin{minipage}[b]{0.5\linewidth}
		\centering
		\includegraphics[scale=0.39]{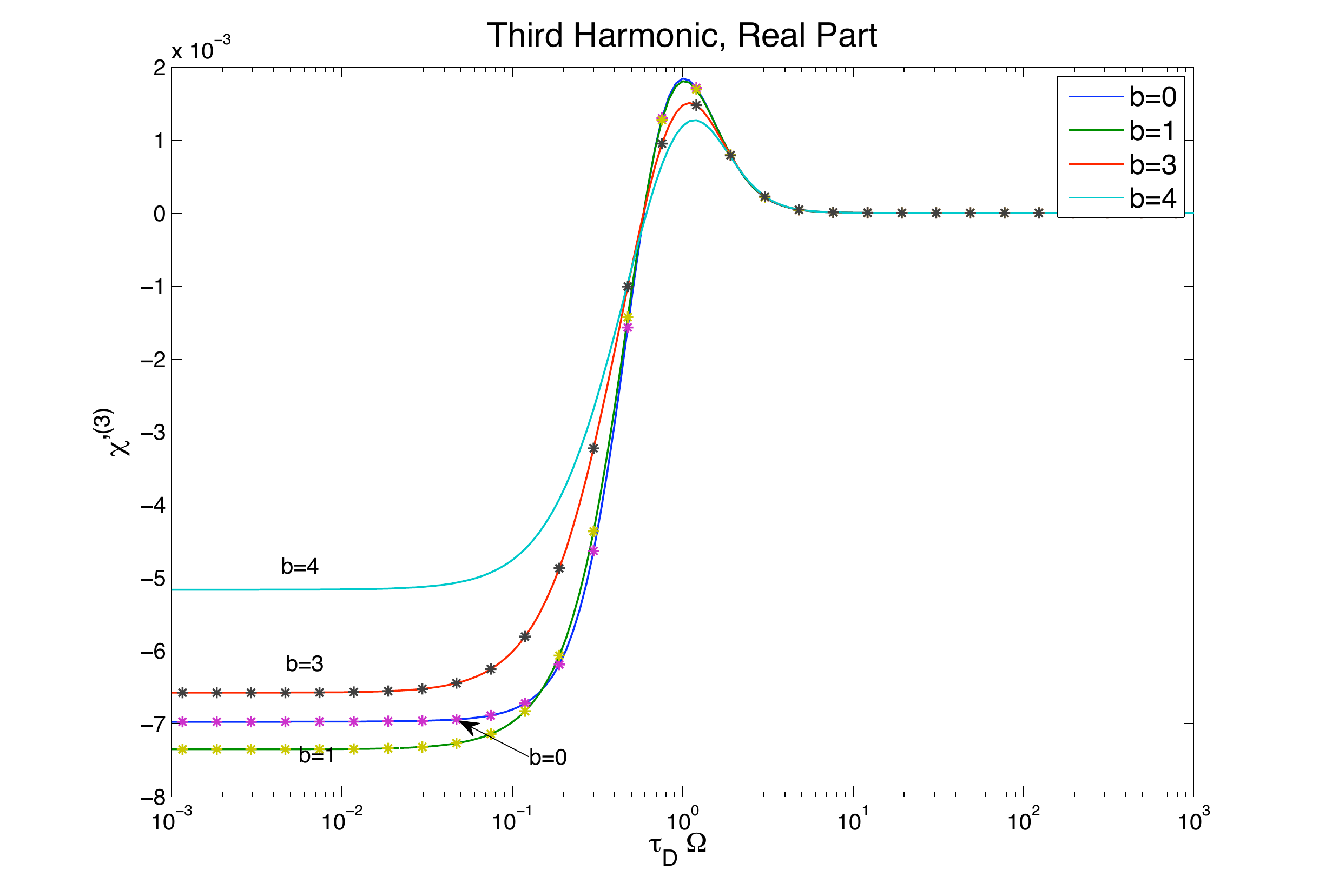}
	\end{minipage}
	\hspace{0.7cm}
	\begin{minipage}[b]{0.5\linewidth}
		\centering
		\includegraphics[scale=0.39]{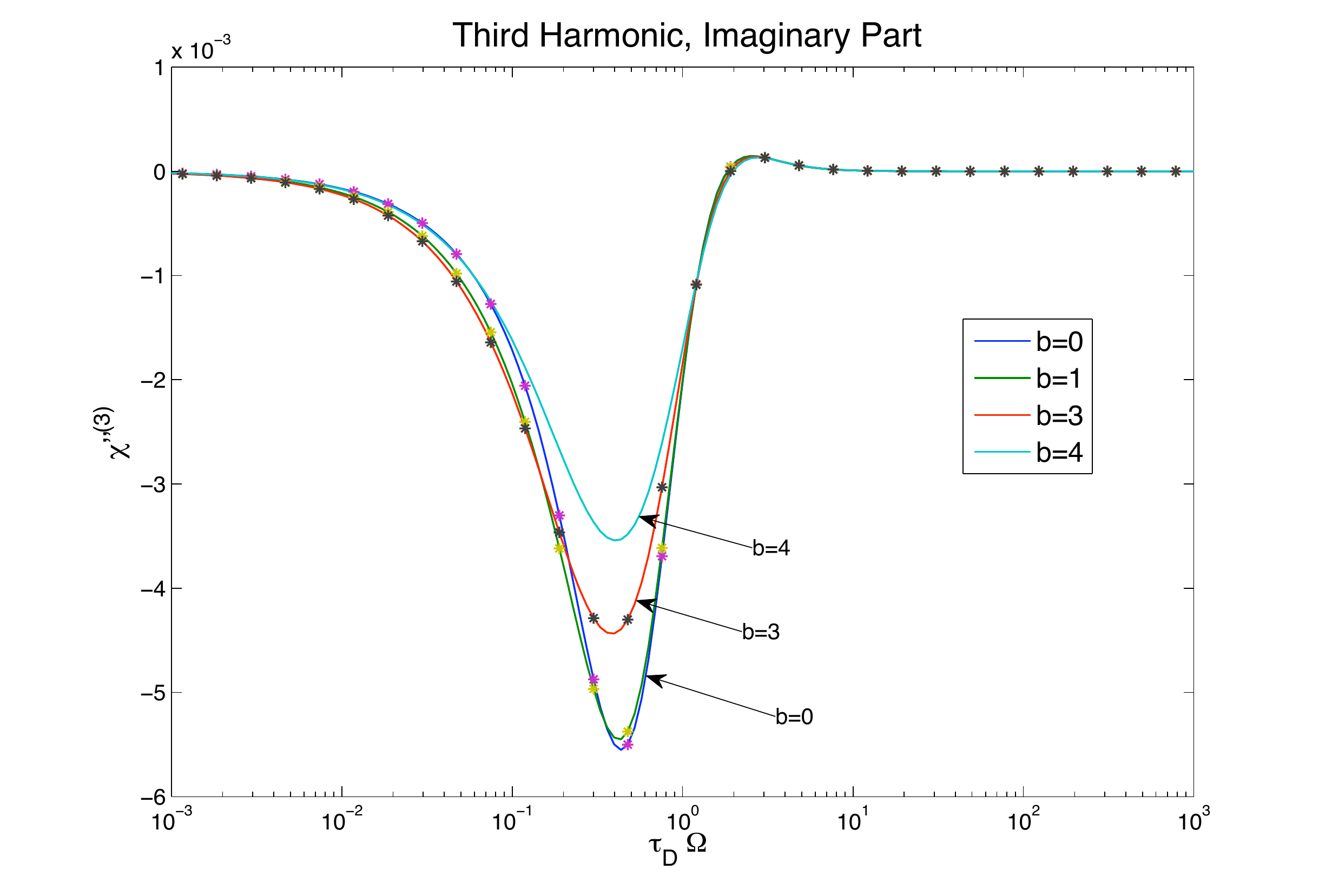}
	\end{minipage}
	\caption{The first three harmonics of the susceptibility at $\alpha_0=1$. The symbols are the results from the perturbative treatment, which fails at $b>3$. The curves of $b=0$ represent the linear response results, and are related to the derivative of the Langevin function.}
	\label{fig:c4_harmonics}
\end{figure}

We also plot the first three harmonics of the susceptibility in Figure
\ref{fig:c4_harmonics}. The susceptibility can no longer be described
by the phenomenological Eq. (\ref{c4_eff_debye}) at large driving (it
only works for $b\rightarrow 0$), as one can see from the
susceptibility curves. As we increase the driving amplitude, the
magnitude of the susceptibility at low frequency (adiabatic)
drops. This is also related to the decreasing slope of the Langevin
function.

The results from the perturbative treatment are plotted in Figure
\ref{fig:c4_harmonics} as a consistency check. The radius of
convergence for the perturbative treatment is about $b\approx3$, which
is large as compared to the previous chapter where $b\approx 0.1$. In
this range, we can already observe significant deviation from the
linear response results ($b=0$) before it breaks down (symbols in
Figure \ref{fig:c4_harmonics}). In the problem of Brownian particle in
a periodic potential, the perturbative expansion fails before we can
observe any significant non-linear effect.

\enlargethispage{1.cm}
%
\section{Discussion}
Here we end our discussion on classical open systems. To conclude, we
have presented the Hamiltonian viewpoint of open systems and two
equivalent ways of studying them: the Langevin and the \FP\
equations. We then used the \CF\ method to solve the \FP\ equations
for particle in a periodic potential and dipole under driving force,
and studied their linear and non-linear responses. We shall devote the
next two chapters to the study of open quantum systems.

\chapter{Quantum Open Systems}

\section{Hamiltonian and Reduced Description}
In quantum open systems, we study the fluctuation and dissipation on a quantum system due to the interaction with the environment. Unlike the classical counterparts, it is difficult to introduce phenomenological equations to describe these effects, because of the unitarity of the quantum dynamics. Previous attempts are plagued with various problems, e.g. violating the uncertainty principle or the superposition principle \cite{weiss}.  
\\

	Therefore, it is natural to view an open quantum system as a system with a few degrees of freedom coupled to a bath with many (infinite) degrees of freedom. The total \Ha, $\totHa$,  describing such a model is written as 
\begin{eqnarray} 
		 \totHa     &=&  \sysHa \otimes I+ I\otimes \bathHa + \CouHa \\ 
		 			&=&  H + \CouHa, \nonumber
\end{eqnarray}
where $H$ is the free Hamiltonian of the system plus bath and $\CouHa$ is their interaction \Ha. 
%
%
%
The combined system is fully described by the total density matrix, $\rho_{\rm{tot}}$, whose dynamics is governed by the von Neumann equation
\begin{eqnarray} 
		 \frac{d \rho_{\rm{tot}}}{dt}= \frac{\iu}{\hbar}[\rho_{\rm{tot}}, \totHa].
\end{eqnarray}
\\

 However, in most cases, we are only interested in the properties and evolution of the system (we do not have much control on the bath). We then introduce the reduced density matrix of the system,  obtained by partial tracing over the bath degrees of freedom, 
\begin{eqnarray} 
		\rho=\mbox{Tr}_B[\rho_{\rm{tot}}].
\end{eqnarray}
The reduced density matrix contains all the information we need in most cases of interest (heat transport is an exception though). In the following sections, we will derive a differential equation to describe its time evolution using the second-order perturbation theory.

\section{Perturbation Theory in System-Bath Coupling}
\subsection{Evolution Operator}
In general, we are not able to obtain an exact \EoM\ for the reduced density matrix similar to the classical equation Eq. (\ref{EoM2}); we have to resort to perturbation theory. We start with the evolution operator, 
\begin{eqnarray} 
		 U(t,t_0)=\mbox{e}^{-\frac{\iu}{\hbar}\totHa\,(t-t_0)}.
\end{eqnarray}
To facilitate the perturbative treatment, we use the identity \cite{kubo} 
\begin{eqnarray} 
	\mbox{e}^{\beta(A+B)}=\mbox{e}^{\beta A}\Big [ 1+ \int^{\beta}_{0} d\lambda \,
												\mbox{e}^{-\lambda A}\, B \,\mbox{e}^{\lambda (A+B)}\Big].
\end{eqnarray}
It can be confirmed by multiplying both sides by $\mbox{e}^{-\beta A}$ and differentiating with respect to $\beta$.  Using this identity, the evolution operator can be expressed in the following Dyson-like form
\begin{eqnarray} 
	U(t,t_{0}) = e^{-\frac{\iu}{\hbar}H(t-t_{0})} \Big[ 1-\frac{\iu}{\hbar} \int^{t-t_{0}}_{0} ds \, e^{\frac{\iu}{\hbar}H \,s}  \CouHa \, e^{-\frac{\iu}{\hbar}(H + \CouHa)\, s}\Big].
\end{eqnarray}
Let us make the transformation $s\rightarrow s-t_0$ and keep up to the first order term in $ \CouHa$, we obtain
\begin{eqnarray} 
	U(t,t_{0}) & \approx &  U_0(t,t_0) \Big[ 1 - \frac{\iu}{\hbar}\int^t_{t_0}ds\,\intCouHa(s) \Big], \label{c5_1st_evo}
\end{eqnarray}
where
\begin{eqnarray} 
	U_0(t,t_{0}) &=& \mbox{e}^{-\frac{\iu}{\hbar}H (t-t_0)}; \\
	\intCouHa(s)   &=& U^\dagger_0(s,t_{0})\CouHa \,U_0(s,t_{0}). \label{c5_free_evo}
\end{eqnarray}
%
\subsection{Heisenberg Equation}
Now instead of looking at the evolution of the reduced density matrix, we look at the Heisenberg equation of an operator acting only on the system's Hilbert space,
\begin{eqnarray} 
	\frac{dA(t)}{dt} &=& \frac{\iu}{\hbar}[\totHa,A(t)]\nonumber \\
						&=&  \frac{\iu}{\hbar}[\sysHa(t),A(t)]+ \frac{\iu}{\hbar}[\CouHa(t),A(t)].
\end{eqnarray}
Other than telling us the evolution of an operator (say momentum or position operator of the system), the above equation also allows us to study the evolution of the reduced density matrix upon choosing an appropriate operator, as what we will do in the next section.  
\\

Making use of Eq.~(\ref{c5_1st_evo}) and the similarity transformation property
\begin{eqnarray} 
	[\CouHa(t),A(t)] &= & U^{\dagger}(t,t_{0})[\CouHa\,,\, A]U(t,t_{0}),
\end{eqnarray}
the Heisenberg equation becomes
\begin{eqnarray} 
	\frac{dA(t)}{dt} &=& \frac{\iu}{\hbar}[H_{sys}(t),A(t)]+ \frac{\iu}{\hbar}[\intCouHa(t),\tilde{A}(t)]\nonumber \\
	                              & &-\frac{1}{\hbar^2}\int^t_{t_0}ds \Big \{ \,\intCouHa(s) \Big[\intCouHa(t),\tilde{A}(t) \Big]\,\, +\,\,\Big [\tilde{A}(t),\intCouHa(t)\Big] \intCouHa(s)\, \Big \}.
\end{eqnarray}
The second order perturbative structure is clear. The first term arises from the free evolution, the second and third terms are the first and second order corrections, respectively. It is worth recalling that we are still in the Heisenberg picture, while the operators with tildes are evolved by its free \Ha\ [cf. Eq.~(\ref{c5_free_evo})].
\\

  We have obtained a generic \EoM\ for a system operator based on perturbation theory. In the next few sections, we will make use of this equation to derive the so-called master equation: the \EoM\ for the reduced density matrix.

%
\section{Quantum Master Equation (Bloch-Redfield)}
\subsection{Hubbard Operators}
We introduce the Hubbard operators $X_{nm}= |n\rangle \langle m|$, where $\{|n\rangle\}$ is the set of energy-eigenstates of the system Hamiltonian,
\begin{eqnarray} 
		\sysHa |n\rangle =\epsilon_{n} |n\rangle.
\end{eqnarray}
The properties of the Hubbard operators can be found in Appendix A.2. The Heisenberg equation of the Hubbard operator is
 \begin{eqnarray} 
	\frac{d X_{nm}(t)}{dt}&= &\frac{\iu}{\hbar}\Delta_{nm} X_{nm}(t) + \frac{\iu}{\hbar}[\intCouHa(t), \tilde{X}_{nm}(t)] +\hat{R}_{nm} ,\nonumber
\end{eqnarray}
where the energy level difference is $\Delta_{nm}= \epsilon_n -\epsilon_m$ and the relaxation term reads
\begin{eqnarray} 
	\hat{R}_{nm} &= & -\frac{1}{\hbar^2}\int_{t_0}^t \,ds\, \Big \{\intCouHa(s) [\intCouHa(t),\tilde{X}_{nm}(t)] + [\tilde{X}_{nm}(t),\intCouHa(t)] \intCouHa(s) \Big \}.
\end{eqnarray}

We introduced the Hubbard operators because they are useful in getting the elements of the reduced density matrix by tracing,
\begin{eqnarray} 
	\rho_{mn}(t) =\mbox{Tr}_s\Big[\rho(t_0)X_{nm} (t)\Big].
\end{eqnarray}
Before we could obtain the master equation, we need to make a few more assumptions.

\subsubsection*{Decoupled Initial Condition}
The factorized initial condition
\begin{eqnarray} 
	\rho_{\rm{tot}}(t_0) =\rho(t_0)\otimes \rho_{\rm{bath}}(t_0),
\end{eqnarray}
is a vital assumption in the process of tracing and getting the matrix elements. We will discuss the decoupled initial condition at the end of this chapter.
\subsubsection*{Coupling Structure and Bath ``Centering"}
The coupling is in standard factorized form (additional summation would give the general case, but this just brings notational changes),
\begin{eqnarray} 
	\CouHa = F\otimes \bathCou.
\end{eqnarray}
The coupling is defined such that the first moment of the bath Hamiltonian vanishes, $\langle \bathCou \rangle =0$. If this were not the case, we redefine the coupling, and lump the non-zero average into the system Hamiltonian [this is equivalent to the zero average of the Langevin force in Eq. (\ref{c2_noise})].  
%
\subsection{Bloch-Redfield Equation}
With these conditions, the \EoM\ obtained after tracing is
\begin{eqnarray} 
	\frac{d \rho_{mn}(t)}{dt}&=&\frac{\iu}{\hbar}\Delta_{nm} \rho_{mn}(t) +R_{mn} \label{c5_EOM_element},
\end{eqnarray}
where the relaxation term is
\begin{eqnarray} 
	R_{mn}=\frac{1}{\hbar^2}\sum_{n'm'}\rho_{m'n'}(t)& \Big[&(\TranUn_{nn'}^{*}+\TranUn_{m'm})F_{n'n}F_{mm'} \\ \nonumber
	 																	     &&-\sum_{l}\delta_{mm'}F_{n'l}F_{ln} \TranUn_{ln'}^{*} \\ \nonumber
																		     	&&-\sum_{l}\delta_{nn'}F_{ml}F_{lm'} \TranUn_{lm'} \Big ],
\end{eqnarray}
in which $\langle n|F|m \rangle =F_{nm}$. The transition rate is
\begin{eqnarray} 
	\TranUn_{nm}&=&\int_{t_0}^{t} ds e^{ \iu \Delta_{nm} (s-t)/\hbar }\,\langle \bathCou(t) \bathCou(s) \rangle \\
	             &=&\int_{0}^{t-t_0} d\tau e^{- \iu \Delta_{nm} \tau /\hbar }\, K(\tau) \label{c5_tran_rate},
\end{eqnarray}
where the bath correlator is $K(\tau)= \langle \bathCou(\tau) \bathCou \rangle$. The real part of the transition rate determines the speed at which the stationary solution is reached. Upon taking the initial time at $t_0\rightarrow -\infty$, the transition rate becomes a half Fourier transfrom of the bath correlator. 
\\

Equation  (\ref{c5_EOM_element}) is frequently called the Bloch-Redfield equation \cite{redfield}, which is widely used in magnetic resonance (nuclear, electron, or ferromagnetic), optical spectroscopy, laser physics, and electron-transfer reactions in molecules and bio-molecules.  This equation is generic, we have not specified our system, bath or the coupling.  All the information of the bath goes into the correlator $K(\tau)$, and we will need to specify the bath coupling, $\bathCou$, the spectral density $J(\omega)$ and its initial state $\rho_{\rm{bath}}(t_0)$. 
\subsection{Ladder Couplings}
Let us consider the couplings of the type
\begin{eqnarray} 
	\langle n|F|m \rangle &=&F_{nm} =L_{m}^{-}\delta_{n,m-1} + L_{m}^{+}\delta_{n,m+1}. 
\end{eqnarray}
It is suited to study the harmonic oscillator problem with coordinate/momentum coupling 
\begin{eqnarray} 
		F=\eta(a+a^\dagger),
\end{eqnarray}
or spin problems with general couplings of the form
\begin{eqnarray} 
		F = \eta_{+}\Big \{ v(S_z), S_{-}  \Big \} + \eta_{-}\Big \{v(S_z), S_{+}  \Big \},
\end{eqnarray}
where $\{\cdot, \cdot\}$ is the anti-commutator and $\eta_{\pm}$ describes the symmetry of the interaction \cite{jose2}. 
\\

With such coupling, the relaxation term becomes
\begin{eqnarray} 
		\hbar^2 R_{mn} &=& -\Big( L_{n}^{+}L_{n+1}^{-} \TranUn^{*}_{n+1, n} +  L_{n-1}^{+}L_{n}^{-} \TranUn^{*}_{n-1, n} 
									+ L_{m}^{+}L_{m+1}^{-} \TranUn_{m+1, m} +  L_{m-1}^{+}L_{m}^{-} \TranUn_{m-1, m}\Big)\, \rho_{mn} \nonumber \\
						&& +\Big(\TranUn^{*}_{n,n-1} + \TranUn_{m,m-1}  \Big)L_{n}^{-} L_{m-1}^{+}\, \rho_{m-1, n-1} \nonumber \\
						&& +\Big(\TranUn^{*}_{n,n+1} + \TranUn_{m,m+1}  \Big)L_{n}^{+} L_{m+1}^{-}\, \rho_{m+1, n+1} \nonumber \\
						&& +\Big(\TranUn^{*}_{n,n-1} + \TranUn_{m,m+1}  \Big)L_{n}^{-} L_{m+1}^{-}\, \rho_{m+1, n-1} \nonumber \\ 		
						&& +\Big(\TranUn^{*}_{n,n+1} + \TranUn_{m,m-1}  \Big)L_{n}^{+} L_{m-1}^{+}\, \rho_{m-1, n+1} \nonumber \\ 		 
						&& - \TranUn^{*}_{n-1,n-2} L_{n-1}^{-} L_{n}^{-}\, \rho_{m, n-2} 
						      - \TranUn^{*}_{n+1,n+2} L_{n}^{+} L_{n+1}^{+}\, \rho_{m, n+2} \nonumber \\
					     && - \TranUn_{m-1, m-2} L_{m-1}^{+} L_{m-2}^{+}\, \rho_{m-2, n} 
						      - \TranUn_{m+1,m+2} L_{m+1}^{-} L_{m+2}^{-}\, \rho_{m+2, n}. 
\end{eqnarray}
%
\subsection{Secular Approximation}
At this stage, one invariably invokes the secular approximation, discarding the terms of the type $L^{+}L^{+}$ and $L^{-}L^{-}$. We basically get rid of the last four lines of the relaxation term above and keep
\begin{eqnarray} 
		\hbar^2 R_{mn} &=& -\Big( L_{n}^{+}L_{n+1}^{-} \TranUn^{*}_{n+1, n} +  L_{n-1}^{+}L_{n}^{-} \TranUn^{*}_{n-1, n} 
									+ L_{m}^{+}L_{m+1}^{-} \TranUn_{m+1, m} +  L_{m-1}^{+}L_{m}^{-} \TranUn_{m-1, m}\Big)\, \rho_{mn} \nonumber \\
						&& +\Big(\TranUn^{*}_{n,n-1} + \TranUn_{m,m-1}  \Big)L_{n}^{-} L_{m-1}^{+}\, \rho_{m-1, n-1} \nonumber \\
						&& +\Big(\TranUn^{*}_{n,n+1} + \TranUn_{m,m+1}  \Big)L_{n}^{+} L_{m+1}^{-}\, \rho_{m+1, n+1}. \label{c5_relaxation}
\end{eqnarray}
 We will justify this approximation in the discussions at the end of the chapter. After the secular approximation, the coupling of the matrix elements becomes simpler. Any matrix element, $\rho_{mn}$, is only coupled to its adjacent diagonal neighbors, $\rho_{m+1,n+1}$ and $\rho_{m-1,n-1}$. This short-ranged coupling simplifies the implementation of the \CF\ method.

\section{Application to the Bath-of-Oscillators Model}
\subsection{Hamiltonian Redux}
In this model, the system is coupled linearly to the coordinates of a bath of oscillators ($\bathCou=\sum_\alpha^N{c_\alpha}x_\alpha$), as in the classical case. The \Ha\ is 
 \begin{eqnarray}
 	\totHa=\sysHa+\sum_{\alpha=1}^{N} \Big[  \frac{p_{\alpha}^2}{2m_{\alpha}} + 	
				\frac{1}{2}m_\alpha \omega_\alpha^2\Big(x_\alpha -\frac{\coupl_\alpha}{m_\alpha \omega_\alpha^2}F \Big)^2   \Big]. \label{c5_tot_H}
\end{eqnarray}
Because of the close resemblance to the classical counterpart, this model gains the name of ``Quantum Brownian Motion". The only difference is that all $x$'s and $p$'s are now quantum operators. For instance, we will substitute the bath coordinates with the bosonic operators, $x_\alpha \propto \frac{1}{\sqrt{2}} \Big(a_\alpha^{\dagger}+a_\alpha \Big)$.
%
%
\subsection{Bath Correlator}
The bath is assumed to be at thermal equilibrium initially,
\begin{eqnarray}
		\rho_{\rm{bath}}(t_0)=\frac{\mbox{e}^{-\beta \,\bathHa}}{Z}.
\end{eqnarray}
The bath correlator thus reads
\begin{eqnarray} 
	K(\tau)=\langle \bathCou(\tau)\, \bathCou \rangle 
			&=& \hbar \,\sum_{\alpha}\frac{c_\alpha^2}{2m_\alpha \omega_\alpha}\langle (a_\alpha^{\dagger} \mbox{e}^{\iu \omega_\alpha \tau}+
					a_\alpha \mbox{e}^{-\iu \omega_\alpha \tau})(a_\alpha^{\dagger}+a_\alpha)\rangle  \\ \nonumber
			&=& \hbar \int^{\infty}_{0}\frac{d \omega}{\pi}J(\omega)
					 \Big[n_{\omega}e^{\iu \omega \tau} +(n_{\omega}+1)e^{-\iu \omega \tau} \Big]\\ \nonumber
	          &=&  \hbar \int^{\infty}_{0}\frac{d \omega}{\pi}J(\omega)\Big[ \coth(\frac{1}{2}\beta \hbar \omega)\cos(\omega \tau)- \iu \sin(\omega\tau)\Big],
\end{eqnarray}

\noindent where the Bose function is
\begin{eqnarray} 
	n_{\omega}=1/(e^{\beta\hbar \omega}-1).
\end{eqnarray}
At high temperature, the real part of the correlator behaves classically, as in Eq. (\ref{damping_kernel}). The imaginary part arises from the non-commutability of the bosonic operators and has no classical analog. 
%
\subsection{Relaxation Coefficients}
\subsubsection*{Counter-Term and Renormalized Rate}
The counter-term $F^2$, can be handled by modifying the transition rate as
\begin{eqnarray} 
	W \rightarrow W+ \iu\, \hbar \gamma(0),
\end{eqnarray}
where $\gamma(0)$ is the damping kernel evaluated at $t=0$ [cf. Eq. (\ref{damping_kernel2})]
 \begin{eqnarray} 
	\gamma(0)= \frac{1}{M}\int^\infty_0\frac{d\omega}{\pi} \frac{J(\omega)}{\omega}.
\end{eqnarray}
\subsubsection*{Drude-Ohmic Damping}
We discussed in the Rubin model (cf. Section 2.3.4) that, in a physical model, the density of the bath modes is cut-off at high frequency. A frequently used model satisfying this condition is the Ohmic spectral density with Drude cut-off,
\begin{eqnarray} 
	J(\omega)=\frac{\gamma M\,\omega}{1+{\omega^2 }/{\omega_D^2}}.
\end{eqnarray}
The damping coefficient, $\gamma$, keeps track of the coupling strength, and is proportional to $c_\alpha ^2$ in the total Hamiltonian (\ref{c5_tot_H}). This should not be confused with the gamma with time argument, which is the damping kernel. 
\\

For the above spectral density, transition rate is found to be 
\begin{eqnarray} 
	\TranN_{mn}					&=& \TranN(\Delta_{mn}); \label{c5_trans-rate} \\
	\mbox{Re}[\TranN(\Delta)]	&=& \frac{\gamma \Delta }{1+(\Delta/ \hbar \omega_D)^2}\, \frac{1}{e^{\beta  \Delta}-1};\nonumber \\
	\mbox{Im}[\TranN(\Delta)] 	&=& \frac{\gamma}{2\hbar\omega_D} \,\,\frac{ \Delta^2}{1+(\Delta/ \hbar \omega_D)^2} 
										+ \frac{\gamma\Delta}{\pi[1+(\Delta/ \hbar \omega_D)^2]}\Big\{ \psi(\frac{\beta \hbar \omega_D}{2 \pi}) 
										-\mbox{Re}[\psi(\frac{\beta  \Delta}{2 \pi}\iu)]+ \frac{\pi}{\beta \hbar \omega_D} \Big\}. \nonumber
\end{eqnarray}
$\psi$ is the Psi function defined as the derivative of the logarithm of gamma function, $\psi(x)=\frac{d}{dx}\mbox{ln}\Gamma(x) $. Some properties of the transition rate are discussed in Appendix A.3.

\section{Discussion of the Approximations}
At this point, we have our master equation ready for use, we just need to specify the spectrum (energy levels) of the system. But before we try to solve for any system, there is a need to discuss some of the subtleties of the master equation.
\subsubsection*{Weak Coupling} 
We need to be more specific on what we mean by weak coupling. We assume $\CouHa$ is small, and thus $R_{mn}$ can be treated as a perturbation to the free evolution. But letting $|t-t_0|\rightarrow\infty$, the integral in the transition rate $W$ [see Eq.~(\ref{c5_tran_rate})] would become very large and the perturbation theory breaks down. In many problems of interest, there exists a correlation time $\tau_{c}$, such that the correlator is negligible $K(\tau_c)\approx 0$ after $\tau_c$. Thus, the integral would not grow as we feared. The weak coupling approximation is then valid in the regime of 
\begin{eqnarray} %
	\gamma \, \tau_c \ll 1. \label{c5:weak_coupl_cond}
\end{eqnarray}
%


\subsubsection*{Secular Approximation} 
The secular approximation is equivalent to the rotating wave approximation (RWA) in quantum optics. Specifically, a coupling of the form, $V\propto F_{-}(a+a^\dagger)+F_{+}(a+a^\dagger)$ is reduced to  $V\propto F_{-}a^\dagger+F_{+}a$. It is argued that the secular term $F_{+}a^\dagger$ is rotating at $\mbox{e}^{\iu(\omega+\Delta)t}$, which is much faster than the term $F_{-}a^\dagger \propto \mbox{e}^{\iu(\omega-\Delta)t}$, and can be averaged out.  In the case of $\Delta$ being negative, one discards the terms $F_{-}a^\dagger$ and $F_{+}a$ instead, since they are the ones which are rotating faster. 

\subsubsection*{Decoupled Initial Condition}
We have chosen the factorized initial condition to facilitate the process of tracing. In the context of condensed matter, the system and bath do not meet at our chosen time, they have always been in contact. Thus, we set the initial time at minus infinity, and hope that the artificial initial condition is forgotten by the time we start to manipulate the system at $t=0$ \cite{weiss}. 
\\

In the next chapter, we will use the master equation to study a damped quantum harmonic oscillator and make comparison with some exact results as bench-marking.

\chapter{Quantum Harmonic Oscillator}

\section{Introduction}
In this chapter, we study the properties of a quantum \HO\ linearly coupled in coordinate to the bath. The \Ha\ is 
 \begin{eqnarray}
 	\totHa=\frac{p^2}{2M}+ \frac{1}{2}M\omega_0^2 x^2+\sum_{\alpha=1}^{N} \Big[  \frac{p_{\alpha}^2}{2m_{\alpha}} + 	
				\frac{1}{2}m_\alpha \omega_\alpha^2\Big(x_\alpha -\frac{\coupl_\alpha}{m_\alpha \omega_\alpha^2}x \Big)^2   \Big].
\end{eqnarray}
The study of the damped  quantum \HO\ is important because this model applies to any system slightly displaced from its stable local potential minimum. On the other hand, this is one of the few problems amenable to exact solution, by the use of path integrals or diagonalization of the Hamiltonian. Thus, we can make comparison and assess the validity of the master equation derived in the previous chapter. This comparison is essential before we solve for systems with no exact solution.
\\

We will first show how to cast the master equation into a set of recurrence relations and how to solve them. After which we will study the equilibrium and driven properties of a damped quantum \HO.

\section{Method of Solution}
\subsection{Coefficients of the Master Equation}
The eigen-energies of a quantum \HO\ are equally spaced,
 \begin{eqnarray}
 	\epsilon_n= (n+\small{\frac{1}{2}})\hbar \omega_0; \,\,\,\,\,\,\,\,\Delta_{nm}= (n-m)\hbar \omega_0.
\end{eqnarray}
The matrix elements of the coupling \Ha\ ($F=x$) entering the relaxation term (\ref{c5_relaxation}) are
\begin{eqnarray} 
	L^{+}_{n} &=&\sqrt{\frac{\hbar}{2M\omega_0}} \, \sqrt{n+1}; \\
	L^{-}_{n}  &=& \sqrt{\frac{\hbar}{2M\omega_0}} \, \sqrt{n}. \\ \nonumber
\end{eqnarray}
\subsection{Casting the Master Equation into Recurrence Form}
We can construct vectors from the columns of the reduced density matrix (truncated at the $N^{\rm{th}}$ level),
%
\[\textbf{c}_n= \left( \begin{array}{ccccccc}
         \rho_{0n}    \\
         \rho_{1n}    \\
         \rho_{2n}    \\
         \vdots           \\
          \rho_{N,n} \\
 \end{array} \right),\] 
 and the master equation acquires the following recurrence form
 \begin{eqnarray}   
		\dot{\textbf{c}}_n =\hat{\textbf{Q}}_{n}^{-}\,\textbf{c}_{n-1}+\hat{\textbf{Q}}_{n}\,\textbf{c}_{n}+\hat{\textbf{Q}}_{n}^{+}\,\textbf{c}_{n+1}\,\,, \label{c6_3RR}
\end{eqnarray} 
where the elements of the matrices $\hat{\textbf{Q}}$'s are
\begin{eqnarray}   
	(\hat{\textbf{Q}}_{n})_{m,m}   \,\,\,\,   &=&  \frac{\iu}{\hbar}\Delta_{nm} -\frac{1}{\hbar^2}\Big( L_{n}^{+}L_{n+1}^{-} \TranUn^{*}_{n+1, n} + 															 L_{n-1}^{+}L_{n}^{-}\TranUn^{*}_{n-1, n} + L_{m}^{+}L_{m+1}^{-} \TranUn_{m+1, m} +  												      	L_{m-1}^{+}L_{m}^{-} \TranUn_{m-1, m}\Big);  \nonumber \\
	(\hat{\textbf{Q}}_{n}^{+})_{m,m+1}  &=& \frac{1}{\hbar^2}\Big(\TranUn^{*}_{n,n+1} + \TranUn_{m,m+1}  \Big)L_{n}^{+} L_{m+1}^{-}; \nonumber \\
	(\hat{\textbf{Q}}_{n}^{-})_{m,m-1}   &=& \frac{1}{\hbar^2} \Big(\TranUn^{*}_{n,n-1} + \TranUn_{m,m-1}  \Big)L_{n}^{-} L_{m-1}^{+}. 
\end{eqnarray} 
%
\subsection{Implementation}
The truncation level $N$ has to be chosen such that ${N\hbar \omega_0/k_B T} \gg 1 $, and the energy levels higher than $N$ become irrelevant. When we solve for systems with finite levels (e.g. spin problems \cite{jose2}), there is no need to perform truncation and the recurrence relation is exact.
\\

We are happy when we see the 3-term reccurrence relation, since we know how to solve it with the \CF\ method. To obtain the solution, we first need the ``seed" $\textbf{c}_0$ so that all other $\textbf{c}_n$'s can be calculated by the equation (cf. Appendix A.1)
\begin{eqnarray}
	\textbf{c}_n=\hat{\textbf{S}}_n \textbf{c}_{n-1}+\textbf{a}_n. \label{c6_ansatz}
\end{eqnarray}
But there is a problem in solving for the stationary solution ($\dot{\textbf{c}}_n=0$). Eq.~(\ref{c6_3RR}) is a set of homogeneous equations where the solution involves a multiplicative constant. Unlike the classical problems, we cannot obtain the ``seed" $\textbf{c}_{0}$, from the normalization condition $ \mbox{Tr}[\rho]=1$, as it involves all the vectors $\textbf{c}_{n}$'s. This problem can circumvented by fixing one of the matrix elements in $\textbf{c}_{0}$. The first vector obeys
 \begin{eqnarray}   
		\hat{\textbf{A}}\,\textbf{c}_0=(\hat{\textbf{Q}}_{0}+\hat{\textbf{Q}}_{0}^{+} \hat{\textbf{S}}_{1})\textbf{c}_{0} =0,
\end{eqnarray} 
where  $\hat{\textbf{S}}_{1}$ is some function of the  $\hat{\textbf{Q}}$'s. We provide an extra equation by requiring the first element of $\textbf{c}_{0}$ to be 1, getting a set of over-determined equations
%
\[ \left( \begin{array}{ccccccc}
         \hat{\textbf{A}} \\
        \textbf{d}  \\
 \end{array} \right)\, \textbf{c}_0=\textbf{b},\\ \]
 where
\[\textbf{d}= ( 
      \,1 ,\,\, 0,\,\, 0, \,\, .\,\, . \,\, .\,\, .\,\, .\,\, . \,\, .\,\, .  \,\, 0\,
),
 \quad \quad
\textbf{b}= \left( \begin{array}{ccccccc}
         0    \\
         0	\\
         \vdots \\
         0 \\
         1   \\
\end{array} \right).\]
We can solve this set of equations by using the least square method. The subsequent $\textbf{c}_n$'s can then be generated by the upward iterations Eq. (\ref{c6_ansatz}) in the \CF\ method. Eventually, we will need to normalize the density matrix by the operation $\rho=\rho/\mbox{Tr}[\rho] $.
\\

As opposed to the classical problems, the indexed structure is automatically given by the energy levels, it saves us the troubles of choosing an appropriate basis function and manipulating to get the recurrence structure. The price to pay is the extra effort in getting the seed. The master equation can actually be solved by inverting a matrix of dimension $N^2 \times N^2$, which involves computational efforts of $O(N^6)$. In using the \CF\ method, all the matrices are of dimension $N\times N$, thus we have reduced the complexity to $O(N^3)$ with $N$ iterations. This allows us to reach the high temperature regime, where high energy levels are excited and $N$ becomes large. In the next few sections, we will use the \CF\ method to study the equilibrium properties and the response to applied fields.

\section{Equilibrium Results: Dispersion of Coordinate and Momentum}
\begin{figure}[h!] 
\begin{center}
	\includegraphics[scale=0.50]{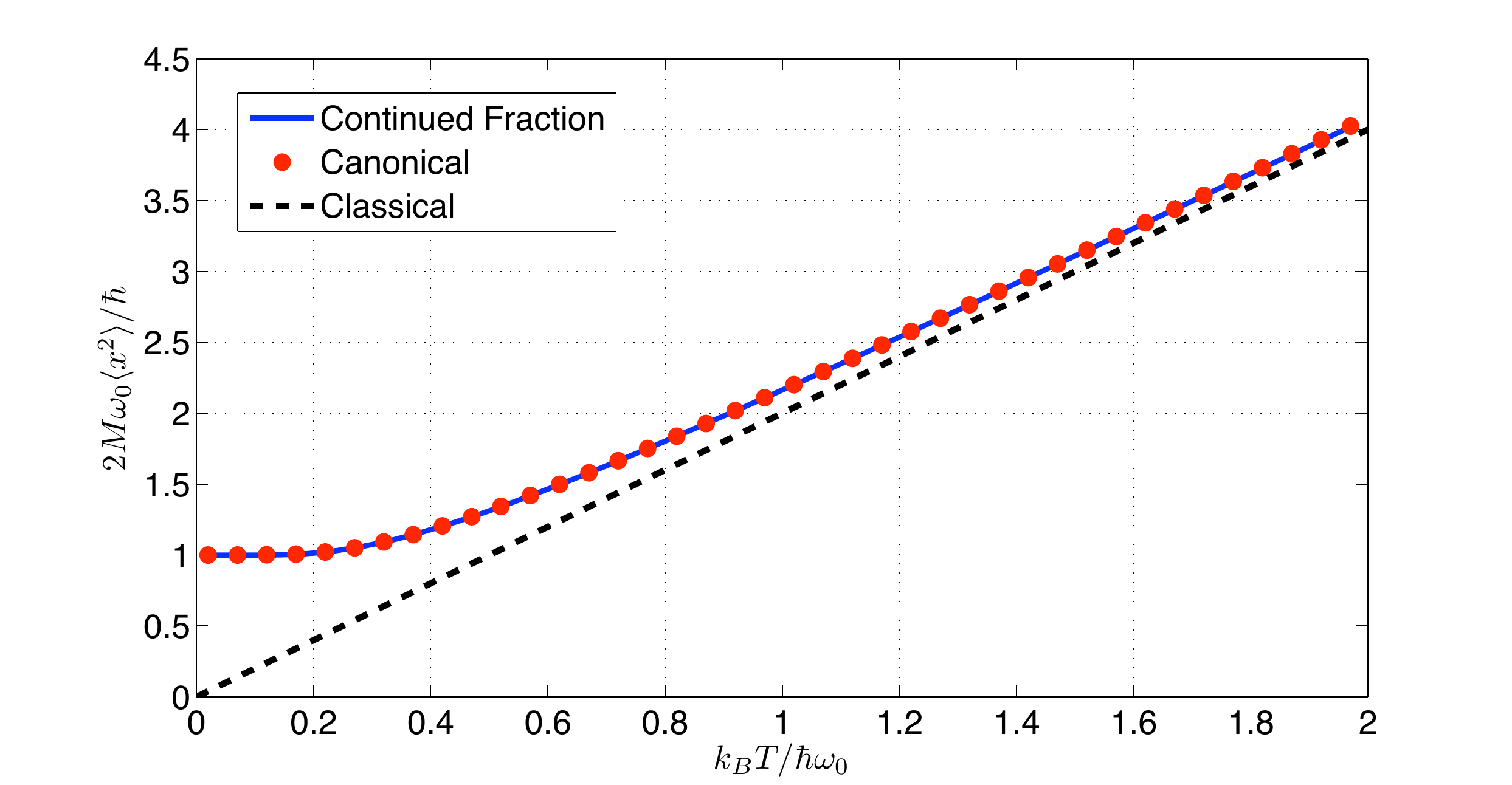}	
	\includegraphics[scale=0.50]{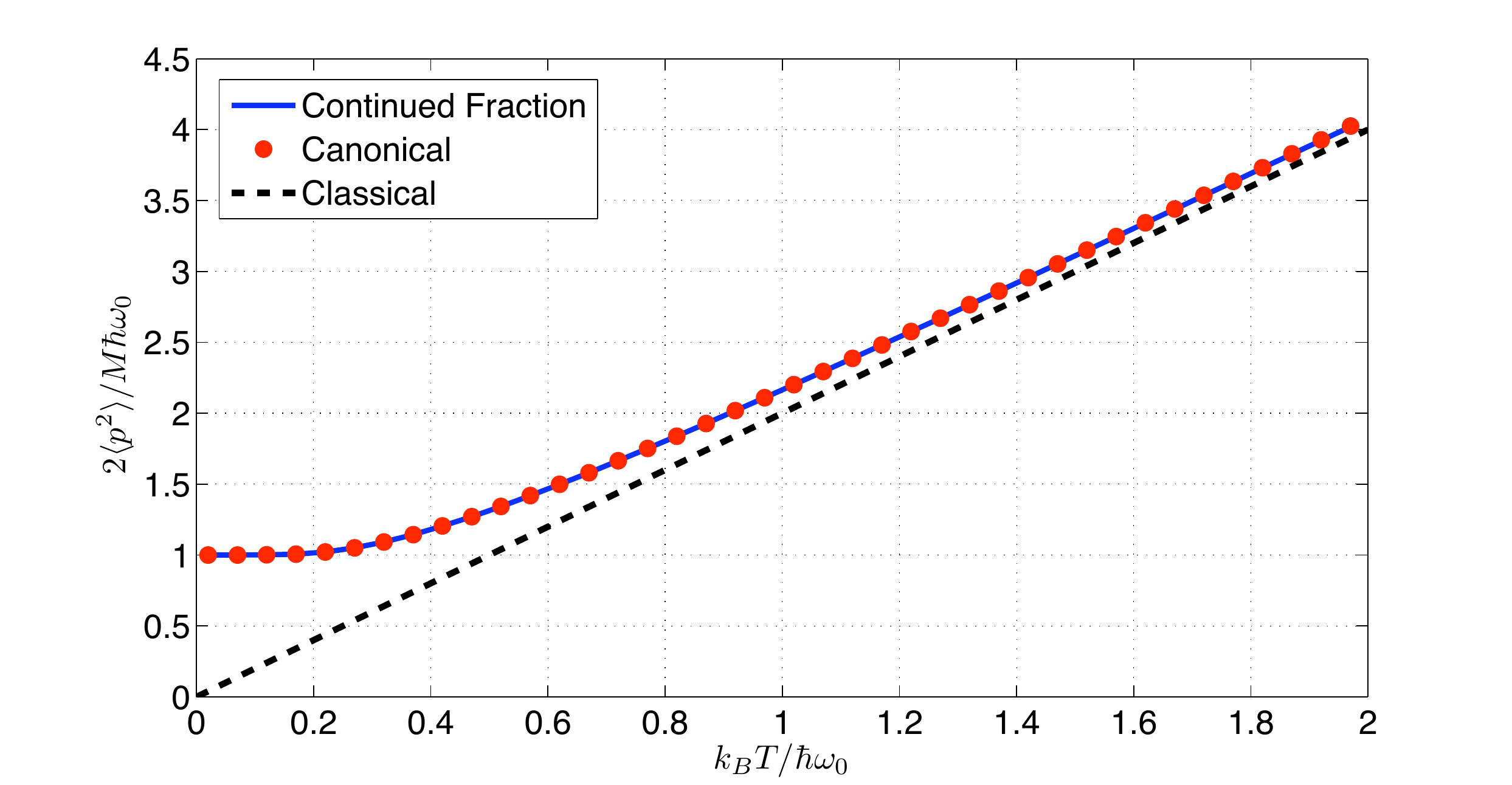}
	\caption{The dispersion curves of the coordinate and momentum. The numerical results coincide with the canonical distribution. At high temperature, both quantities approach the classical limit. }
	\label{fig:c6_dispersion}
\end{center}
\end{figure}

The equilibrium solution ($\dot{\textbf{c}}_n=0$) is found to be the canonical distribution
 \begin{eqnarray}   
		\rho_{\rm{eq}}= \frac{\mbox{e}^{-\beta \,\sysHa}}{Z},
\end{eqnarray}
independent of the coupling strength $\gamma$\footnote{ 
Though the numerical solution does not depend on the coupling strength, the exact solution says otherwise \cite{weiss}: it is canonical only in the limit of $\gamma\rightarrow 0$. For any finite $\gamma$, the exact solution is different from the canonical distribution, and the correction is of the order of $\gamma$ \cite{david}. To explain the discrepancy, one should recall the approximations we have made: the secular approximation and the weak coupling approximation. Because of these two approximations, we always obtain the canonical distribution as the stationary solution. However, the correction is small within the weak coupling regime.
}.
 Some of the interesting thermal-equilibrium quantities are the mean square of the coordinate and momentum, plotted in Figure \ref{fig:c6_dispersion}. 
The results from the canonical distribution are
\begin{eqnarray} 
	\frac{2M \omega_0}{\hbar}\langle x^2\rangle=\frac{2}{M\hbar \omega_0}\langle p^2\rangle=\coth(\beta \hbar \omega_0/2).
\end{eqnarray} 
 At high temperature, both quantities grow linearly with the temperature. The system behaves classically at high temperature and follows the equipartition law. At $T=0$, they reach the standard quantum limit $\sqrt{\langle x^2\rangle\langle p^2\rangle}=\hbar/2$, as opposed to zero in the classical case, a manifestation of the zero point energy.

Therefore, we have achieved our goal of reproducing the dispersion curves by solving the quantum master equation with the \CF\ method. The results are in full agreement with the statistical mechanics, using a truncation level of $N=30$.

\section{Response to Static Field}
The master equation after the secular approximation is of Pauli type: it involves only the diagonal terms. In order to check our handling of the off-diagonal structure, we apply a static force, $F_0$, to the harmonic oscillator. We assume the applied force is small (of the order of $\gamma$), such that we can ignore the change to the relaxation term.  $R_{mn}$ is already of the order of $\gamma$, thus, any modification is at least of the order of $\gamma^2$ and can be discarded. We just have to alter the free \Ha, the \EoM\ of the Hubbard operator then reads
\begin{eqnarray} 
	\frac{d X_{nm}}{dt} =\frac{\iu}{\hbar}\Big[H-x\,F_0, X_{nm}\Big]+\hat{R}_{nm}\,\,.  \label{c6_Heisenberg_forced}
\end{eqnarray}
The corresponding master equation is
\begin{eqnarray} 
	\frac{d \rho_{mn}}{dt}&=&\frac{\iu}{\hbar}\Delta_{nm} \rho_{mn} +D_{mn} +R_{mn}\,\, ,  \label{c6_master_forced}
\end{eqnarray}
where $R_{mn}$ remains the same as Eq.~(\ref{c5_relaxation}) and 
\begin{eqnarray} 
		D_{mn}=\frac{\iu \,F_0}{\sqrt{M\hbar \omega_0}} &\Big[&\sqrt{\frac{(m+1)}{2}}\rho_{m+1,n} + \sqrt{\frac{m}{2}}\rho_{m-1,n}\\ \nonumber 
			          	&&-\sqrt{\frac{(n+1)}{2}}\rho_{m,n+1} - \sqrt{\frac{n}{2}}\rho_{m,n-1}	\Big]		
\end{eqnarray}
gives the off-diagonal structure. The addition of the DC force does not break the short-ranged coupling of the matrix elements, we are still able to solve Eq.~(\ref{c6_master_forced}) with the \CF\ method.
\\

\begin{figure}[h!] 
\begin{center}
	\includegraphics[scale=0.45]{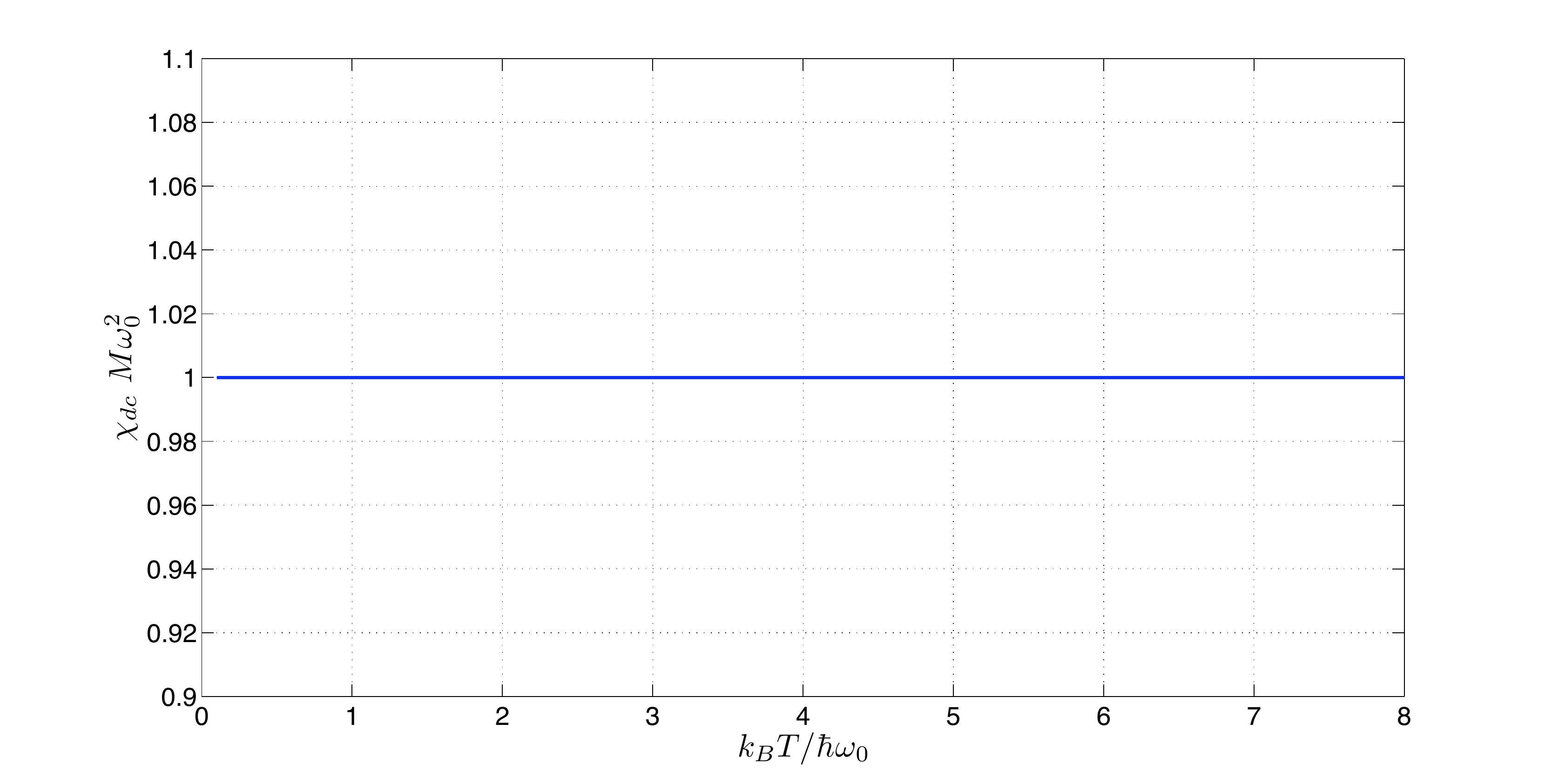}	
	\caption{The DC susceptibility at $\frac{F_0}{\sqrt{M\hbar \omega_0^3}}=0.0001$, $\gamma/ \omega_0=0.01$ and $\omega_D/ \omega_0=10$. It is independent of the temperature.}
	\label{fig:c6_dc_sus}
\end{center}
\end{figure}

The DC response is characterized by the DC susceptibility defined as
 \begin{eqnarray}   
		\chi_{dc}= \langle x\rangle/F_0. 
\end{eqnarray}
Let us first see how this quantity behaves classically. The Langevin equation of a forced oscillator is 
\begin{eqnarray} 
		\ddot{x}+\gamma \dot{x} = \xi(t) +F_0-M\omega_0^2\,x. 
\end{eqnarray}
At equilibrium, both terms on the left hand side vanish. Taking average, one finds
\begin{eqnarray} 
		\chi_{dc}=  \langle x\rangle/F_0= \frac{1}{M\omega_0^2}.
\end{eqnarray}
which is independent of the temperature. As the system is linear, we expect this to hold in the quantum problem. Figure \ref{fig:c6_dc_sus} shows that this is indeed the case, and it serves as a check of our handling of the off-diagonal structure of the quantum master equation.
%
\section{Linear Response to Time-Dependent Field}
\subsection{Perturbative Chain of Equations}
As we did in the classical problems, we apply a small AC field, $F_d \cos(\Omega t)$, and study the system's response characterized by the AC susceptibility
\begin{eqnarray} 
		 \langle x\rangle =  \frac{F_d}{2}\Big(\chi_{ac}\, \mbox{e}^{+\iu\Omega t} +\chi_{ac}^{*}\, \mbox{e}^{-\iu\Omega t}  \Big). \label{c6_sus}
\end{eqnarray}
To obtain the new master equation, we just need to replace $F_0$ in  Eq.~(\ref{c6_Heisenberg_forced}) and Eq.~(\ref{c6_master_forced}) by $F_d \cos(\Omega t)$. Assuming the driving force is small, the system is only slightly perturbed from its equilibrium state. We can split the matrix elements into time-independent and time-dependent parts
\begin{eqnarray} 
		\rho_{mn}= \rho_{mn}^{(0)}+\frac{F_d}{2} \Big( \rho_{mn}^{(+1)} e^{+\iu \Omega t} +   \rho_{mn}^{(-1)} e^{-\iu \Omega t}\Big), 
\end{eqnarray}
and solve the recurrence relations for zeroth order and first order in $F_d$ sequentially.
\begin{eqnarray} 
		\mbox{Zeroth Order:}&&\,\,\,\,\,\,\,\,\,\,\,\, \frac{\iu}{\hbar} \Delta_{nm}\, \rho_{mn}^{(0)} +R_{mn}^{(0)} =0,  \\
		\mbox{First Order:}&&\,\,\,\,\,\,\,\,\,\,\,\, \iu \Big(\Delta_{nm}/\hbar -\Omega \Big)\rho_{mn}^{(1)}+R_{mn}^{(1)} = -D_{mn}^{(0)}/F_d.
\end{eqnarray} 
The superscript in $R_{mn}$ and $D_{mn}$ indicates the order of density matrix element that should be used in the expression. The solution of the zeroth order enters into the first order equation, on its right hand side. Therefore, these equations have to be solved sequentially. 
%
\subsection{AC Susceptibility Curves (Dispersion and Absorption)}
\begin{figure}[h] 
\begin{center}
	\includegraphics[scale=0.57]{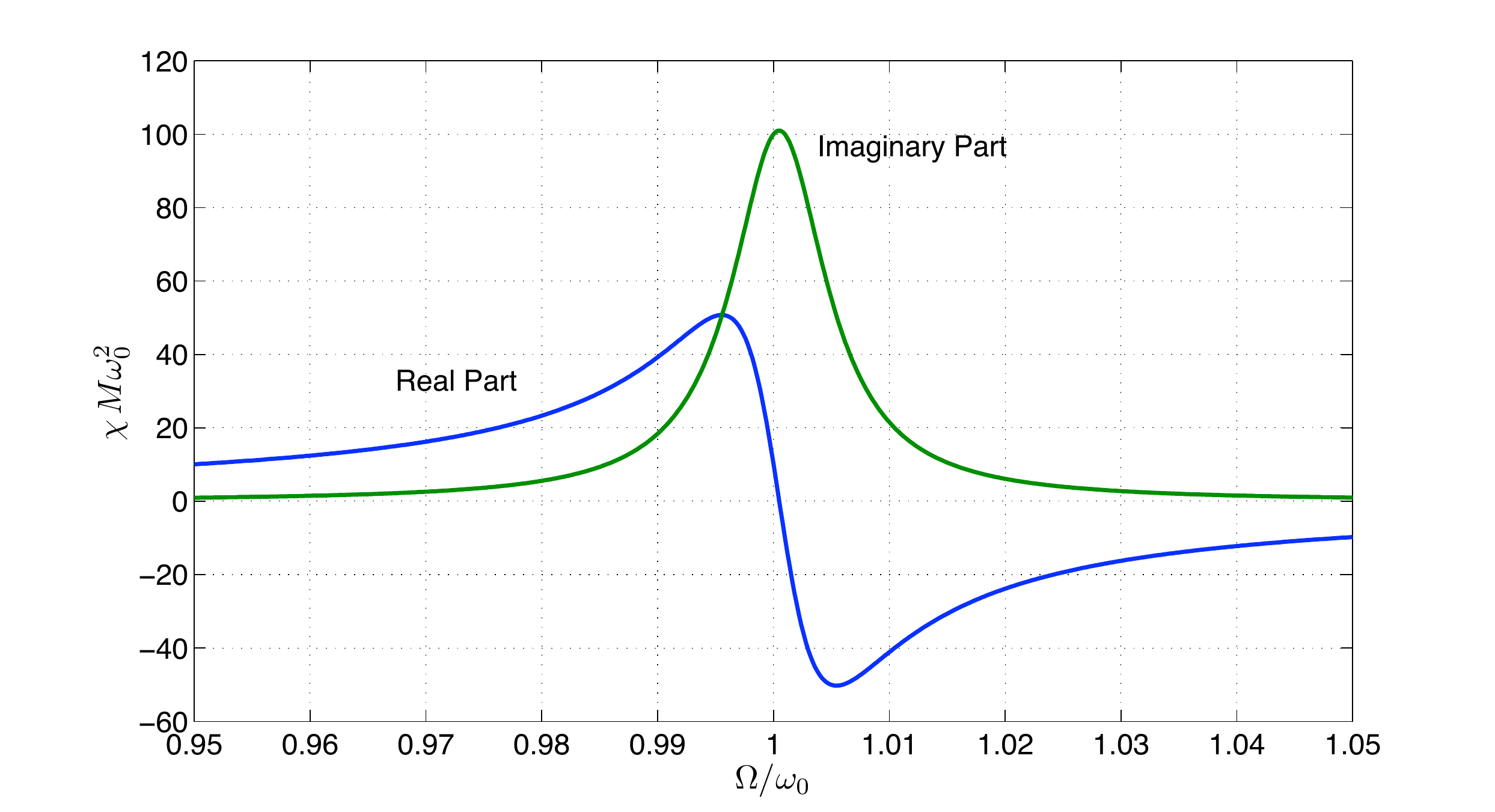}	
	\caption{The AC susceptibility at $\frac{k_BT}{\hbar \omega_0}=1$, $\gamma/\omega_0=0.01$ and $\omega_D/\omega_0=10$. Resonant behavior occurs around $\Omega/\omega_0 \approx1$.}
	\label{fig:c6_ac_sus}
\end{center}
\end{figure}

\begin{figure}[h!] 
\begin{center}
	\includegraphics[scale=0.57]{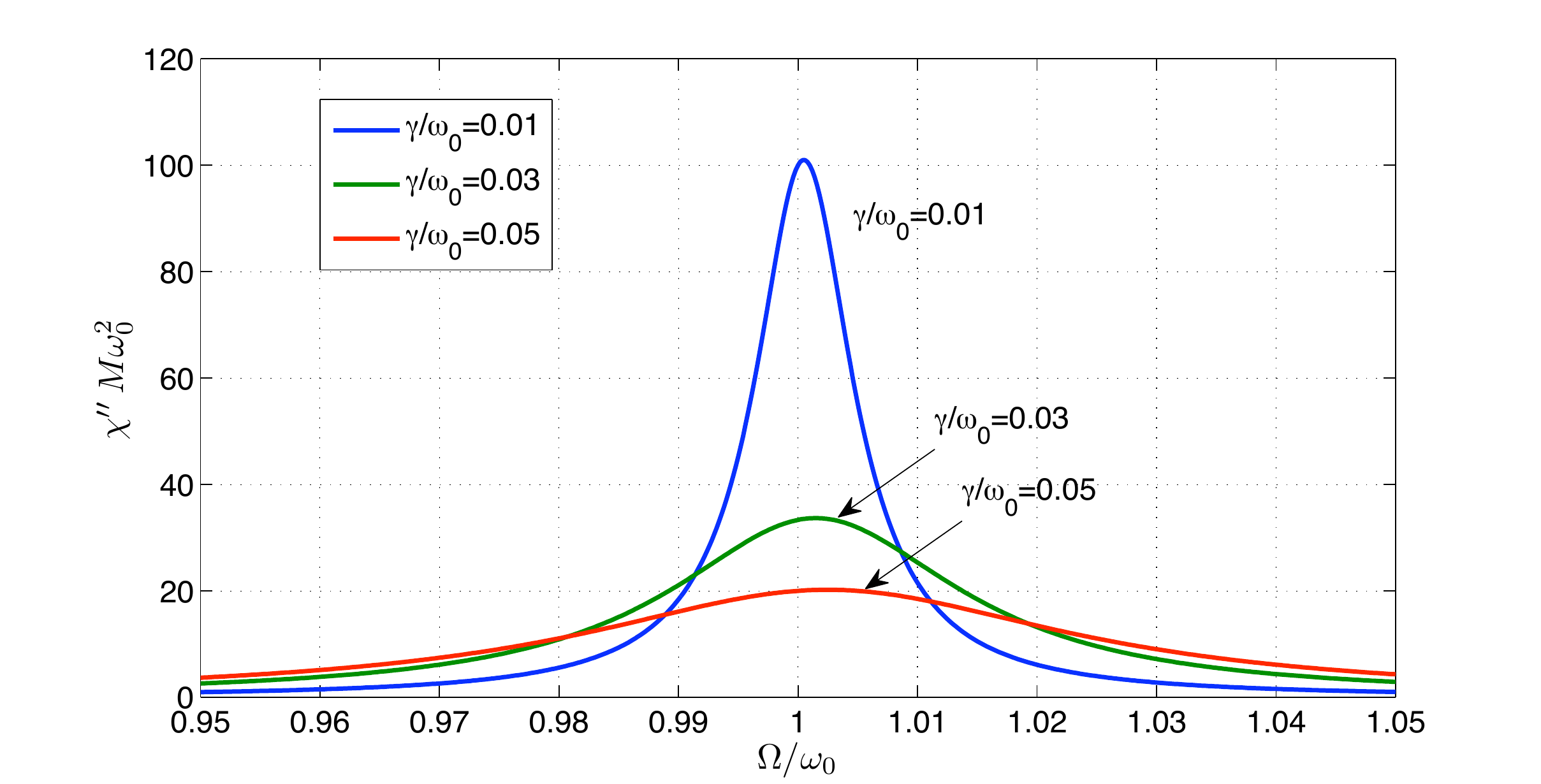}	
	\caption{The imaginary part of the AC susceptibility at $\frac{k_BT}{\hbar \omega_0}=1$ and $\omega_D/\omega_0=10$. The absorption peaks flatten when the damping is increased.}
	\label{fig:c6_ac_sus_gamma}
\end{center}
\end{figure}

The AC susceptibility is plotted in Figure \ref{fig:c6_ac_sus} and we observe resonant behavior at $\Omega \approx \omega_0$. The imaginary (absorptive) part is of Lorentzian shape with peak at around $\omega_0$. At resonant frequency, the real part becomes zero because the system is oscillating  completely out of phase with the driving ($ \langle x\rangle \propto \sin{\Omega t}$),  as one can see from Eq. (\ref{c6_sus}) by setting $\chi'=0$. The dependence of the absorptive part on the damping strength can be seen in Figure \ref{fig:c6_ac_sus_gamma}. The peaks become flattened and shifted to the right when the damping strength is increased. This dependence is useful if we wish to estimate the damping strength by measuring the height or the full width half maximum of the absorption peaks. 

These results are in good agreement with the analytical result obtained by stochastic modeling \cite{weiss}
\begin{eqnarray} 
		\chi=\frac{1}{M}\frac{1}{\omega_0^2 -\omega^2-\iu\,\omega \,\tilde{\gamma}(\omega)}, \quad \,\,\,  
		\tilde{\gamma}(\omega)=\frac{\gamma}{1-\iu \,\omega /\omega_D}.
\end{eqnarray}
Since the agreement is nearly perfect, we have not shown the analytical curves in the figures above.
%

\subsection{Secular Approximation Revisited}
As one can see from the susceptibility curves, the dominant contribution comes from the region around the resonant frequency. We can relate it with the secular approximation we have made earlier. We keep the terms with resonant frequency and discard the off-resonant terms since their contribution is minute. As one increases the damping, the curves flatten and the contribution from the region away from the resonant frequency becomes important as well. This means the secular approximation is less accurate at large damping. Actually, the error introduced by the secular approximation is of the order of $\gamma^2$ \cite{hanggi}. A thorough investigation of the secular approximation in the Jaynes-Cummings model   (two-level system in one-mode electromagnetic field) in a dissipative bath can be found in Ref. \cite{rau}.

\section{Discussion}
To conclude, we have used the master equation of Chapter 5 to solve for the damped quantum \HO\ and discussed some of the implementation issues. We studied its thermodynamical properties with and without constant field. After which we investigated the response to a time-dependent field and obtained the absorption-dispersion curves. The results are consistent with the exact results, provided $\gamma/\omega_0 \ll 1$.
\\

This agreement gives us confidence on our handling of the quantum master equation and our implementation of the \CF\ method. It is important as this efficient method is independent of the approximations used to get the quantum master equation, so our implementation could be extended to master equation with improved approximations. Possible extensions include the removal of the secular approximation or the inclusion of higher order terms in the system-bath coupling. These will break the 3-term recurrence structure, and more efforts are needed to solve them. The investigation is still ongoing and we shall leave them out of the thesis. 

\chapter{Summary}
To end the thesis, let us summarize our main results:

\begin{itemize}

	\item In the classical regime, we reviewed how the fluctuation and dissipation of an open system can be explained by modeling the bath as a set of harmonic oscillators. Two equivalent approaches were presented to study classical open systems: the trajectory approach (Langevin equation) and the distribution function approach (\FP\ equation). The use of the \CF\ method in solving \FP\ equations was also discussed.
	
	\item We then demonstrated the use of the \CF\ method in solving the \FP\ equations for particle in a periodic potential and classical dipole, both in high friction limit. We solved for both equilibrium and time-dependent solutions. The time dependent solution is obtained at arbitrary DC and AC fields. At large AC field, we observed significant deviation from the linear response results. The non-linear effects are reflected in the distortion of the shapes of the dynamical hysteresis loops.
\\
	
	Previous studies of the particle problem focused on the zeroth and first harmonic susceptibilities; and for the dipole the first few harmonics but in the limit of weak driving. Here we obtained all the harmonics, for any driving strength and biasing DC field. We interpreted their characteristic features and presented all the information in a compact way with the dynamical hysteresis loops;  this was not discussed in the literatures before. The assessment of the perturbative approach versus exact approach is also new for both problems. This assessment can be of valuable methodological interest. 
	
	\item In the quantum regime, we reviewed the derivation of the quantum master equation for the reduced density matrix, and its application to the bath-of-oscillators model.  Approximations and subtleties of the master equations were also discussed. 
	
	\item We went on to solve the master equation of a damped quantum \HO\ using the \CF\ method. This problem allows us to make comparison with exact results and assess the validity of the master equation. We investigated both the equilibrium and time-dependent solutions. Driven systems are more challenging in the quantum case, so we are content with implementing a linear response treatment. These results are in good agreement with the exact results (when available) under the condition $\gamma/ \omega_0 \ll 1$. 
\\
	
This showed that we successfully implemented the efficient \CF\ method in solving the master equation. This method reduces the computational complexity significantly and allows us to solve for systems with many levels. In fact, the problem of damped quantum \HO\ constitutes the first attempt in using the \CF\ method to solve for a mechanical system (it was originally proposed and tested for spin systems with finite number of levels). This widens the application range of this approach, and opens doors for further studies.

	
\end{itemize} 
\appendix
\chapter{Appendices}
\section{Continued-Fraction Method}
Here we give a summary on solving the 3-term recurrence relation of the form
\begin{eqnarray} 
	Q_n^{-}c_{n-1}+Q_nc_{n}+Q_n^{+}c_{n+1}= -f_n\,, \,\,\,\,\,\,\,\, n=0,1,2,3,..... \,\,\, , \label{a1_3rr}
\end{eqnarray}
with the continued fraction method. The coefficients $Q$'s and the inhomogeneous part $f$'s are some known constants. Risken \cite{risken} introduced the following ansatz
\begin{eqnarray} 
	c_n=S_n c_{n-1}+a_n, \label{a1_ansatz}
\end{eqnarray}
and obtained the relations
\begin{eqnarray}  
	S_{n} = -\frac{Q^{-}_{n}}{Q_{n}+Q^{+}_n S_{n+1}}; \,\,\,\, a_{n} = -\frac{Q_n^{+} a_{n+1}+f_n}{Q_n+Q^{+}_n S_{n+1}}. \label{a1_down_ite}
\end{eqnarray}
For finite recurrence, $c_{n \ge N}=0$ for some $N$. We can enforce this by setting $S_N=0, \,\, a_N=0$, and generate all the other $S_n$'s and $a_n$'s by the downward iteration Eq. (\ref{a1_down_ite}). To obtain all $c_n$'s, we only need the ``seed'' $c_0$, which can be obtained by solving
\begin{eqnarray} 
	(Q_0 +Q_1^{+}S_1)c_{0}= -(f_0+Q^{+}_{0}a_1).
\end{eqnarray}
Other $c_n$'s can then be generated by the relation Eq. (\ref{a1_ansatz}). In the case of homogeneous equation ($f=0$), we have to obtain $c_0$ by other means, i.e. normalization of distribution.
\\

As for the name of the method, note that $S_n$ is expressed in terms of $S_{n+1}$ in the denominator, which can be in turn written in terms of $S_{n+2}$ and so on. It produces the continued-fraction structure
\begin{eqnarray} 
	S=\frac{p_1}{q_1+\frac{p_2}{q_2+ \cdots}}.
\end{eqnarray}

When the quantities in the recurrence relation are scalar, we call this the \emph{scalar \CF} method. However, this method also applies to vectors recurrence relation, we then talk about \emph{matrix \CF}. In such case, $c_n$, $f_n$ and $a_n$ become vectors, while $Q_n$ and $S_n$ are matrices. The inversion in Eq. (\ref{a1_down_ite}) then becomes matrix inversion from the left ($A/B = B^{-1}A$). 
\\

In fact, the recurrence relation can be treated as a set of linear equations, and solved by inverting a matrix of dimension $N\times N$ (for scalar recurrence relation). The operation involves complexity of $O(N^3)$. The use of \CF\ method reduces the complexity to $O(N)$, and allows us to handle a much larger system of equations. 

\section{Hubbard Operators}
Here we discuss some of the properties of the Hubbard operators  $X_{nm}=|n \rangle \langle m |$.  They form a complete set, and one can think of $X_{nm}$ as a matrix with zeros everywhere, except 1 at the position $(n,m)$. Some of the useful properties are

 \begin{itemize}
 
	\item Any operator can be expressed in terms of the Hubbard operators:
		\begin{eqnarray}	
			A=\sum_{nm}A_{nm} X_{nm}, \,\,\,\,\, \mbox {where}\,\,\, A_{nm}=\langle n|A|m\rangle.
		\end{eqnarray}
		
	\item Equal-time relation
		\begin{eqnarray}	
			X_{nk}X_{lm}=\delta_{kl}X_{nm}.
		\end{eqnarray}
	 
	 \item Commutator
		\begin{eqnarray}	
			[X_{nk},X_{lm}]= \delta_{kl}X_{nm}- \delta_{mn}X_{lk}.
		\end{eqnarray}
		
	 \item Adjoint
		\begin{eqnarray}	
			(X_{nm})^{\dagger}= X_{mn}.
		\end{eqnarray}
	
	\item Relation with the density matrix 
		\begin{eqnarray}	
			\text{Tr}[\rho \,X_{nm}]=\langle  X_{nm} \rangle =\rho_{mn}.
		\end{eqnarray}
\end{itemize}

Using the eigenstates of the system, the Heisenberg equation of the Hubbard operator becomes
\begin{eqnarray}	
			\frac{d X_{nm}}{dt}= \frac{\iu}{\hbar}\Delta_{nm} X_{nm},
\end{eqnarray}
where ${\Delta_{nm}=\epsilon_n - \epsilon_{m}}$.

\section{Transition Rate}
To obtain the expression Eq. (\ref{c5_trans-rate}), one needs the identity
\begin{eqnarray}	
			\int^\infty_{0} \frac{x\,\coth(\frac{1}{2}\tau x)}{(x^2+A^2)(x^2+B^2)}dx =
			\frac{1}{A^2-B^2}\Big[\psi\Big(\frac{\tau A}{2\pi}\Big)-\psi\Big( \frac{\tau B}{2\pi}\Big)\Big] -\frac{\pi}{\tau} \frac{1}{AB(A+B)}, \label{app_psi_iden}
\end{eqnarray}
in evaluation of the imaginary part. The Psi function, defined as $\psi(x)=\frac{d}{dx}\mbox{ln}\Gamma(x)$,  has the following properties \cite{table-integral}:
\begin{eqnarray}	
			\psi(x+1)&=&\psi(x)+\frac{1}{x};\\
			\psi(1)	&=& -C \,\,\,\,\,\,\, \mbox{where the Euler number}\,\,\, C=0.577 215;\\
			\psi(x)&=&\mbox{ln}\,x-\frac{1}{2x}-2\int^\infty_{0} \frac{t}{(t^2+x^2)(\mbox{e}^{2\pi t}-1)}dt.
\end{eqnarray}
From the last expression with an integral, we obtain Eq.~(\ref{app_psi_iden}) upon a partial fraction expansion of the denominator
\begin{eqnarray}	
		\frac{1}{(x^2+A^2)(x^2+B^2)}=\frac{1}{B^2-A^2}\Big[\frac{1}{x^2+A^2}-\frac{1}{x^2+B^2}     \Big].
\end{eqnarray}
\\

\begin{figure}[h!]  
	\centering
		\includegraphics[scale=0.45]{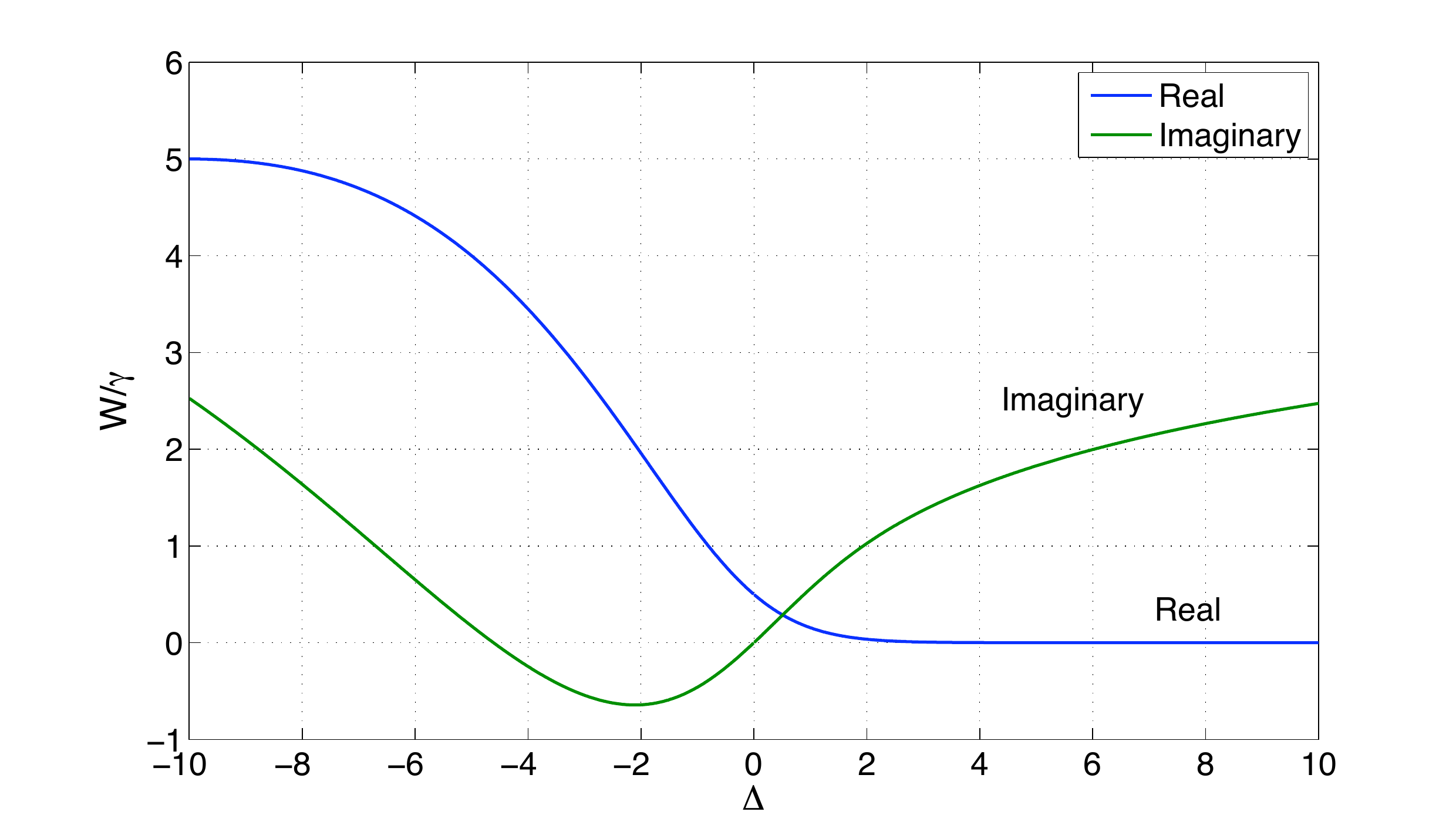}
		\includegraphics[scale=0.45]{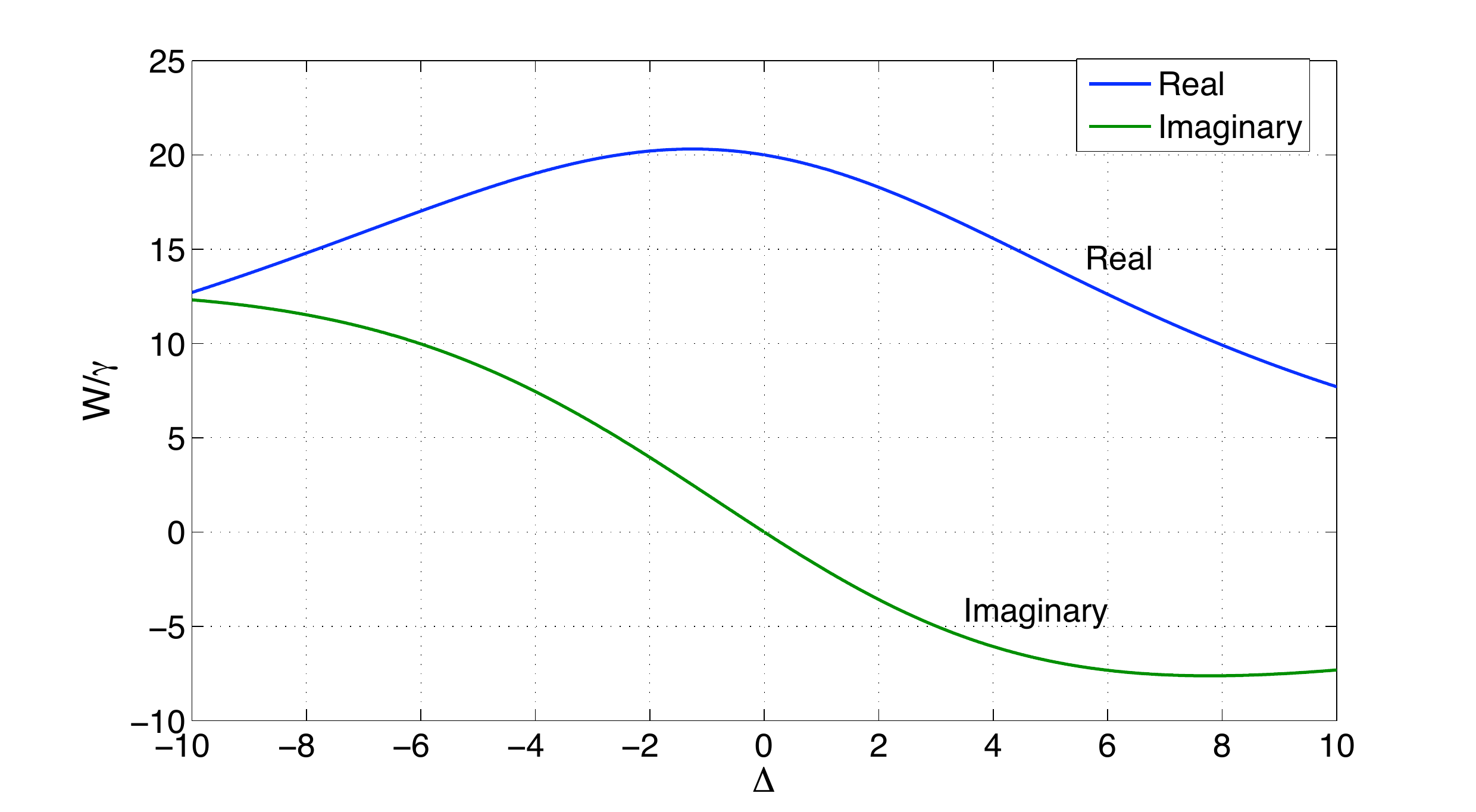}
		\caption{In arbitrary unit, the transition rates at low temperature (top, $\beta=2$) and high temperature (bottom, $\beta=0.05$) with cut-off frequency $\hbar \omega_D=10$.}
	\label{fig:append-trans-rate}
\end{figure}

The real part and imaginary part of the transition rate are plotted in Figure \ref{fig:append-trans-rate}. At low temperature, the real part decreases monotonically and is negligible at positive $\Delta$, indicating that the process is dominated by de-excitation. At high temperature, the contribution from both excitation and de-excitation are important.

\bibliographystyle{prsty}	
\bibliography{myrefs}		
\end{document}